\documentclass{aa}
\usepackage{graphicx}
\usepackage{xcolor}
\usepackage[position=top]{subfig}
\usepackage[fleqn]{amsmath}
\DeclareMathOperator{\sign}{sign}
\usepackage[utf8]{inputenc}
\usepackage{epsfig,amsmath,amssymb}
\usepackage[varg]{txfonts}
\usepackage{multirow}

\usepackage{tcolorbox}

\definecolor{myblue}{rgb}{0.00, 0.0, 0.9}
\definecolor{myred}{rgb}{0.90, 0.0, 0.0}
\definecolor{mygreen}{rgb}{0.0, 0.7, 0.0}
\usepackage[breaklinks,  citecolor=myblue, linkcolor=myred, urlcolor=purple, colorlinks=true, debug, baseurl=' ']{hyperref}
\usepackage{orcidlink}

\definecolor{vincent}{rgb}{0.1, 0.0, 0.7}
\definecolor{raphael}{rgb}{0.7, 0.1, 0.5}

\def \HI{H\,{\sc i}}

\titlerunning{Tomography of the dust polarization sky}

\authorrunning{Pelgrims et al.}

\begin{document}

%

%
\title{
The first degree-scale starlight-polarization-based tomography map of the magnetized interstellar medium
\thanks{Table 1 is only available in electronic form at the CDS via anonymous ftp to cdsarc.u-strasbg.fr (130.79.128.5) or via http://cdsweb.u-strasbg.fr/cgi-bin/qcat?J/A+A/.}}

\author{V. Pelgrims
      \inst{1,2,3}\fnmsep\thanks{vincent.pelgrims@ulb.be}\orcidlink{0000-0002-5053-3847},
      N.~Mandarakas\inst{2,3}\orcidlink{0000-0002-2567-2132},
      R.~Skalidis\inst{4}\orcidlink{0000-0003-2337-0277},
      K. Tassis\inst{2,3}\orcidlink{0000-0002-8831-2038},
      G.~V.~Panopoulou\inst{5}\orcidlink{0000-0001-7482-5759},
      V.~Pavlidou\inst{2,3}\orcidlink{0000-0002-0870-1368},
      D.~Blinov\inst{2,3}\orcidlink{0000-0003-0611-5784},
      S.~Kiehlmann\inst{2,3}\orcidlink{0000-0001-6314-9177},
      S.~E.~Clark\inst{6,7}\orcidlink{0000-0002-7633-3376},
      B.~S.~Hensley\inst{8}\orcidlink{0000-0001-7449-4638},
      S.~Romanopoulos\inst{2,3}\orcidlink{0000-0002-2897-2448},
      A.~Basyrov\inst{9},
      H.~K.~Eriksen\inst{9},
      M.~Falalaki\inst{2,3},
      T.~Ghosh\inst{10,11},
      E.~Gjerl{\o}w\inst{9},
      J.~A.~Kypriotakis\inst{2,3},
      S.~Maharana\inst{12}\orcidlink{0000-0002-7072-3904},\\
      A.~Papadaki\inst{2,3,13},
      T.~J.~Pearson\inst{4}\orcidlink{0000-0001-5213-6231},
      S.~B.~Potter\inst{12,14}\orcidlink{0000-0002-5956-2249},
      A.~N.~Ramaprakash\inst{1,4,11},\\
      A.~C.~S.~Readhead\inst{4}\orcidlink{0000-0001-9152-961X},
      and
      I.~K.~Wehus\inst{9}
}
          
\institute{
Universit{\'e} Libre de Bruxelles, Science Faculty CP230, B-1050 Brussels, Belgium
\and
Institute of Astrophysics, Foundation for Research and Technology-Hellas, N. Plastira 100, Vassilika Vouton, GR-71110 Heraklion, Greece
\and
Department of Physics, and Institute for Theoretical and Computational Physics, University of Crete, Voutes University campus, GR-70013 Heraklion, Greece
\and
Owens Valley Radio Observatory, California Institute of Technology, MC 249-17, Pasadena, CA 91125, USA
\and
Department of Space, Earth \& Environment, Chalmers University of Technology, SE-412 93 Gothenburg, Sweden
\and
Department of Physics, Stanford University, Stanford, CA 94305, USA
\and
Kavli Institute for Particle Astrophysics \& Cosmology, P.O. Box 2450, Stanford University, Stanford, CA 94305, USA
\and
Jet Propulsion Laboratory, California Institute of Technology, 4800 Oak Grove Drive, Pasadena, California, U.S.A
\and
Institute of Theoretical Astrophysics, University of Oslo, P.O. Box 1029 Blindern, NO-0315 Oslo, Norway
\and
National Institute of Science Education and Research, An OCC of Homi Bhabha National Institute, Bhubaneswar 752050, Odisha, India
\and
Inter-University Centre for Astronomy and Astrophysics, Post bag 4, Ganeshkhind, Pune, 411007, India
\and
South African Astronomical Observatory, PO Box 9, Observatory, 7935, Cape Town, South Africa
\and
Institute of Computer Science, Foundation for Research and Technology-Hellas, GR-71110 Heraklion, Greece
\and
Department of Physics, University of Johannesburg, PO Box 524, Auckland Park 2006, South Africa
}

\date{Received: 19 December 2023 / Accepted: 4 February 2024}

\abstract{
We present the first degree-scale tomography map of the dusty magnetized interstellar medium (ISM) from stellar polarimetry and distance measurements.
We used the RoboPol polarimeter at Skinakas Observatory to conduct a survey of the polarization of starlight in a region of the sky of about four square degrees.
We propose a Bayesian method to decompose the stellar-polarization source field along the distance to invert the three-dimensional (3D) volume occupied by the observed stars.
We used this method to obtain the first 3D map of the dusty magnetized ISM. Specifically, we produced a tomography map of the orientation of the  plane-of-sky component of the magnetic field threading the diffuse, dusty regions responsible for the stellar polarization.
For the targeted region centered on Galactic coordinates $(l,\,b) \approx (103.3^\circ,\, 22.3^\circ)$, we identified several ISM clouds. Most of the lines of sight intersect more than one cloud. A very nearby component was detected in the foreground of a dominant component from which most of the polarization signal comes and which we identified as being an intersection of the wall of the Local Bubble and the Cepheus Flare. Farther clouds, with a distance of up to 2~kpc, were similarly detected. Some of them likely correspond to intermediate-velocity clouds seen in \HI\ spectra in this region of the sky.
We found that the orientation of the plane-of-sky component of the magnetic field changes along distance for most of the lines of sight.
Our study demonstrates that starlight polarization data coupled to distance measures have the power to reveal the great complexity of the dusty magnetized ISM in 3D and, in particular, to provide local measurements of the plane-of-sky component of the magnetic field in dusty regions.
This demonstrates that the inversion of large data volumes, as expected from the \textsc{Pasiphae} survey, will provide the necessary means to move forward in the modeling of the Galactic magnetic field and of the dusty magnetized ISM as a contaminant in observations of the cosmic microwave background polarization.
The 3D map obtained in this paper can be visualized online\thanks{\url{https://pasiphae.science/visualization}}.
}


\maketitle


\section{Introduction}
The polarization of starlight is a powerful probe of the magnetized interstellar medium (ISM). Starlight acquires a polarization due to dichroic absorption by aspherical interstellar dust grains, which align their minor axis with the magnetic field (e.g., \citealt{Davis1951}; \citealt{Andersson2015}). The polarization position angle of starlight is parallel to the plane-of-sky (POS) component of the magnetic field, and the maximum degree of polarization is proportional to the column density of the polarizing dust through which the light beam passes.
Since its discovery (\citealt{Hiltner1949}; \citealt{Hall1949}), the polarization of starlight has contributed significantly to the study of the magnetic field in our Galaxy and to our understanding of its role as an agent of the Galactic ecosystem, from the smallest to the largest scales (e.g., \citealt{Spoelstra1972}; \citealt{Ellis1978}; \citealt{Goodman1990}; \citealt{Heiles1996}; \citealt{Heyer2008};
\citealt{Nishiyama2010}; \citealt{Li2013}; \citealt{Berdyugin2014}; \citealt{Doi2023}).
Once stellar distances are known, starlight polarization could provide information on the properties of the magnetized ISM directly in three dimension (3D) \citep{Pan2019a}, and with good spatial resolution given the high stellar density.

In recent years, the {\it Gaia} satellite (\citealt{Gaia2016}) has provided the data necessary for the precise localization in 3D space of more than a billion stars in our Galaxy (e.g., \citealt{Bailer-Jones2021}; \citealt{GaiaEDR32021}; \citealt{Lindegren2021}; \citealt{GaiaDR32023}). By combining stellar parallax and reddening data, several teams have been successful in reconstructing 3D tomography maps of the dust density distribution in large volumes centered on the Sun, up to around 3~kpc in the Galactic disk and up to around 1.2~kpc in the halo (\citealt{Green2019}; \citealt{Lallement2019}; \citealt{Leike2019}; \citealt{Leike2020}; \citealt{Lallement2022}; \citealt{Vergely2022}; \citealt{Edenhofer2023}), or in more focused areas. This formidable community effort has revolutionized our view of the 3D structure of dust distribution in the ISM, and it has already enabled the modeling and better understanding of some of the main structures in our cosmic neighborhood and their history (e.g., \citealt{Pelgrims2020}; \citealt{Alves2020}; \citealt{Das2020}; \citealt{Bialy2021}; \citealt{Zucker2022}; \citealt{Grossschedl2018}; \citealt{Marchal2023}; \citealt{Ivanova2021}; \citealt{Tahani2022b}). Such a 3D mapping, with the additional knowledge of magnetic field properties, would be certain to enable breakthroughs and discoveries in several research topics, as discussed below and in \cite{Pelgrims2023}.

For example, a tomographic view of the magnetized ISM would offer new avenues to address open questions such as the role of the magnetic field in star formation (see, e.g., \citealt{MTK2006}; \citealt{MO2007}), and the search for the sources of ultra-high energy cosmic rays (see, e.g., \citealt{Boulanger2018}; \citealt{Magkos2019}; \citealt{Tsouros2024}).
Likewise, by providing measurements of the polarization properties for each individual dust cloud in 3D space, such a tomography map would enable significant progress in the modeling and characterization of the dusty magnetized ISM as a contaminating foreground to observations of the polarization of the cosmic microwave background (\citealt{Tassis2015}; \citealt{Martinez2018} ; \citealt{Pelgrims2021}).
Hence, it would help clear the path for an unbiased study of one of the first moments of the Universe's history.
This holds true, although less directly, for the characterization of the Galactic synchrotron emission as demonstrated by \cite{Panopoulou2021}.
Combined with multiwavelength observations of the polarized emission of dust in the submillimeter, knowledge of the polarization properties of each individual dust cloud would also enable us to advance our modeling of astrophysical dust (e.g., \citealt{Hensley2021}), to better determine its composition and understand its interaction with its cosmic environment, and to better assess its role as a building block of life.

\smallskip

Mapping the dust distribution in 3D over large volumes (kiloparsecs) required millions of stellar extinction and distance measurements. A similar amount of data is likely required for magnetic field mapping in such a volume. Currently, the availability of stellar polarization data is limited (\citealt{Panopoulou2023}), with only the inner Galaxy having millions of stars with measured near-IR polarization (GPIPS survey, \citealt{GPIPS2020}). In the future, planned optical polarimetric surveys will deliver millions of measurements throughout the entire sky (\citealt{Magalhaes2005,Magalhaes2012}; \citealt{Tassis2018}; \citealt{Covino2020}). It is therefore timely to develop techniques to analyze these forthcoming datasets in such a way that would make it possible to extract from the data most of the information on the properties of the dusty magnetized ISM in 3D.

The observed polarization of each single star is the integrated effect of the dichroic absorption from all ISM clouds lying between us and the star. This line-of-sight (LOS) integration needs to be inverted in order to derive the complex 3D structure of the magnetized ISM from starlight polarization and distance measurements.
Recently, we have developed the first standalone method to perform this LOS inversion through Bayesian modeling (\citealt{Pelgrims2023}). This method, which takes into account all sources of uncertainties in polarization and parallax measurements, works on a per LOS basis. It has the advantage that it can be easily automated to run on a large set of sightlines. We have extensively tested the method and its performance based on simulated data. We further applied it to existing data for two sightlines and demonstrated that this new method leads to results that are fully consistent with previously obtained results but within a robust Bayesian framework and, therefore, put on a more solid footing.

\smallskip

\begin{figure*}
    \centering
    \includegraphics[trim={0.2cm 11.4cm 0.4cm 0.07cm},clip,width=.98\linewidth]{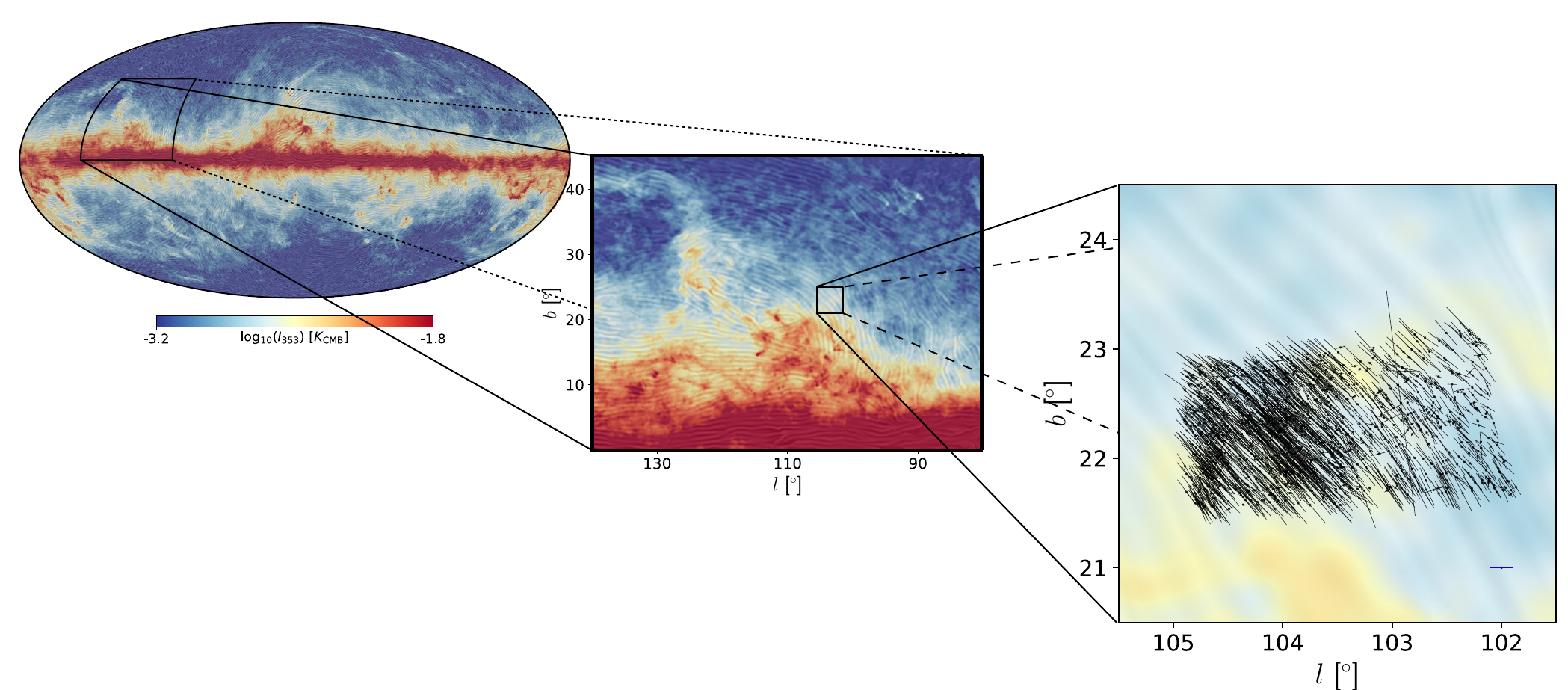}\\[-1.2ex]
    \caption{
    Sky location of the surveyed area of about four square degrees. Left: Full-sky map of the dust emission as seen by {\it Planck} at 353~GHz (\citealt{PlanckXII2020}). The color represents the intensity of dust emission { on} logarithmic scale. The line-integral-convolution texture shows the polarization angle of the dust emission rotated by 90 degrees. Middle: A zoom-in of the map toward the surveyed regions, which includes part of the North Celestial Pole Loop on the East of the map. Right: A closer view of the surveyed region. Black segments indicate the polarization orientation from the stars in our survey and from \cite{Pan2019a}. The segments are scaled according to the polarization fraction. Unpolarized stars appear as dots. The blue, horizontal segment in the bottom right corner shows the scale for a 1\% polarized star. Outlier candidates (see Sect.~\ref{sec:outliers}) are not shown.
    }
    \label{fig:TestPatchOnSky}
\end{figure*}
In this paper we continue our preparation for the large datasets to come.
We aim to develop a pipeline capable of inverting measurements of parallax and starlight polarization for an actual, extended volume of 3D space in order to derive the properties of the dusty magnetized ISM in 3D.
We develop a first pipeline based on our LOS-inversion method and apply it to observations taken for an extended region of the sky.
Our goal is to obtain, with as few assumptions as possible, a first tomography map of the POS component of the magnetic field in dusty regions from which 3D properties of the magnetized ISM can be accessed.

For the purposes of this work, we carried out a survey of starlight polarization for a continuous region of the sky covering about four square degrees. We present our survey and the resulting dataset in Sect.~\ref{sec:data}. We devote Sect.~\ref{sec:method} to the description of our data analysis pipeline and its application to our data to obtain the first degree-scale tomography map of the magnetized ISM from starlight polarization and parallax measurements.
Section~\ref{sec:results} presents our main results and how we produce and visualize our 3D map of the POS component of the magnetic field in dusty region from the posterior distributions output by our Bayesian analysis. Our 3D map extends up to 3~kpc distance and covers a sky region of about four square degrees. We discuss our results in comparison with other probes of the (magnetized) ISM in Sect.~\ref{sec:discussion} and conclude in Sect.~\ref{sec:conclusion}.

\section{Dataset}
\label{sec:data}
To demonstrate the feasibility of starlight-polarization-based tomography of the diffuse magnetized ISM, \cite{Pan2019a} obtained dense optical polarization measurements of stars for two circular beams of 9.6~arcmin radius in a sky region that they identified as being likely to exhibit complexity along the distance axis based on inspection of \HI\ velocity data and polarized dust emission maps.
These data indeed suggest possible variations of the polarization signal both in the POS and along distance as several components can be identified.
As they showed through the analysis of their polarization data along distance, and as we recently confirmed (\citealt{Pelgrims2023}), one of the two beams likely intersects at least two dust clouds while the other one likely intersects only one. Showing complexity both along the LOS and in the POS, this region therefore seems to be well suited to develop and test tomography methods to reconstruct the magnetized ISM in 3D. Hence, we carried out a survey to expand that region of the sky with optical starlight polarization data.
The final surveyed region, comprising about four square degrees, and its location in the sky is shown in Fig.~\ref{fig:TestPatchOnSky}.

\subsection{Survey strategy and observation plan}
\label{sec:surveyStrategy}

We aimed at obtaining polarization measurements for a large, continuous region centered on Galactic coordinates $(l,\,b) \approx (103.5^\circ,\,22.25^\circ)$ with complete star samples limited in magnitude. Due to limited observing time, it was infeasible to perform a deep, photometrically uniform survey for the entire area. Hence, we relied on \HI\ observations to gauge a priori which part of the targeted region likely intersects dust clouds at large distances, in order to increase the number of stellar polarization data accordingly.

\begin{figure*}
    \centering
    \includegraphics[trim={0.8cm 3.4cm .8cm 3.4cm},clip,width=.98\linewidth]{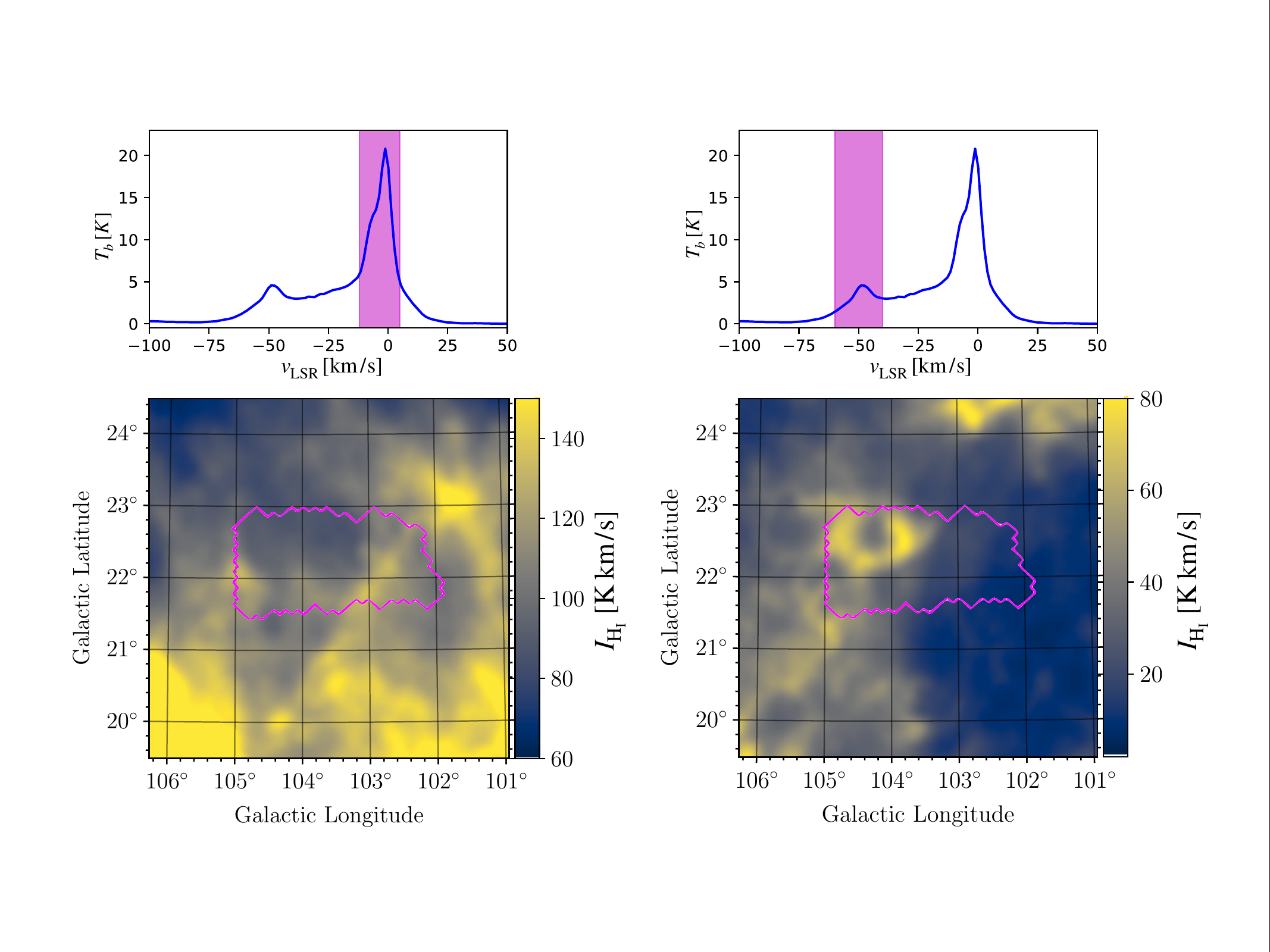}\\[-1.5ex]
    \caption{\HI\ velocity data from the HI4PI survey in the sky area toward the region surveyed in starlight polarization. The top panels show the brightness temperature as a function of the gas velocity measured in the LSR. The blue spectrum corresponds to the velocity spectrum averaged over the full region. Two dominant peaks with velocities in the LVC ($\approx -1$~km/s) and IVC ($\approx -50$~km/s) ranges are seen. The bottom panels show the column density maps resulting from the integration of the velocity spectra in the ranges depicted by shaded regions marked in the top panels in the LVC (left) and IVC (right) ranges. The maps reveal a high degree of complexity in morphological structures. The magenta outline indicates the sky region of 3.8 sq. deg. for which starlight-polarization-based tomography is obtained in this work (see Sect.~\ref{sec:subsamples}).
    }
    \label{fig:average_HI_spectrum}
\end{figure*}

As discussed by \cite{Pan2019a}, and also shown in Fig.~\ref{fig:average_HI_spectrum}, the averaged \HI-velocity spectrum (measured in the Local Standard of Rest, LSR) in this region of the sky shows the existence of two very distinct components. The dominant component is centered on $v_{\rm{LSR}} \approx -2.5\, {\rm{km/s}}$ and we refer to it as the low velocity cloud (LVC). The second component, with a lower amplitude, is centered on $v_{\rm{LSR}} \approx -50\, {\rm{km/s}}$ and we refer to it as the intermediate velocity cloud (IVC). This velocity component may be a southern extension of the IV Arch as identified by \cite{Kuntz1996} who determined that it is located at large distance ($\gtrsim 1\,{\rm{kpc}}$).
We used the HI4PI survey (\citealt{HI4PI2016}) to look at the spatial distribution of the intensity maps obtained from the integration of the velocity spectra in channels corresponding to the peaks observed in the averaged velocity spectrum.

These data are shown in Fig.~\ref{fig:average_HI_spectrum} where we also show the outline of the region for which we obtain a tomography reconstruction in this work.
As can be seen clearly in the right-hand bottom panel of Fig.~\ref{fig:average_HI_spectrum}, the \HI\ velocity spectra show power in the IVC range in the eastern part of the region (with $l \gtrsim 103.5^\circ$). Thus, we concluded that this region is more likely to contain dust clouds at larger distances than the other (western) part of the region. Hence, as the stellar magnitude generally increases with stellar distance, we conduct a deeper survey in that part of the sky (with a limiting magnitude $R \lesssim 15.5$~mag, compared to $R \lesssim 14$ mag in the remaining area); to increase the number density of data points, in particular at large distances.
Despite this choice that resulted in more sparse stellar polarization measurements on the right part of the region, a hint of distant clouds is still found there (see Sects.~\ref{sec:results} and~\ref{sec:discussion}). This indicates that even with a shallower polarization survey distant clouds could still be recovered - though see Sect.~\ref{sec:discussion} where we discuss the reliability of cloud detection.

We have conducted our survey so that the entire region is photometrically complete in the $R$ band up to 14~mag (shallow survey), and up to 15.5~mag for the part of the region located at $l \gtrsim 103.5^\circ$ (deep survey).
The upper panel of Fig.~\ref{fig:starSample_skyAndMu} shows the locations of stars on the sky for both the shallow and the deep survey. We used the $R$-band magnitude and sky coordinates from the USNO-B1.0 catalog (\citealt{Monet2003}) to plan our observations.

\subsection{Observations and data reduction}
\label{sec:ObsAndData}
Observations of optical polarization were conducted using the 1.3-m telescope situated at Skinakas Observatory\footnote{\url{http://skinakas.physics.uoc.gr}} in Crete (1750 meters above the sea level, 24\degr 53\arcmin 57\arcsec E, 35\degr 12\arcmin 43\arcsec N). The telescope is equipped with the RoboPol polarimeter, which consists of two adjacent half-wave retarders, with their fast-axes rotated by 67.5$^\circ$, relative to each other, followed by two Wollaston prisms with orthogonal fast-axes \citep{Ramaprakash2019}.
This setup splits each incident ray into four rays with different polarization states on a single CCD, which provide information on the $q = Q/I$ and $u = U/I$ normalized Stokes parameters in the instrument's reference frame (see Eq.~1 in \citealt{King2014}) with only one exposure. In this way, the use of a fixed instrument configuration eliminates random and systematic errors resulting from changes in the sky, imperfect alignment, and non-uniformity of rotating optical elements, as the instrument has no moving parts aside from the filter wheel. To enhance the signal-to-noise ratio (S/N), a special mask was placed in the center of the telescope focal plane where systematic uncertainties have been estimated to be lower than 0.1\% in the degree of polarization (\citealt{Skalidis2018, Ramaprakash2019}).

The observations were carried out star-by-star over three observation seasons, from May 2019 to November 2022. Each star was measured in the mask. Optical polarization measurements were obtained for each star in the Johnsons-Cousins $R$ band.
For each star, the observation exposure time was estimated on-the-fly so as to guarantee that photon-noise-driven uncertainties fall below the estimated uncertainties from instrumental calibration (which turned out not always to be the case as discussed below).
Zero- and highly polarized polarimetric standards were observed every observing night to monitor the instrumental polarization and polarization angle zero-point through time, and to estimate the corresponding uncertainties as described in \cite{Blinov2021} and  \cite{Blinov2023}. We obtained instrumental uncertainties in both $q$ and $u$ at the level of 0.1\%.

\smallskip

Data reduction was performed with the standard RoboPol pipeline (\citealt{King2014}; \citealt{Panopoulou2015}; \citealt{Blinov2021}). For any given source, we produced the stacked image of all observations and deduced the linear Stokes $q$ and $u$ parameters and their photon-noise-driven uncertainties through differential aperture photometry as outlined in \cite{Ramaprakash2019}. Then we removed the contribution from instrumental polarization and added in quadrature observational and instrumental uncertainties as:
\begin{align}
    & q = q^{\rm{measured}} - q^{\rm{instr}} \;, \\
    & \sigma_q = \sqrt{\left(\sigma_q^{\rm{measured}}\right)^2 + \left(\sigma_q^{\rm{instr}}\right)^2} \;,
\end{align}
and similarly for Stokes $u$, where the superscripts ``measured'' and ``instr'' refer to the values obtained from differential photometry and from estimation of the instrumental polarization, respectively.
The polarization data are given in the equatorial coordinate system and follow the IAU convention for the polarization position angle (zero at north, increasing toward east). In total, we obtained reliable optical polarization data for 1530 stars, spending approximately 153 telescope hours.

\subsection{Cross-match with {\it Gaia} and quality cuts}
\label{sec:Data_fullSample}
For the purpose of this work, we complemented our new polarization measurements with data from \citet{Pan2019a} for 192 stars.
As we need estimates of stellar parallaxes and their corresponding uncertainties to perform the tomography decomposition along distance, we cross-matched our sample with polarization measurements with the {\it Gaia} DR3 catalog (\citealt{GaiaDR32023}).

We used a cross-match radius of 5 arcseconds around the USNO-B1.0 coordinates of the targets. We chose to use the USNO-B1.0 coordinates as the astrometric accuracy of the RoboPol pipeline output varies with position on the field of view and can be limited to a few arcseconds in some cases as a result of distortions and the 4-spot pattern of the stars which are modeled within the software\footnote{The limited astrometric accuracy of RoboPol is primarily due to uncertainties in the joint modeling of the 4-spot pattern and field distortions. New generation wide-field polarimeters such as the Wide Area Linear Optical Polarimeters (e.g., \citealt{Maharana2020}) will enable improved astrometric accuracy with wider coverage and the absence of the 4-spot pattern.}. 
For most sources, there was a unique {\it Gaia} source found within the search radius, to which we assigned the match. Some sources had multiple matches with the {\it Gaia} catalog, as a result of the proximity of the target star with another star.
The cross-match was not always straightforward mainly due to the lack of precise astrometry in the USNO-B1.0 catalog. This problem mostly happened when, in the USNO-B survey, adjacent stars (approximately less than $2\arcsec$ spatial separation) were blended\footnote{In the USNO-B1.0 catalog, blended objects have a single identifier and the magnitude of the star represents the total photometry of both stars.}.
We visually inspected all cases where multiple matches were found by comparing the raw RoboPol images with the USNO-B1.0 and {\it Gaia} catalogs.
We also checked for possible misidentifications by comparing the $R$- and $G$-magnitudes of the sources after the cross-match, correcting for a couple cases where a fainter {\it Gaia} star had been mistakenly associated with a bright USNO-B1.0 star. Finally, we note that we observed several faint stars, which exceed the photometric magnitude limits of our survey ($R > 15.5$ mag), because they were close (around $1\arcsec$ distance) to the target stars. These stars happened to lie within the mask of RoboPol during the observations. The resulting obtained S/N in their degree of polarization is usually low.
In some cases, this proximity led to photometrically blended measurements, which we disregarded in our analysis.

The final cross-match was successful for 1698 stars for which we retrieved their {\it Gaia} identifier, their $G$-band photometric magnitude, their parallax and corresponding uncertainty, and their renormalized unit weight error (${\rm{RUWE}}$). The latter measures the quality of the single-star model to account for astrometric {\it Gaia} observations, and must be used as a quality criterion to guarantee the reliability of the solution, and thus of the parallax estimates. We used stars with ${\rm{RUWE}} < 1.4$ (as recommended) to avoid unreliable measurements that could occur because of blended sources for example (\citealt{Lindegren2018}; \citealt{Lindegren2021}).
Among the successfully cross-matched stars, 24 stars do not have parallax information and 226 stars have ${\rm{RUWE}} \geq 1.4$ or are unknown.
Consequently, after applying the quality criterion we have 1448 stars with both reliable optical polarization measurements and reliable parallax estimates. This is the star sample that we use for our tomography of the magnetized ISM.

\smallskip

\begin{figure}
    \centering
    \includegraphics[trim={0.3cm 0cm 0.2cm 0cm},clip,width=.98\linewidth]{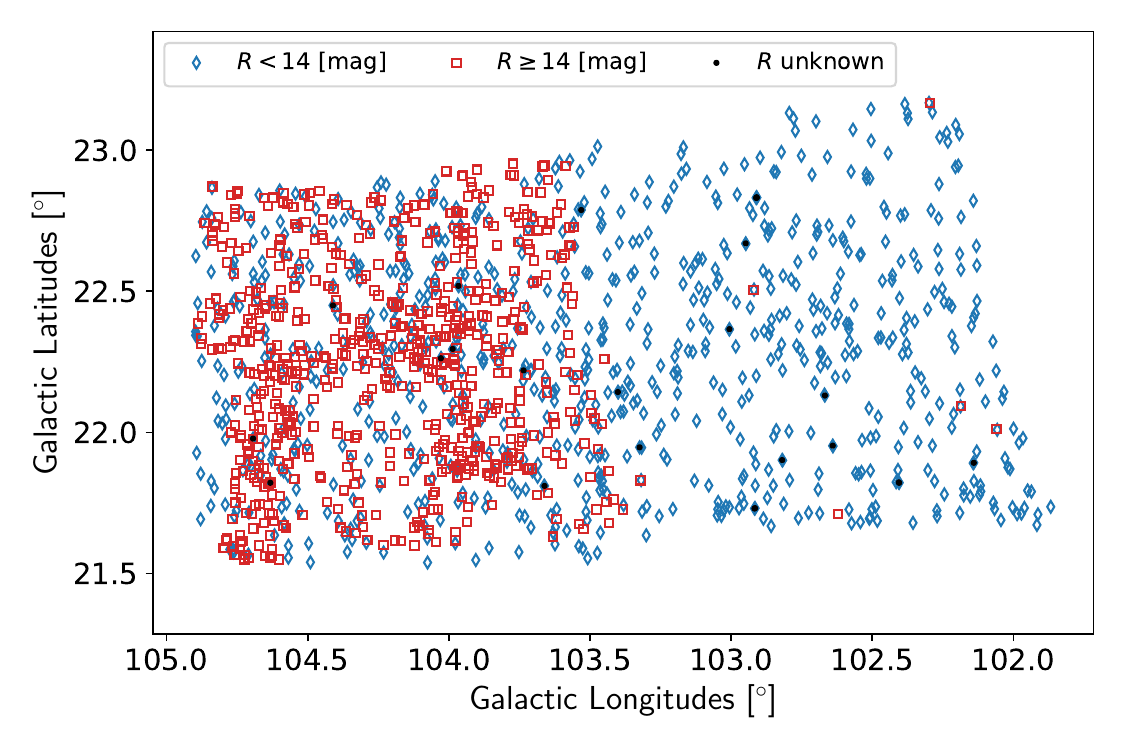}\\
    \includegraphics[trim={0.0cm 0cm 0.2cm 0cm},clip,width=.98\linewidth]{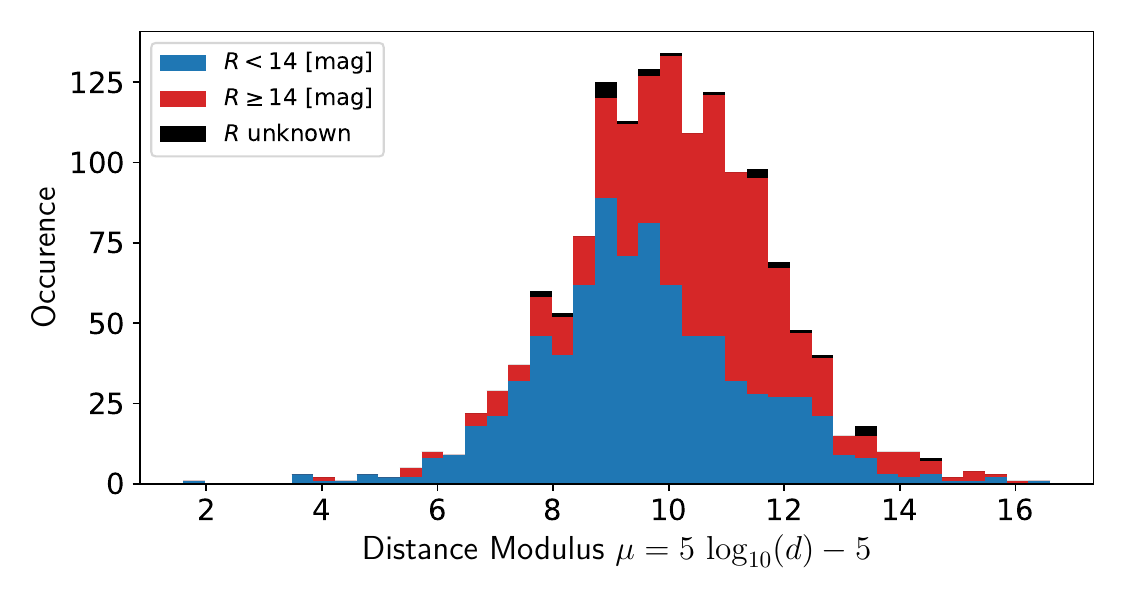}\\[-1.5ex]
    \caption{Sky distribution (top) and distance (modulus) distribution obtained by inverting the parallax (bottom) of our stellar sample with reliable parallax estimates. The sample is divided according the USNO-B $R$-band magnitude. Blue, red and black correspond to stars with $R < 14$ (shallow survey), $R \geq 14$ (deep survey) and unknown values due to identification mismatch.
    The histograms on the bottom panel are stacked.
    }
    \label{fig:starSample_skyAndMu}
\end{figure}

\begin{figure}
    \centering
    \includegraphics[trim={.5cm 0cm .6cm .4cm},clip,width=.98\linewidth]{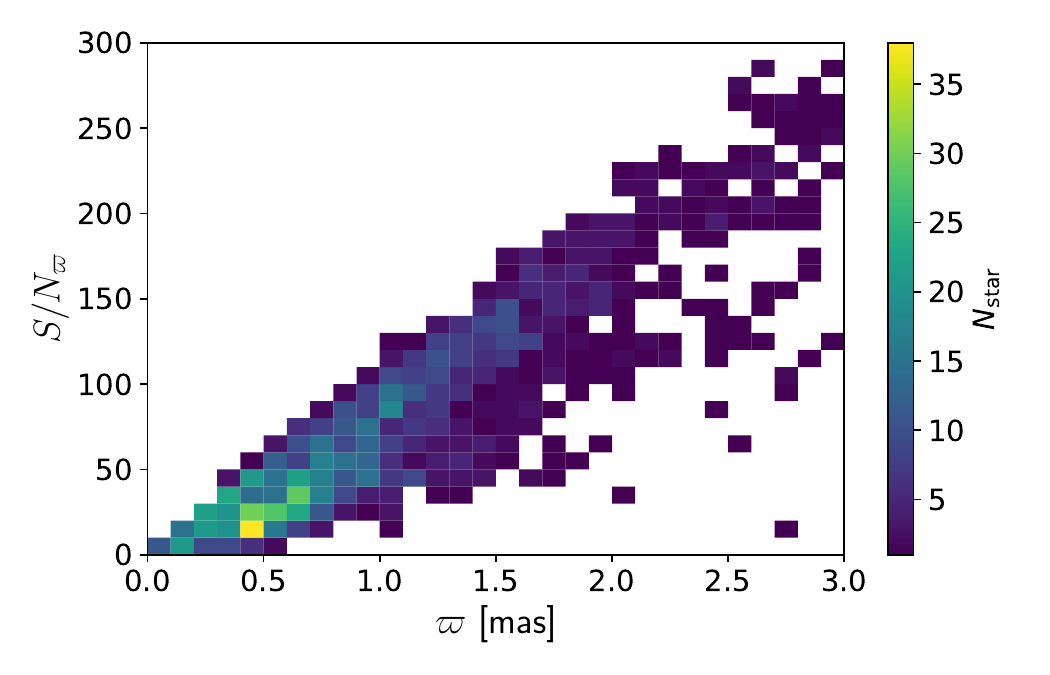}\\[-1.5ex]
    \caption{2D distribution of the number of stars in our catalog in the parallax - S/N plane.
    For better visualization, a dozen stars with large parallax values ($\varpi$) and or large parallax S/N are not included in this plot.
    }
    \label{fig:starSample_plx}
\end{figure}

\begin{figure}
    \centering
    \includegraphics[trim={0cm .4cm 0cm .8cm},clip,width=.98\linewidth]{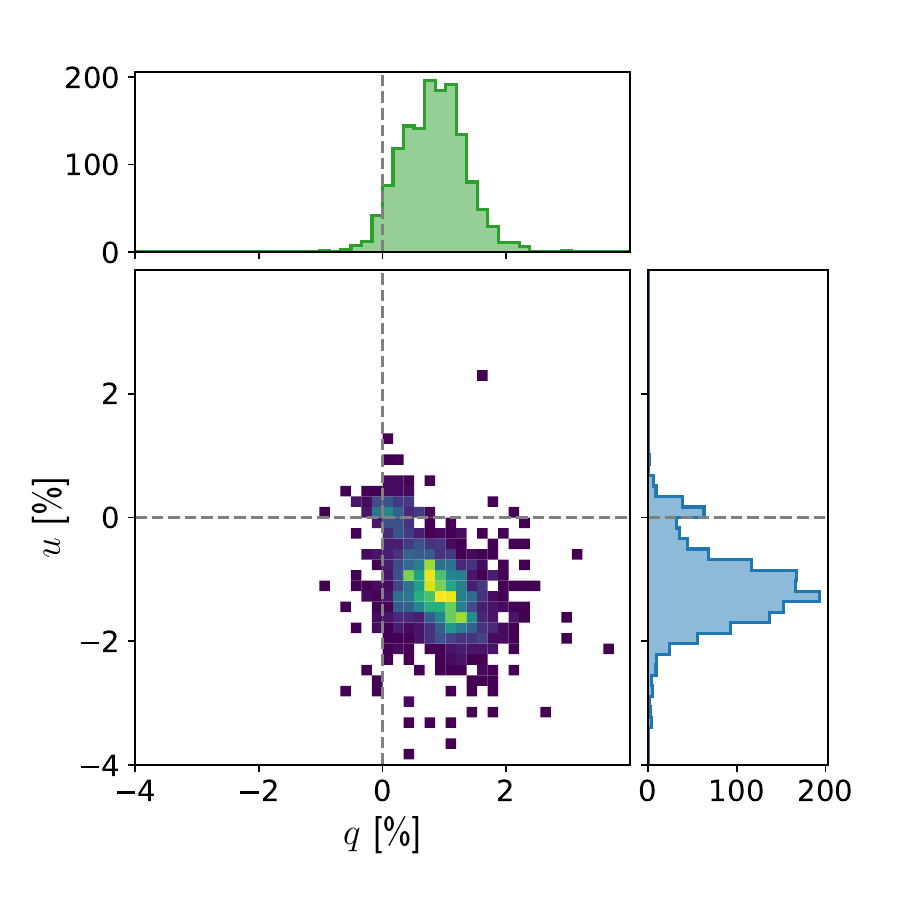}\\
    \includegraphics[trim={0cm .4cm 0cm .8cm},clip,width=.98\linewidth]{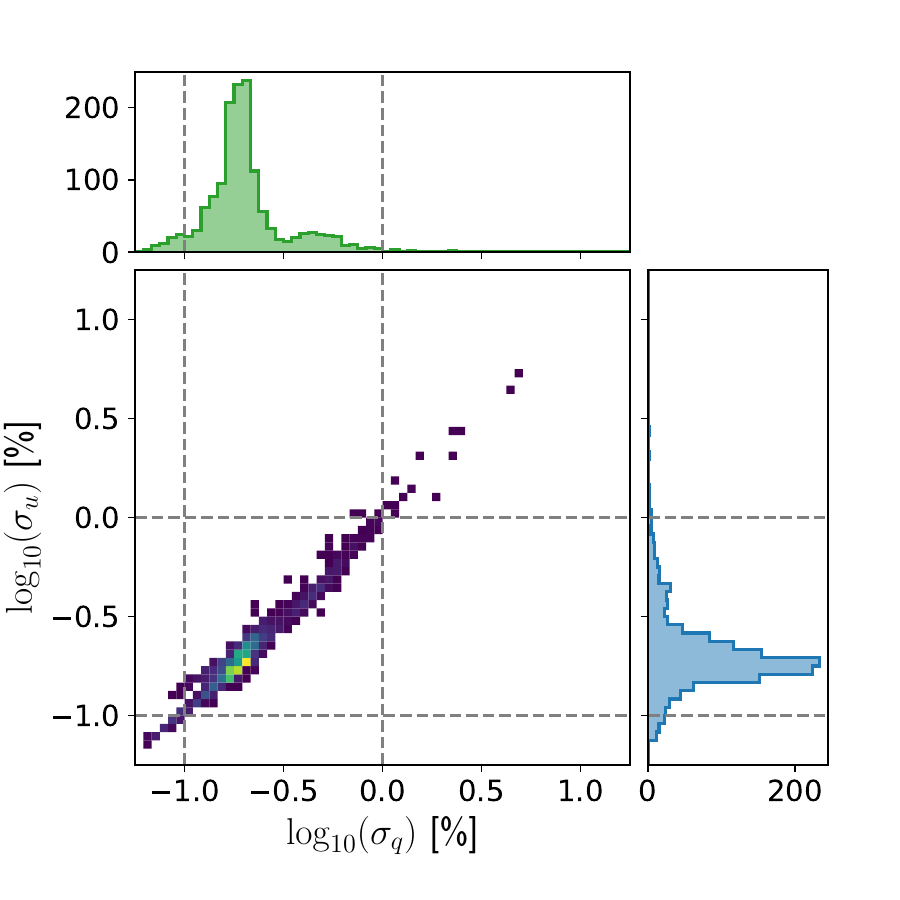}\\[-1.5ex]
    \caption{Distribution of the polarization properties for the stellar sample with a reliable parallax estimate. Distributions of the Stokes parameters ($q$'s and $u$'s) and of the logarithm of their uncertainties ($\log_{10}(\sigma_q)$ and $\log_{10}(\sigma_u)$) are shown in the top and bottom panels, respectively.
    The horizontal and vertical dashed lines are used for visual reference. They indicate the $q = 0$ and $u = 0$ loci in the top panel and indicate the values of 0.1\% and 1\% in polarization uncertainties in the bottom panel.
    }
    \label{fig:starSample_polDist}
\end{figure}

Some properties of our final dataset of 1448 stars are given in Figs.~\ref{fig:starSample_skyAndMu},~\ref{fig:starSample_plx} and~\ref{fig:starSample_polDist}, which illustrate the non-homogeneous character of our sample.
The different cuts in $R$-band magnitude used to design our survey based on \HI\ complexity are clearly seen in the sky distribution shown in the top panel of Fig.~\ref{fig:starSample_skyAndMu}. The histogram in the bottom panel of the same figure clearly shows that fainter stars ($R \geq 14$ mag) generally have larger distances (obtained simply by taking the inverse of the parallax) than brighter ones. Consequently, the density of stars, in particular at large distances, is much larger in the eastern half of the observed region than in the western one. We also discuss this in Sect.~\ref{sec:subsamples}.
Figure~\ref{fig:starSample_plx} shows the distribution of parallax and parallax S/N in our sample. Most of our targets (with ${\rm{RUWE}} < 1.4$) have parallax S/N higher than 10, demonstrating the reliability of our distance markers through the ISM.
The distributions of the Stokes parameters and the (total) uncertainties of the entire sample with reliable estimate of the parallax are shown in Fig.~\ref{fig:starSample_polDist}. It is seen from the 2D distribution of the Stokes parameters that a certain number of stars have low polarization while the majority have polarization degree at the per cent level or higher\footnote{The polarization as a function of distance is shown in Sect.~\ref{sec:outliers} and we study it starting from Sect.~\ref{sec:LOSdeco}.}. The uncertainty on the Stokes parameters is at the level of 0.19\% in both $q$ and $u$ for a large fraction of our sample, and therefore dominated by systematic uncertainty from the instrument calibration, while the photon noise contributes significantly for a subset of measurements. We decided to keep all measurements since we are confident in their uncertainty estimates.

\subsection{Coordinate system conversion}
The Stokes parameters are measured in the equatorial celestial coordinate system. We construct our 3D map of the POS component of the magnetic field in the Galactic coordinate system, as this seems natural in the context of Galactic tomography.
Although we could perform the tomographic decomposition in the equatorial coordinate system and then convert the resulting 3D map in the Galactic coordinate system, we prefer to first convert the starlight polarization data and then perform our analysis to obtain the tomography result directly in the Galactic coordinate system. Because of the change of coordinate system, the values of the polarization position angles, and therefore of the Stokes parameters, change to account for the change of the orientation of the meridian from one reference frame to the other at the location of the stars. The change of coordinate system thus involves a rotation of the polarization plane which is implemented using the rotation matrix (e.g., \citealt{Tegmark2001})
\begin{align}
    \rm{R} &= \begin{pmatrix}
    \cos(2\psi_{\rm{R}}) & \sin(2\psi_{\rm{R}}) \\
    - \sin(2\psi_{\rm{R}}) & \cos(2\psi_{\rm{R}})
    \end{pmatrix}, 
\end{align}
where the rotation angle ($\psi_{\rm{R}}$) is defined locally. It depends on the celestial coordinates (right ascension and declination) of a star $(\alpha_\star,\,\delta_\star)$ and on the celestial coordinates of the North Galactic pole $(\alpha_{\rm{NGP}},\,\delta_{\rm{NGP}})$, and is given by (e.g., \citealt{Hutsemekers1998}) 
\begin{align}
    \psi_{R} = \rm{arctan2}\lbrace & \cos(\delta_{\rm{NGP}}) \sin(\alpha_{\rm{NGP}} - \alpha_\star), \nonumber \\
    & \phantom{a} \sin(\delta_{\rm{NGP}})\cos(\delta_\star)  \nonumber \\
    & \phantom{a} - \sin(\delta_\star) \cos(\delta_{\rm{NGP}}) \cos(\alpha_{\rm{NGP}} - \alpha_\star)
    \rbrace \;,
\end{align}
where we use the two arguments arctangent function to place the resulting angle in the correct trigonometric quadrant.
The Stokes parameters of a star in the Galactic coordinate system are thus obtained from the Stokes parameters in equatorial coordinate system through the following:
\begin{align}
    \begin{pmatrix}
    q_{\rm{gal}} \\
    u_{\rm{gal}}
    \end{pmatrix}
    =
    \rm{R}
    \begin{pmatrix}
    q_{\rm{eq}} \\
    u_{\rm{eq}}
    \end{pmatrix} \; .
\end{align}

Similarly, it can be shown (e.g., see Appendix~A of \citealt{PlanckXIX2015}) that the noise covariance matrix of the Stokes parameters in the Galactic reference frame is obtained from the rotation matrix and the noise covariance matrix in the equatorial coordinate system through
\begin{align}
    \rm{C}^{\rm{gal}} &= \rm{R}\, \rm{C}^{\rm{eq}} \rm{R}^T \;,
\end{align}
where $\rm{R}^T$ is the transpose of the rotation matrix.
Introducing $a = \cos(2\psi_{\rm{R}})$ and $b = \sin(2\psi_{\rm{R}})$, the elements of the noise covariance matrix are thus obtained as follows:
\begin{align}
    {\rm{C}}_{qq}^{\rm{gal}} & =  a^2 \, {\rm{C}}_{qq}^{\rm{eq}} + 2 \, a \, b \, {\rm{C}}_{qu}^{\rm{eq}} + b^2 \, {\rm{C}}_{uu}^{\rm{eq}} \nonumber \\
    {\rm{C}}_{qu}^{\rm{gal}} & =  \left(a^2 - b^2\right) \, {\rm{C}}_{qu}^{\rm{eq}} + a \, b \, \left({\rm{C}}_{uu}^{\rm{eq}} -{\rm{C}}_{qq}^{\rm{eq}}\right) \\ \nonumber
    {\rm{C}}_{uu}^{\rm{gal}} & =  b^2 \, {\rm{C}}_{qq}^{\rm{eq}} - 2 \, a \, b \, {\rm{C}}_{qu}^{\rm{eq}} + a^2 \, {\rm{C}}_{uu}^{\rm{eq}} \; .
\end{align}
The noise covariance matrix of the Stokes parameters remains unchanged from one coordinate system to the other only when there is no noise covariance between $q$'s and $u$'s and when the noise uncertainties in both $q$'s and $u$'s are equal. In any other situation, the noise covariance matrix changes and, most notably, a non-zero off-diagonal term arises simply due to the change of coordinates.
The use of the exact analytical formalism given above allows us to avoid issues related to the polarization bias in low S/N regime and to avoid the need for estimating the noise covariance matrix through Monte Carlo treatment.

\subsection{Stellar polarization catalog}
The polarization catalog is made publicly available through CDS. The columns contained in the catalog are described in Table~\ref{tab:polcat}. The coordinates given are from {\it Gaia} DR3 at epoch 2016. We include all stellar polarization measurements, including those that were identified as outliers and those that did not return a reliable distance (due to no match with {\it Gaia} or high RUWE values). We distinguish between sources used in the tomographic reconstruction in the catalog with a usage flag.
The usage flag is assigned a value of
0 for stars that did not have a reliable distance estimate, 1 for stars that are used in the reconstruction, and 2 for stars identified as outliers (see Sect.~\ref{sec:outliers}).

\begin{table}[]
    \centering
    \caption{Polarization catalog (abbreviated).}
    \small{
    \begin{tabular}{c c c}
    \hline\hline \\[-0.5ex]
        Column name & Unit & Description\\ \\[-.5ex]
        \hline \\[-1.ex]
       {\it Gaia}\_source\_ID  & $-$ & source identifier in {\it Gaia} DR3\\
       RA   & degrees &  Right Ascension (2016)\\
       Dec  & degrees & Declination (2016)\\
       q    & \% & Relative Stokes $q$ \\
       e\_q & \% & Uncertainty in $q$\\
       u    & \%  & Relative Stokes $u$\\
       e\_u & \% & Uncertainty in $u$\\
       p    & \% & Degree of polarization \\
       e\_p & \% & Uncertainties in $p$ \\
       EVPA & degrees & Polarization angle \\
       e\_EVPA & degrees & Uncertainties in EVPA \\
       date & $-$ & Observation date \\
       G    & mag & $G$-band magnitude \\
       plx  & mas & Parallax \\
       e\_plx & mas & Uncertainty in parallax\\
       d\_Maha & $-$ & Post tomography outlier flag \\ \\
       usage\_flag & $-$ & \begin{tabular}{l}
            0: missing or unreliable parallax\\
            1: used in tomography\\
            2: outlier (sigma-clipping)
       \end{tabular}
       \\ \\[-1.5ex]
        \hline
    \end{tabular}
    }
    \label{tab:polcat}
\end{table}

\section{Method and data analysis}
\label{sec:method}
In this paper we aim to obtain a 3D tomographic decomposition of the dusty magnetized ISM for an extended region of the sky, from stellar measurements of optical polarization and distance.
As there is no currently available method to perform a 3D inversion of an actual volume of stellar polarization data, we design an analysis pipeline based on the LOS-inversion method that we presented in \cite{Pelgrims2023} and which is implemented in the \texttt{BISP-1} (Bayesian Inference of Starlight Polarization in 1D) Python code\footnote{\url{https://github.com/vpelgrims/Bisp_1}}.

\subsection{\texttt{BISP-1}}
Our maximum-likelihood method, implemented in \texttt{BISP-1}, makes it possible to reconstruct the dusty magnetized ISM along a single LOS using starlight polarization and distance (parallax) only.
It assumes that the effect of multiple clouds along the LOS is well approximated by the vector addition of the linear Stokes parameters induced by each cloud separately - which holds for the typically low polarization levels in the diffuse ISM (\citealt{Martin1974}; \citealt{Pat2010}; \citealt{Pan2019a}).
Relying on the nested sampling method (\citealt{Skilling2004}) implemented in the code \texttt{dynesty} (\citealt{Speagle2020}), our algorithm is able to determine the number of components (dust clouds) along the LOS and to determine the distance and polarization properties of each component using six parameters per component: cloud distance ($d_{\rm{C}}$), cloud mean polarization ($q_{\rm{C}},\,u_{\rm{C}}$), and three parameters to characterize the covariance matrix $C_{\rm{int}}$ encoding the intrinsic scatter of stellar polarizations arising as a result of ISM turbulence.
As such, the method makes it possible to recover the stellar-polarization source field which directly informs on the local orientation of the POS component of the magnetic field (the position angle) and on the local degree of polarization. The latter is related to the dust grain density, the dust grain polarization efficiency and on the angle made by the magnetic field lines with the POS, as explained in (e.g. \citealt{Pelgrims2023}).
In this picture, it is implicitly assumed that the aspherical dust grains always align their shortest axis with the local orientation of the magnetic field, at least statistically. This assumption is expected to be true for the diffuse ISM and agrees with current state of dust alignment theory (e.g., \citealt{Draine2021b}; \citealt{Hensley2021}).

The model that we use to reconstruct the dusty magnetized ISM assumes that the dust clouds can be represented as thin layers distributed along the sightline. That is, we consider that the typical extent of dust clouds along the LOS is smaller than the typical separation of stars or, in practice, smaller than the distance range spanned by the number of stars needed to allow for the detection of a cloud given the amplitude of the polarization it induces and all sources of scatter in the polarization data.
We expect this assumption to hold at high and intermediate Galactic latitudes and for clouds of the cold neutral medium and molecular clouds (\citealt{Heiles1976}; \citealt{Zucker2021}; \citealt{Marchal2023}). It is worth noting that the thin-layer approximation also seems to hold  in the denser regions of the ISM, at least for some sightlines (\citealt{Doi2023}).

Our method (\texttt{BISP-1}) further assumes that all stellar measurements trace the dusty magnetized ISM only (as opposed to having intrinsic polarization), and it does not implement spatial variation in the POS.
To circumvent these limitations while benefiting from the strengths of our LOS-inversion method, we embed it in a multistep process to form our 3D inversion pipeline, as described below.

\subsection{Design of the 3D inversion pipeline}
\begin{figure}
    \centering
    \includegraphics[trim={2.6cm 1.cm 2.6cm 1.4cm},clip,width=.98\columnwidth]{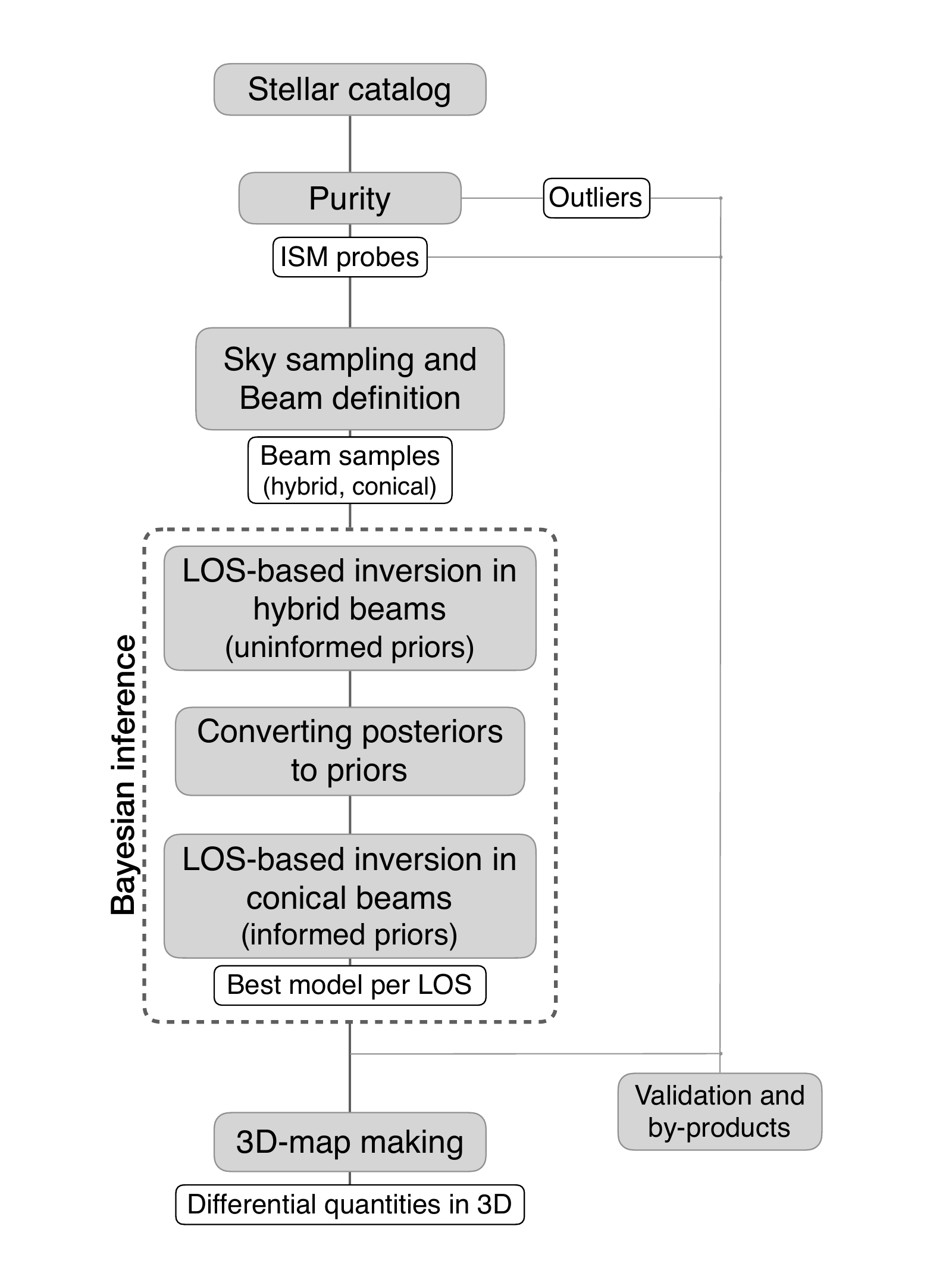}
    \caption{Flowchart for the 3D inversion pipeline, from the stellar catalog with reliable parallax estimates (top) to the 3D maps of differential quantities (bottom).}
    \label{fig:flowchart}
\end{figure}

The broad outline for the workflow of our 3D inversion pipeline is illustrated in the flowchart given in Fig.~\ref{fig:flowchart}.
From the original sample of stars in the extended region, we first identify stars that likely trace the magnetized ISM (as opposed to the intrinsically polarized candidates).
Then, for a set of sightlines that span the observed region, we create two sets of overlapping subsamples centered on each LOS, as detailed in Sect.~\ref{sec:subsamples}. The first is meant to capture ISM properties at small distances despite the low stellar density; the second corresponds to the highest angular resolution we can achieve. We then perform the LOS inversion for each LOS; first on the subsamples which connect different sightlines at small distances, and then at higher resolution taking into account the information obtained in the previous step. Finally, we produce the 3D map of the magnetized ISM from the posterior distributions obtained for each LOS individually.
Each step is detailed in this section and in Sect.~\ref{sec:results}, and illustrated by applying the method to our dataset.

\subsection{Identification of outliers}
\label{sec:outliers}
Some stars may exhibit intrinsic polarization, possibly due to the existence of a circumstellar disk or other asymmetries in the object (e.g., \citealt{Fadeyev2007};  \citealt{Cotton2016}; \citealt{Gontcharov2019}).
These stars usually show either  a higher degree of polarization or an unrelated polarization position angle as compared to their neighboring stars, and sometimes they display both.
As \texttt{BISP-1} assumes that all starlight-polarization data points trace the dusty magnetized ISM, the first step is to remove from the original sample all stars for which their polarization is unlikely to be of only interstellar origin. This includes intrinsically polarized stars and also any other outliers.

To identify these stars, we adopt recursive sigma-clipping in groups of neighboring stars.
Based on the sky coordinates and distance estimate of every star, we identify the group of $N$ neighbors of each star in 3D space using their heliocentric Cartesian coordinates, ignoring the distance uncertainties from parallax measurements.
We estimate the weighted-average Stokes parameters ($\hat{\bar{\mathbf{s}}}$) and associated covariance ($\hat{C}$) in the ($q,\,u)$ plane from the neighbors, and then compute the Mahalanobis distance of the polarization of the central star ($\mathbf{s}_\star$) as compared to the 2D bivariate distribution from the neighbors as
\begin{align}
    d_{\rm{Maha}}^\star = \sqrt{(\mathbf{s}_\star - \hat{\mathbf{\bar{s}}})^\dagger \, {\Sigma_\star}^{-1} \, (\mathbf{s}_\star - \hat{\mathbf{\bar{s}}})} \; ,
    \label{eq:DmahaNeighb}
\end{align}
where the noise covariance matrix of the polarization of the central star ($C_{\rm{obs}}^\star$) is added to the covariance matrix from the bivariate distribution ($\Sigma_\star = C_{\rm{obs}}^\star + \hat{C}$) to compute the Mahalanobis distance.
\begin{figure}
    \centering
    \includegraphics[trim={1.0cm .4cm 0cm 0cm},clip,width=1.\columnwidth]{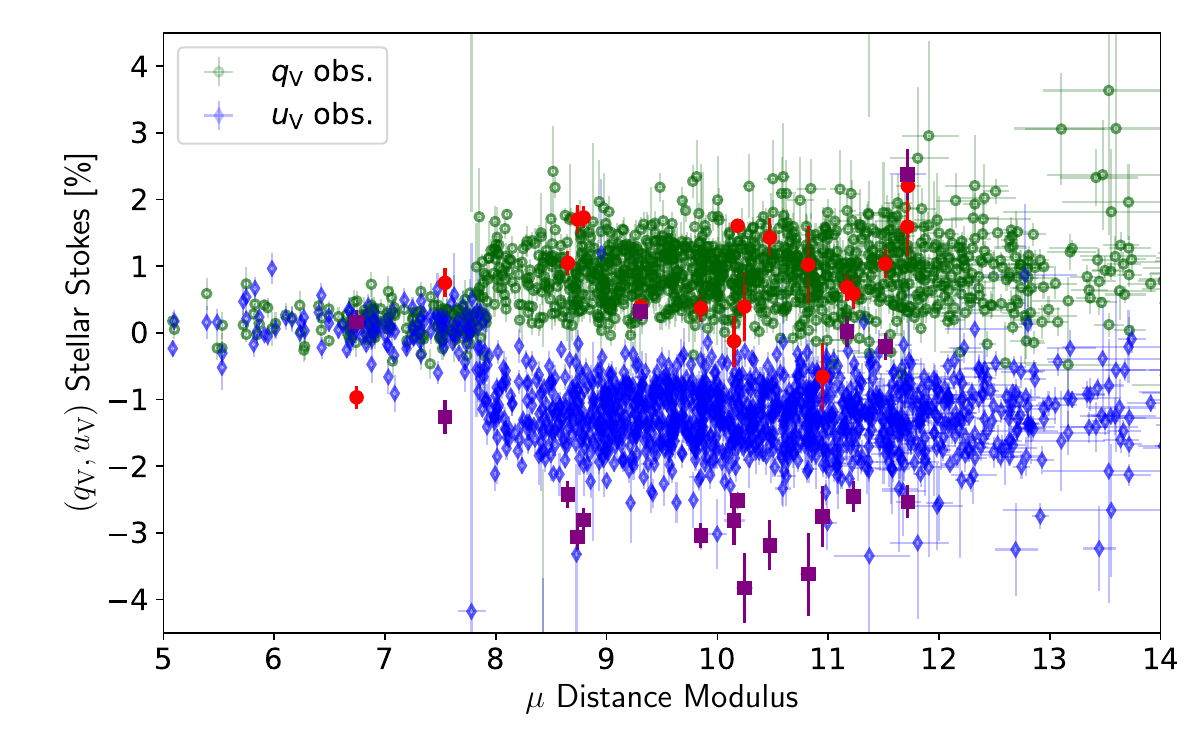} \\[-1.5ex]
    \caption{Stellar relative Stokes parameters ($q$, $u$) versus the distance modulus for the full sample. The  $q_{\rm{V}}$ and $u_{\rm{V}}$ (in the Equatorial coordinate system) are shown by green circles and blue diamonds, respectively. The vertical error bars indicate the uncertainties (noise and systematic) on the Stokes parameters and the horizontal error bars represent the uncertainties on the star distance modulus converted from the 1$\sigma$ parallax uncertainties.
    Outliers identified from the iterative sigma-clipping approach in groups of nearest neighbors are highlighted with red-filled circles for $q$ and purple-filled squares for $u$. To facilitate the visual clarity of the plot we restrict the range of ($q,\,u$) and $\mu$ values. Ten stars have $\mu <5$ and 30 stars have $\mu>14$.}
    \label{fig:outliers_qumu}
\end{figure}
If the Mahalanobis distance exceeds some threshold, the probability that the polarization of the central star is drawn from the same parent distribution as the neighbors (assumed to be representative of the magnetized ISM) is small and, therefore, the central star must be identified as an outlier. Repeating this process for all the stars in the original sample and running the whole process until no additional outliers are identified yields a catalog of outliers and a ``clean'' sample of stars whose polarization likely traces the dusty magnetized ISM. Only the clean sample of stars is then considered to infer the dusty magnetized ISM.
The exact list of outliers depends on our specific choice for the size of the neighbor groups and the adopted threshold in significance level. In addition, the sensitivity of the ``clean sample'' membership to these parameters should also ideally be tested against distance uncertainties. However, if we can recover the 3D dusty magnetized ISM from the clean sample of stars, we can test a posteriori the hypothesis that the polarization of a given star is given by the magnetized ISM only, as we do in Sect.~\ref{sec:IntrPolStar}.
This has the potential to lead to a somewhat more robust list of candidate targets for being intrinsically polarized stars, or at least outliers, and can trigger follow-up observations.
However, we notice that caution has to be made for the choice of the significance-level threshold. A too strong selection criterion would discard stars that merely pick up fluctuations of the magnetized ISM. This would subsequently lead to an underestimation of the turbulence-induced intrinsic scatter which we want to avoid. This point is further discussed in Sect.~\ref{sec:IntrPolStar}.

We apply the recursive sigma-clipping approach to the full sample of stars discussed in Sect.~\ref{sec:Data_fullSample}. We adopt a size of $N=30$ to build the groups of neighbors and we choose to flag every star as an outlier if its polarization shows a probability of less than 1\% for it to be drawn from the same parent distribution as the neighbors. In our case, the number of outliers remained constant after three iterations. Using these parameters, 18 stars out of 1448 are identified as outlier candidates. This represents a fraction of 1.2\% of our full sample.
In Fig.~\ref{fig:outliers_qumu} we show the stellar polarization data plotted against the distance modulus ($\mu = 5\,\log_{10}(d) - 5$) and highlight the outliers. These outliers are not considered in the analysis discussed below.

\subsection{Definition of subsamples}
\label{sec:subsamples}
\texttt{BISP-1} assumes that all the stars in a sample lie along a narrow, one dimensional beam. It consequently returns the structure of the dusty magnetized ISM averaged in the POS over the sky region spanned by the input sample, which we refer to our ``beam''.
The method also relies on the dust-layer model that we have introduced (\citealt{Pelgrims2023}) and which we expect to hold true as long as the magnetic field and dust density do not vary appreciably in our beam. The validity of these assumptions is tested by the data in the following.
Depending on the geometry of the volume filled by the star sample, the averaging scale in the POS may depend on distance. For example, if a conical geometry is chosen to define the star subsamples, the averaging scale at small distances is much smaller than the one at large distances. A cylindrical geometry would keep the averaging scale constant.
The number density of constraints (i.e., of stars) as a function of distance also depends on the chosen geometry of the beam, the specific size of the volume encompassed by the data, and the actual 3D spatial distribution of stars.
Hence, there is a trade-off between having a sufficient number of data points to constrain the model and the achieved resolution. The resolution also needs to be sufficiently good to minimize the POS variations of ISM properties in the beam and thus ensure that the intrinsic scatter, which we fit for, is not dominated by POS variations of the mean ISM properties.

We find that running the \texttt{BISP-1} decomposition solely on subsamples defined according to conical beams is not appropriate. The reason is that the spatial distribution of nearby stars is sparse and that, unless a very large opening angle is chosen, nearby clouds may be missed or their parameters loosely determined due to the absence of a sufficient number of constraints in the required range of distances. If a large opening angle is chosen in order to be able to capture those nearby clouds, then the signal of any faraway structure would be averaged out and therefore likely missed.
Alternatively, a cylindrical geometry for our beam would lead to a constant resolution in the transverse direction to the distance axis. However, at high and intermediate Galactic latitudes the number density of stars decreases substantially after approximately 1~kpc. Faraway clouds would then be missed, due to the sparsity of data points, unless a large cylinder radius is chosen. A cylindrical geometry would also imply that any detail at small distances would be averaged out over large angular scales. Thus, a cylindrical geometry alone is also not appropriate to define our beam samples.

\begin{figure}
    \centering
    \includegraphics[trim={.5cm .5cm .5cm .5cm},clip,width=.98\columnwidth]{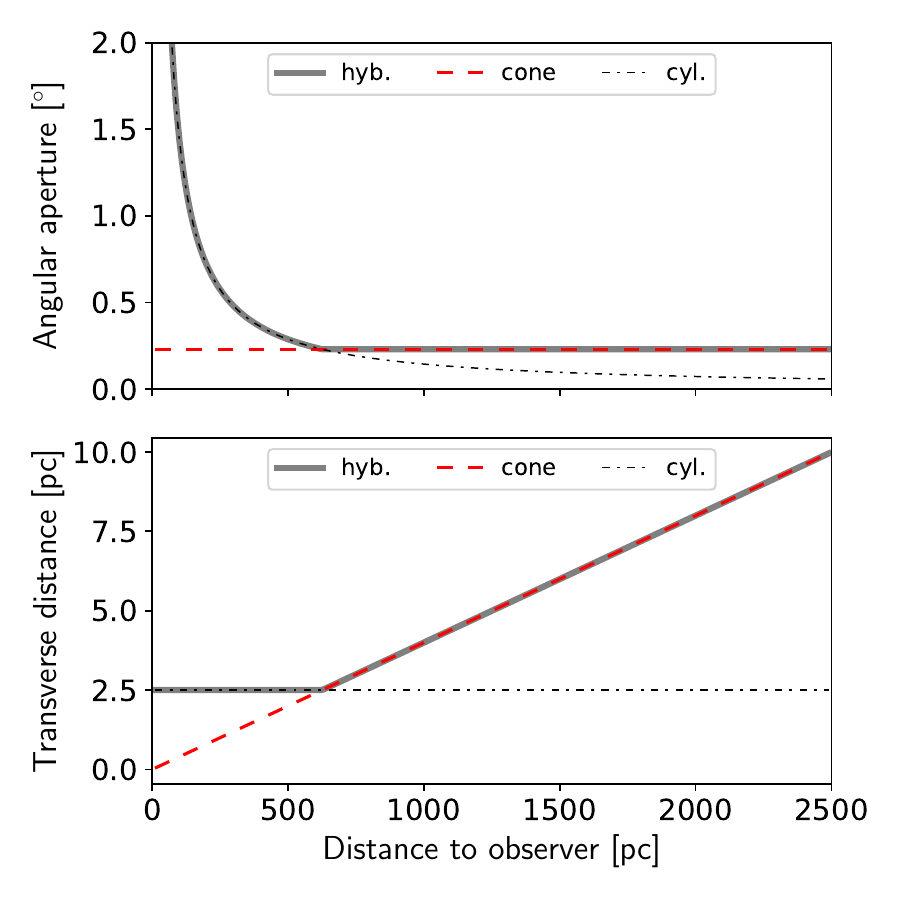}\\[-1.5ex]
    \caption{Beam sizes as a function of distance from the observer.
    (top): Angular aperture radius of our hybrid beam (continuous gray) resulting from the combination of a conical beam (red dashed) and cylindrical beam (dot-dashed).
    (bottom): Extent of the beams in the transverse direction to the LOS. Same line convention as in the top panel.
    In this example the cylinder radius is fixed at 2.5~pc and the angular radius of the cone at 13.74 arcminutes. For the hybrid beams, the cylindrical geometry prevails at distances smaller than $\approx$ 630~pc, while the conical geometry prevails at larger distances.}
    \label{fig:hybridBeam}
\end{figure}
To guarantee a good angular resolution at all distances, and to avoid missing clouds or only placing loose constraints on their distance and polarization properties, we adopt a two-step hierarchical decomposition process.
The first step is performed on samples defined according to a hybrid geometry for our beam: the beam follows a cylindrical geometry at low distances and a conical geometry at large distances.
For the second step, the star samples are defined in a beam with conical geometry only. The idea is to perform the \texttt{BISP-1} decomposition in the second step using priors defined from the posterior distributions obtained in the first step. In this way, constraints on distances and polarization properties of nearby clouds are obtained from the decomposition on hybrid-beam samples (with larger number of stars at lower distances) while good angular resolution is achieved, even at low distances, from the conical-beam samples. We use the same opening angle for the conical part of the hybrid beam and the purely conical beam so that they match at large distances.

In Fig.~\ref{fig:hybridBeam} we show the distance-dependent angular radius of our beams along with the corresponding physical scale in the POS for the specific choice of opening angle and cylinder radius that we use.
Our hybrid beam centered on a given LOS has an angular size that depends on distance. Any star is considered as part of a given subsample if its angular separation to the LOS is lower than the maximum between the opening angle of the cone and the distance-dependent angular size of the cylinder evaluated at the star distance. For the conical beam, the angular separation cutoff is constant.

\smallskip

To cover an extended region of the sky, such as the one we observed, we adopt a moving-window strategy. That is, we sample the observed region with a large number of sightlines. For each LOS, we define a subsample of stars according to our choice for the beam geometries. The star samples of neighboring sightlines are overlapping and, in the first step, the overlap extends to larger angular scale for nearby stars than that for distant stars. This strategy ensures a continuous scan of the observed region and also implies that the results of the LOS decomposition of neighboring sightlines will not be independent.

\smallskip

\begin{figure}
    \centering
    (a) \\[-.3ex]
    \includegraphics[trim={-.3cm .5cm 1.2cm 0.3cm},clip,width=1.\columnwidth]{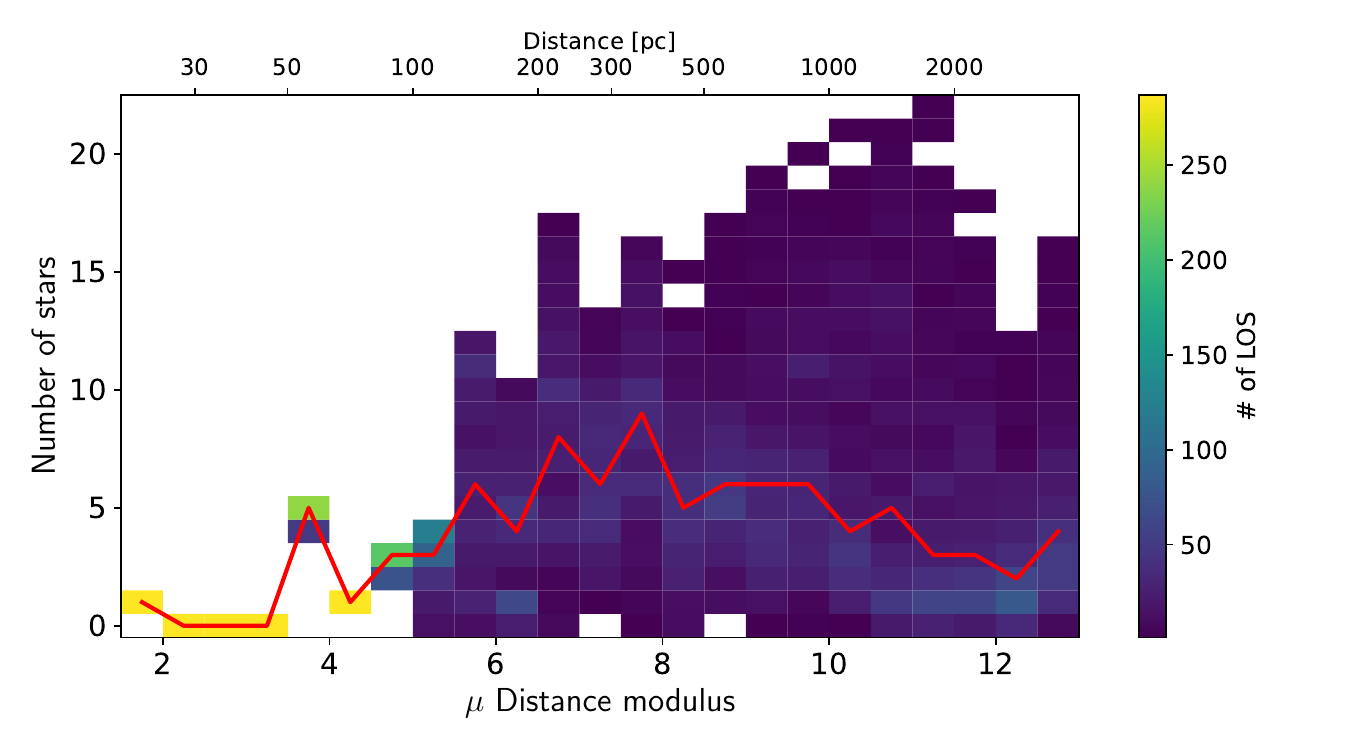}\\[.5ex]
    (b) \\[-.3ex]
    \includegraphics[trim={-.3cm .5cm 1.2cm 0.3cm},clip,width=1.\columnwidth]{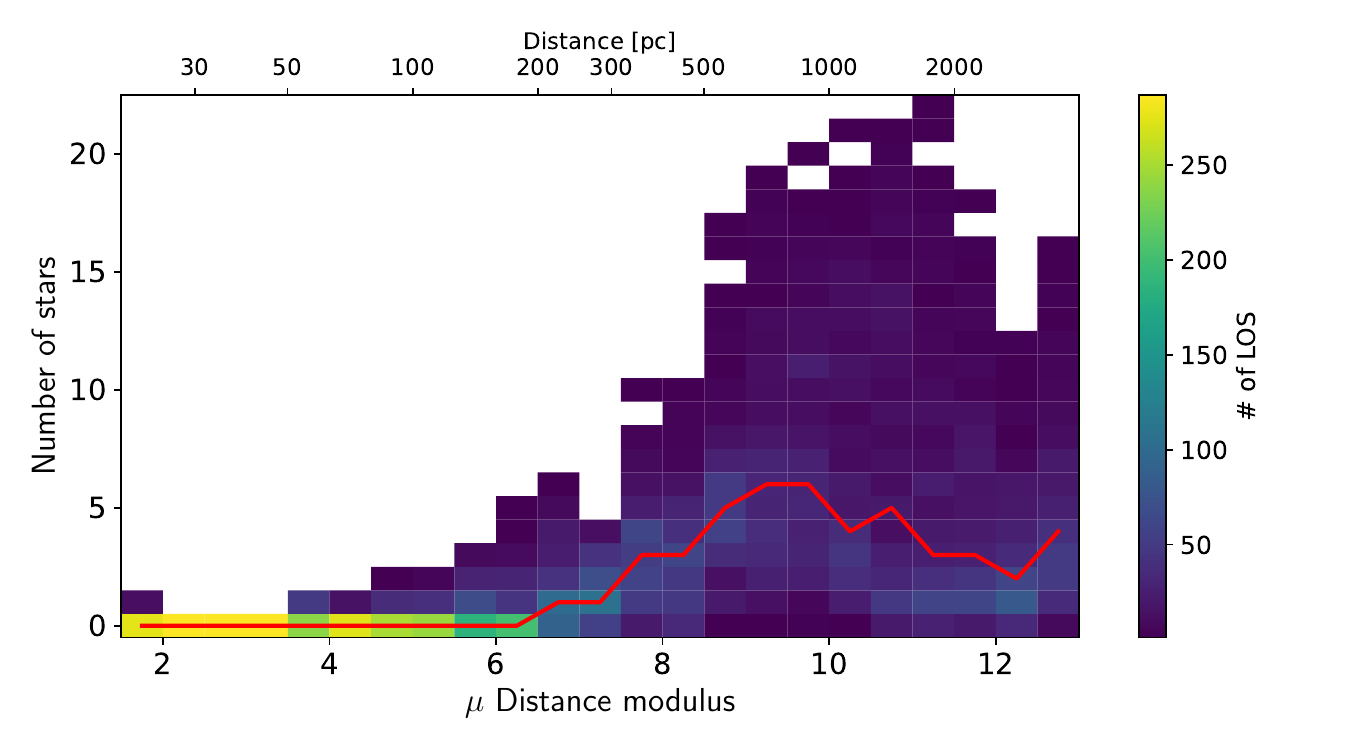}\\[.5ex]
    (c) \\[-.3ex]
    \includegraphics[trim={-.3cm .5cm 1.2cm 0.3cm},clip,width=1.\columnwidth]{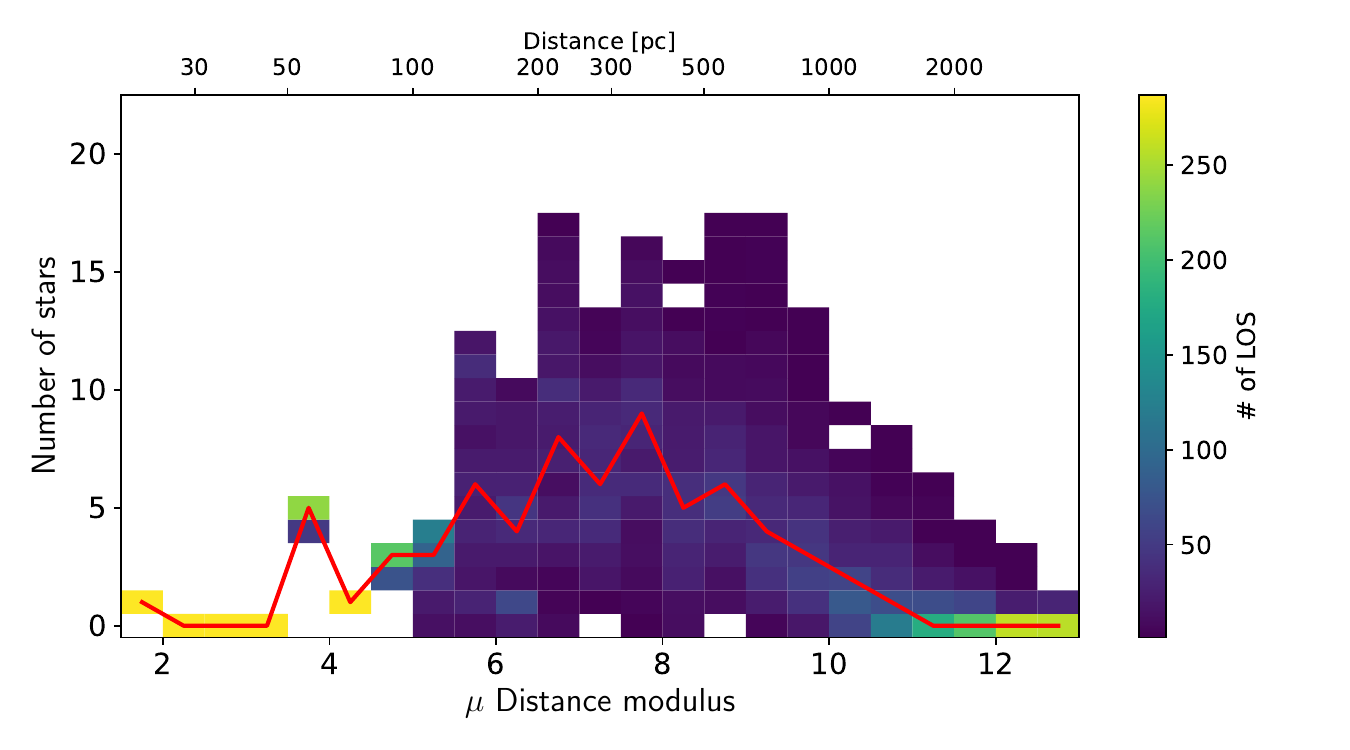}\\[-1.5ex]
    \caption{Distribution of the number of stars in beam samples.
    \textbf{(a)} Number of stars per bin of distance modulus for the 287 sightlines sampling the observed regions when the beam geometry is hybrid. The colors indicate the number of sightlines for which a specific number of stars in a given distance bin is observed. The red-continuous line indicates the median number of stars per bin of distance for the entire set of sightlines.
    \textbf{(b)} Same as for (a) but for the conical-beam geometry.
    \textbf{(c)} Same as for (a) but for the cylindrical-beam geometry.}
    \label{fig:NstarDistribution}
\end{figure}
\begin{figure}
    \centering
    \phantom{(a)} \\[-.3ex]
    \includegraphics[trim={.0cm .5cm 0cm 0.4cm},clip,width=1.\columnwidth]{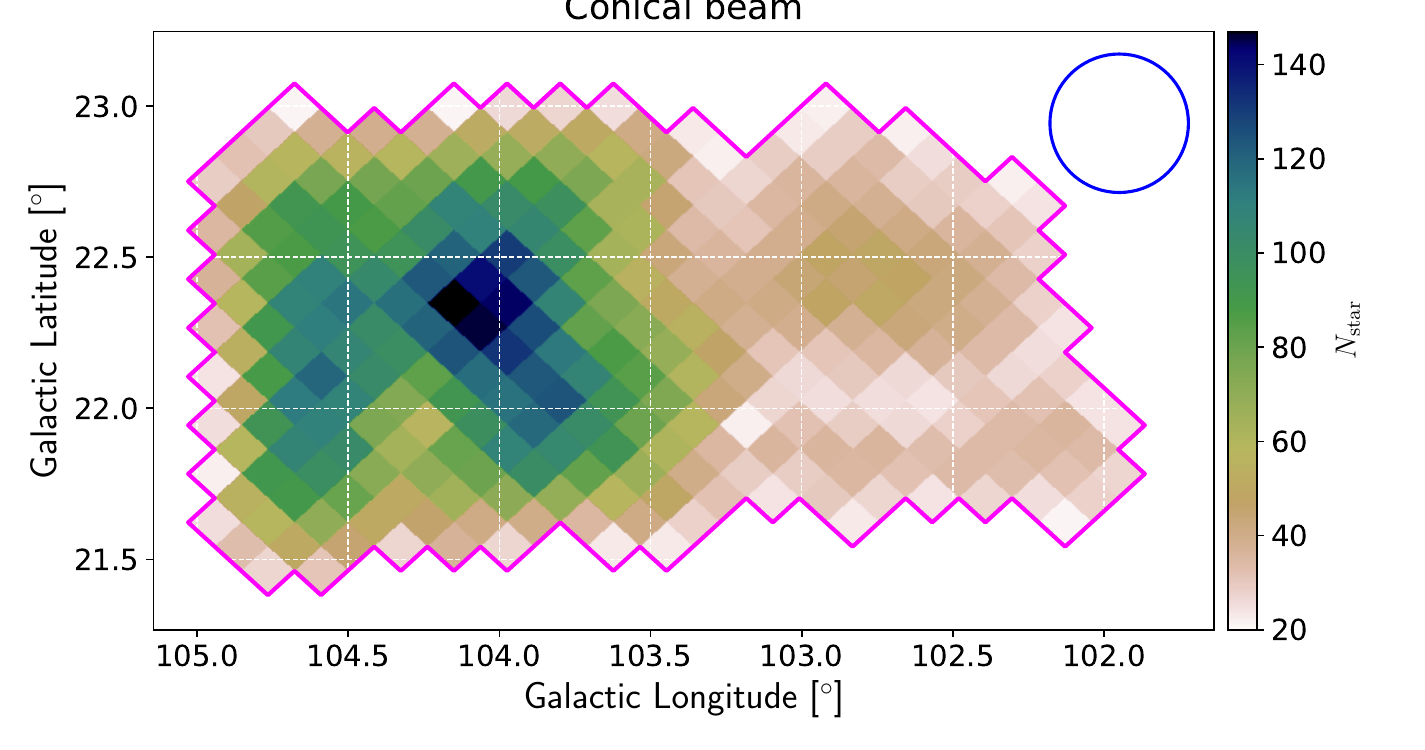}\\[-1.5ex]
    \caption{Number of stars per conical beam. The blue circle in the top right corner indicates the size of our conical beam with an angular radius of 13.74 arcmin. The magenta contour surrounds all HEALPix pixels whose center coincides with the center of the beams defining our star subsamples.}
    \label{fig:NstarMap}
\end{figure}
For our stellar polarization measurements spanning a region of about four square degrees, we sample the sky according to a HEALPix tessellation (\citealt{Gorski2005}; \citealt{Zonca2019}) with the resolution parameter $N_{\rm{side}} = 512$. Each pixel center defines a LOS which we take as the symmetry axis of our beam. The angular separation of neighboring sightlines is about 6.9 arcmin. To define our hybrid beam, we adopt a value of 2.5~pc for the cylinder radius and an angular radius of 13.74 arcmin for the cone. The conical beam has the same value of 13.74 arcmin for the angular radius. Each conical beam spans a sky area of about 0.16 square degrees.
We choose the angular radius of the conical beam so that (i) at least 20 stars are contained within each LOS and (ii) every pixel has at least two neighbors to ensure continuity at the edges of the map.
The sky region is covered by an ensemble of 287 sightlines.
Panels (a), (b), and (c) of Fig.~\ref{fig:NstarDistribution} show the number of stars per bins of distance modulus for all the 287 sightlines covering the regions of the hybrid beams, of the conical beams, and of the cylindrical beams, respectively.
We see that the density of stars is generally very low at distances smaller than 300~pc for the conical beams and at distances larger than 1~kpc for the cylindrical beams. This justifies the use of the hybrid beams in extracting information of nearby and distant ISM clouds. With our choice of beam parameters, the geometry of the hybrid beam transitions from cylindrical to conical at a distance of about 630~pc ($\mu \approx 9$).
Figure~\ref{fig:NstarMap} shows the number of stars per beam samples for the conical beam geometry. The higher density for points at $l \gtrsim 103.5^\circ$ is due to the aforementioned survey strategy (Sect.~\ref{sec:surveyStrategy}).

\subsection{LOS decomposition of the dusty magnetized ISM}
\label{sec:LOSdeco}
Once the star subsamples are defined for the entire observed region, we independently apply the \texttt{BISP-1} code to each subsample, first on the hybrid-beam samples and then on the conical-beam samples.

\subsubsection{Step~1 -- LOS decomposition in hybrid beams}
\label{sec:Step1}
For each LOS, we test the layer-model for one to four clouds along the distance with uniform priors on all the model parameters.
The limits of the priors on cloud parallaxes are set so that the minimum allowed distance is 20~pc and the maximum allowed distance of the farthest cloud is the minimum between 3.5~kpc and the maximum distance of the stars in the analyzed sample. The upper distance limit may thus vary from one LOS to another. The value of 3.5~kpc (corresponding to a distance modulus of $\mu \approx 12.7$) originates from the fact that beyond this distance our data sample generally become very scarce as also seen in Fig.~\ref{fig:starSample_skyAndMu}.
In addition, we make sure that there are at least five stars between clouds. This is required by the \texttt{BISP-1} code (\citealt{Pelgrims2023}).

The uniform priors for the cloud polarization parameters are set as follows. For the mean polarization parameters ($q_{\rm{C}}$ and $u_{\rm{C}}$), we compute the maximum of the absolute values of both the stellar Stokes parameters in the star samples. This value is used to define the limits of the top-hat priors on both $q_{\rm{C}}$ and $u_{\rm{C}}$. This definition of the prior limits, while not fully general, is valid in our case as we know that the extinction is dominated by nearby components as shown for example by comparing {\it Planck} dust column-density map and 3D star extinction maps (e.g., \citealt{OCallaghan2023}).
The diagonal elements of the intrinsic-scatter covariance matrix are positive definite and we require $C_{{\rm{int}},qq},\, C_{{\rm{int}},uu} \in [0,\,10^{-4}]$. This is a very loose range for possible values as it allows for a spread in stellar Stokes parameter due to turbulence as high as 1\% in degree of polarization. The off-diagonal element $C_{{\rm{int}},qu}$ is initially bounded by $\pm 10^{-4}$ but is further constrained to verify $|C_{{\rm{int}},qu}| < (C_{{\rm{int}},qq}\, C_{{\rm{int}},uu})^{1/2}$ inside \texttt{BISP-1} to ensure that the covariance matrix is invertible.

We use \texttt{BISP-1} to run the nested sampling experiment using 1000 live points and sample the parameter space until an uncertainty of around 0.01 is achieved on the log of the model evidence. Typically, this requires 15,000 to 80,000 nested sampling iterations, corresponding to several hundred million calls of the log-likelihood function of each model.

\smallskip

For each of the tested models, \texttt{BISP-1} leads to estimates of the log of the evidence, the maximum log-likelihood value and to the estimated posterior distributions of all model parameters.
As already demonstrated and pointed out in (\citealt{Pelgrims2023}), the posterior distributions of the cloud parallaxes have generally complex shapes, mainly due to the sparse and uneven distribution of stars along distance, and might also be piled up on the far edge of the prior domains. The latter case happens whenever the data are not enough to determine the cloud properties or if there is no cloud to be found.

To deal with this peculiarity, we follow the same idea as in (\citealt{Pelgrims2023}) and rely on the following automated analysis of the marginalized posterior distribution on the cloud parallax.
The idea is that the value at maximum-likelihood must belong to (one of) the main mode(s) of the marginalized posterior distribution on the cloud parallax and that this mode must not be piled-up on the lower limit of the prior. If both criteria are verified, then we qualify the fit as valid. We then select from the $6 \times N_{\rm{C}}$ dimensional posterior distribution (where $N_{\rm{C}}$ is the number of layer in the model) all samples that belong to this mode for the remainder of the analysis.
In practice, we analyze the marginalized posterior distribution using the peak-finder algorithm \texttt{find\_peaks} of the SciPy Python library (\citealt{Virtanen2020}) which identifies all local maxima through simple comparison of neighboring values. We thus find local maxima of the marginalized posterior distribution and the range of parallax values corresponding to the extent of the corresponding peaks. We compute the fraction of the posterior distribution which corresponds to each peak and we consider that it is (one of) the dominant peak(s) if this fraction exceeds the threshold of 30 per cent. Our results do not depend strongly on this choice.

We consider all solutions for which the posterior distribution on the cloud parallax of the farthest cloud passes this selection criterion to be valid. Any solution which does not satisfy this criterion is discarded in this first step. 
Finally, for each tested model $j$, we compute the Akaike Information Criterion (AIC) as
\begin{align}\label{eq:AIC}
    {\rm{AIC}}_j = 2 \, M - 2 \, \log(\hat{\mathcal{L}}_j) \;,
\end{align}
based on the estimated maximum likelihood ($\hat{\mathcal{L}}_j$) and the number of parameters in the model ($M$).
We then compare the model performances by computing the probability
\begin{align}\label{eq:ModelProb}
    P_{j|\{m\}} = \exp \left( (\min_m \{{\rm{AIC}}_m\}  - {\rm{AIC}}_j )/2 \right) \; ,
\end{align}
that, among the tested models $\{m\}$, each model $j$ is actually the one that minimizes the loss of information (\citealt{Boisbunon2014}), as in (\citealt{Pelgrims2023}). The best model is the one with $P_{j|\{m\}} = 1$.
We note that, according to the selection procedure explained above, we are giving some chances to solutions with large number of clouds (up to four) to be considered as the best model while the data itself might not be enough to place strong constraints on the farthest cloud. However, we checked all the solutions and found no evidence of spurious detections that might have resulted from bad fits.

\begin{figure}
    \centering
    \vspace{.3cm}
    \includegraphics[trim={-.1cm .5cm 1.cm 0.4cm},clip,width=1.\columnwidth]{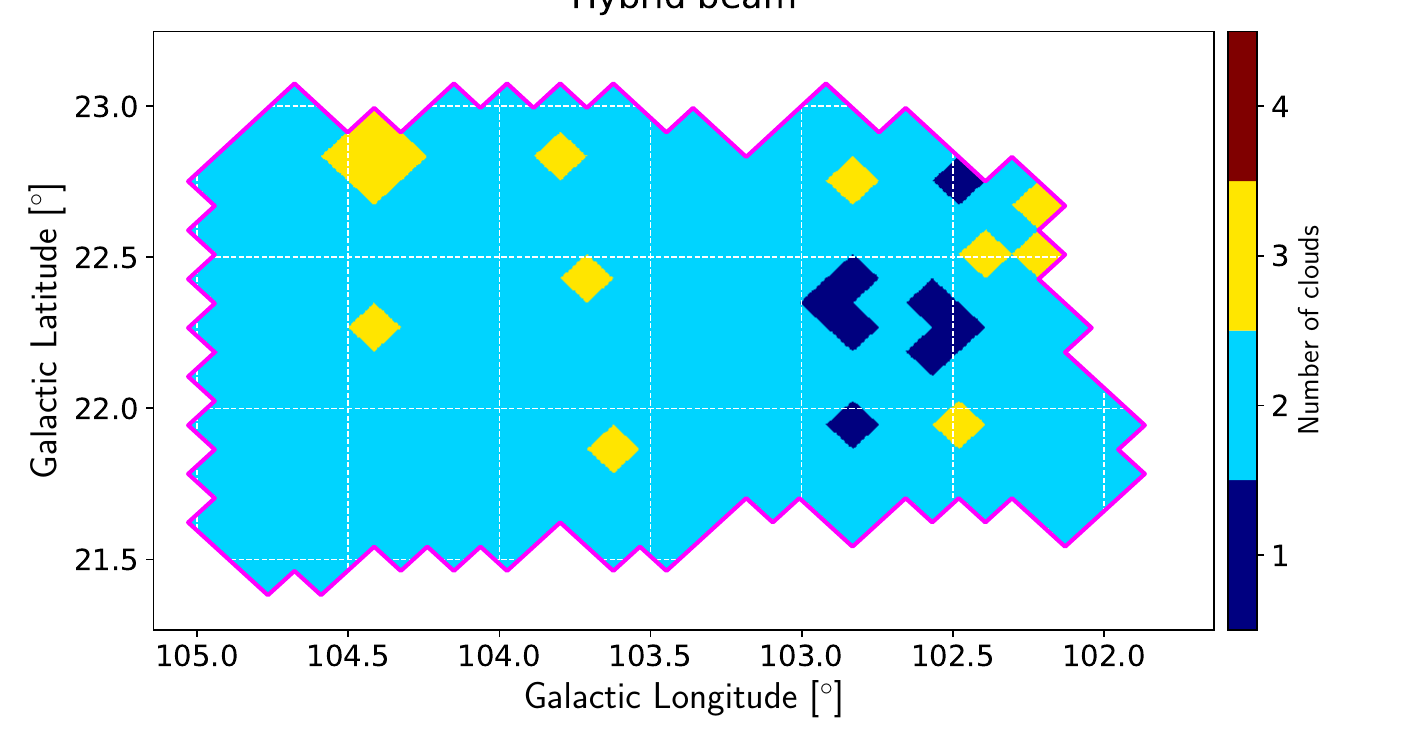}\\[-1.5ex]
    \caption{Map of the number of dust clouds identified at the end of Step~1 from the analysis of the hybrid-beam samples obtained for each beam centered on the pixel centers.
    }
    \label{fig:Ncloud_HybridBeam}
\end{figure}

A map of the number of clouds per LOS determined in this first step from the analysis of the hybrid-beam samples is shown in Fig.~\ref{fig:Ncloud_HybridBeam}.
For most of the surveyed area, we find evidence of two clouds per LOS. The maximum number of clouds is three, and a few sightlines show only one cloud.

\subsubsection{From posteriors to priors}
We now aim at informing the maximum-likelihood analysis of data in conical beams from the results of the analysis of hybrid beam samples.
To do so, we would ideally want to resample the posterior distributions obtained in Step~1, at least for the cloud-parallax parameters. This is however currently not possible using \texttt{BISP-1} which does not make it possible to input prior samples as it relies on the Python nested sampling software \texttt{dynesty} which does not have this feature.
For the purpose of the this work, we thus work around this limitation as explained below and leave the development of improved solutions for the future.
We choose to ignore the possible correlations between model parameters and only extract (and propagate) information on cloud parallax and cloud mean-polarization properties from the marginalized posterior distributions of these parameters.
By ignoring possible correlations, we know we are losing part of the information gained in Step~1.

For each of the valid models selected in Step~1 and their corresponding estimated (marginalized) posteriors, we define priors on the parameters of the models to be constrained by the data in Step~2. For the cases discussed above where the marginalized posterior distribution on the cloud parallax of the farthest cloud is multimodal, only the ``valid'' subset of the posterior samples is used to define the priors.

Neither a Gaussian nor a top-hat distribution generally represent the posterior distributions of the cloud parallax well; however, we find that in our case, they are sufficiently effective to impose constraints on the cloud's parallax.
We choose to construct Gaussian or top-hat priors on cloud parallaxes from the estimated posteriors as follows. We denote $\{ \varpi_{\rm{C}}\}$ the posterior sample of parallax for a given cloud.
For a given cloud, the estimated mean ($\hat{\bar{\varpi}}_{\rm{C}})$ and standard deviation ($\hat{\sigma}_{\varpi_{\rm{C}}}$) of the Gaussian prior, and the estimated minimum ($\hat{\varpi}_{\rm{C}}^{\rm{min}}$) and maximum ($\hat{\varpi}_{\rm{C}}^{\rm{max}}$) limits of the top-hat prior are obtained from percentiles of the posterior sample of the cloud parallax as
\begin{align}
    &\hat{\bar{\varpi}}_{\rm{C}} = \{\varpi_{\rm{C}}\}_{50} \nonumber \\
    &\hat{\sigma}_{\varpi_{\rm{C}}} = (\{\varpi_{\rm{C}}\}_{84} - \{\varpi_{\rm{C}}\}_{16})/2 \nonumber \\
    &\hat{\varpi}_{\rm{C}}^{\rm{min}} = \{\varpi_{\rm{C}}\}_{0.01} \nonumber \\
    &\hat{\varpi}_{\rm{C}}^{\rm{max}} = \{\varpi_{\rm{C}}\}_{99.99} \;,
\end{align}
where $\{\varpi_{\rm{C}}\}_{\rm{X}}$ denotes the $\rm{X}$-th percentile of the sample distribution.
The prior distributions on the cloud parallax are then defined as follows:
\begin{align}
    &\mathcal{P}_{\rm{G}}(\varpi_{\rm{C}}) = \frac{1}{\sqrt{2\pi}\hat{\sigma}_{\varpi_{\rm{C}}}} \, \exp\left\lbrace - \frac{(\varpi_{\rm{C}} - \hat{\bar{\varpi}}_{\rm{C}})^2}{2\,{\hat{\sigma}_{\varpi_{\rm{C}}}}^2}\right\rbrace \\
    &\mathcal{P}_{\rm{H}}(\varpi_{\rm{C}}) =
    \begin{cases}
        \frac{1}{\hat{\varpi}_{\rm{C}}^{\rm{max}} - \hat{\varpi}_{\rm{C}}^{\rm{min}}} & \text{if } \varpi_{\rm{C}} \in [ \hat{\varpi}_{\rm{C}}^{\rm{min}},\,\hat{\varpi}_{\rm{C}}^{\rm{max}} ] \\
        0 & \text{otherwise} \; ,
    \end{cases}
\end{align}
for the Gaussian and the top-hat respectively.

For our data, we find that, even for the cases where the three-layer model is favored, only the two nearest clouds are located in the distance range where the beam geometry is cylindrical. Therefore, we do not need to update the priors of the most distant cloud parameters and simply left it as is for Step~1. That is, we inform the modeling of the data in conical beams (Step~2) from the fit in hybrid beams only for clouds that are found in the distance range where the hybrid and conical beams differ.
If $\hat{\bar{\varpi}}_{\rm{C}}/\hat{\sigma}_{\varpi_{\rm{C}}} > 2$ and if there are at least five stars with parallaxes lower than $\hat{\varpi}_{\rm{C}} - \hat{\sigma}_{\varpi_{\rm{C}}}$, then we adopt the Gaussian prior. In all other cases, we adopt a top-hat prior for this parameter.
This is to avoid unphysical negative parallax values and to allow for the possibility of a farther away cloud.

Defining the priors on the cloud mean polarization from the corresponding posteriors is easier than for the parallax as the marginalized posteriors are generally close to Gaussian.
To build the Gaussian priors on the mean Stokes parameters of the cloud, we consider the means of the posterior distribution samples and both the standard deviations of those samples (taken individually) and the mean of the posterior distribution of the corresponding element of the intrinsic-scatter covariance matrix.
If $\hat{\bar{q}}_{\rm{C}}$ and $\hat{\sigma}_{q_{\rm{C}}}$ are the estimated mean and standard deviation of the posterior distributions on $q_{\rm{C}}$ (the mean $q$ Stokes parameter of the cloud), and if $\hat{C}_{{\rm{int}},qq}$ is the estimated mean of the posterior on the $qq$ element of the covariance matrix of the intrinsic scatter, then the Gaussian prior on $q_{\rm{C}}$ is defined with a mean of $\hat{\bar{q}}_{\rm{C}}$ and a standard deviation of $({{\hat{\sigma}_{q_{\rm{C}}}}^2 + \hat{C}_{{\rm{int}},qq}})^{1/2}$. The same applies for $u_{\rm{C}}$.
We include the terms from the intrinsic scatter in the definition of the priors on the mean Stokes parameters to account for the fact that, in the conical beams (higher angular resolution at lower distance), the ``mean'' polarization may, in general, pick a local fluctuation of the turbulent magnetized ISM defined at larger angular scales in the hybrid beams.
Accordingly, we do not modify the priors for the elements of the intrinsic-scatter covariance matrix because, without further assumptions, we do not know how the intrinsic scatter evolves as a function of physical and angular scales. Consequently, we keep the same uniform priors defined above throughout the analysis.

\subsubsection{Step~2 -- LOS decomposition in conical beams}
\label{sec:Step2}
For each LOS, we have obtained a LOS decomposition in the hybrid beam, we have selected the best model, and we have defined priors from the estimated posterior distributions of this model.
In this second step, we use \texttt{BISP-1} to decompose the starlight polarization along distance for the star samples defined in the conical beams using the information gained above.
For each LOS, we test the models with two, three, and four layers and use the informed priors for the two nearest clouds only. For the cases where the one-layer model was favored in Step~1, we test this model and a higher number of layers using the informed priors only on the nearest layer. In all cases we force the additional cloud(s) to be located at larger distances.

We use \texttt{BISP-1} to run the nested sampling experiment using 1000 live points and sample the parameter space until an uncertainty of around 0.01 is achieved on the log of the model evidence. The required number of nested-sampling iterations was similar or higher than in Step~1.

\begin{figure}
    \centering
    \vspace{.1cm}
    \includegraphics[trim={-.1cm .5cm 1.cm 0.4cm},clip,width=1.\columnwidth]{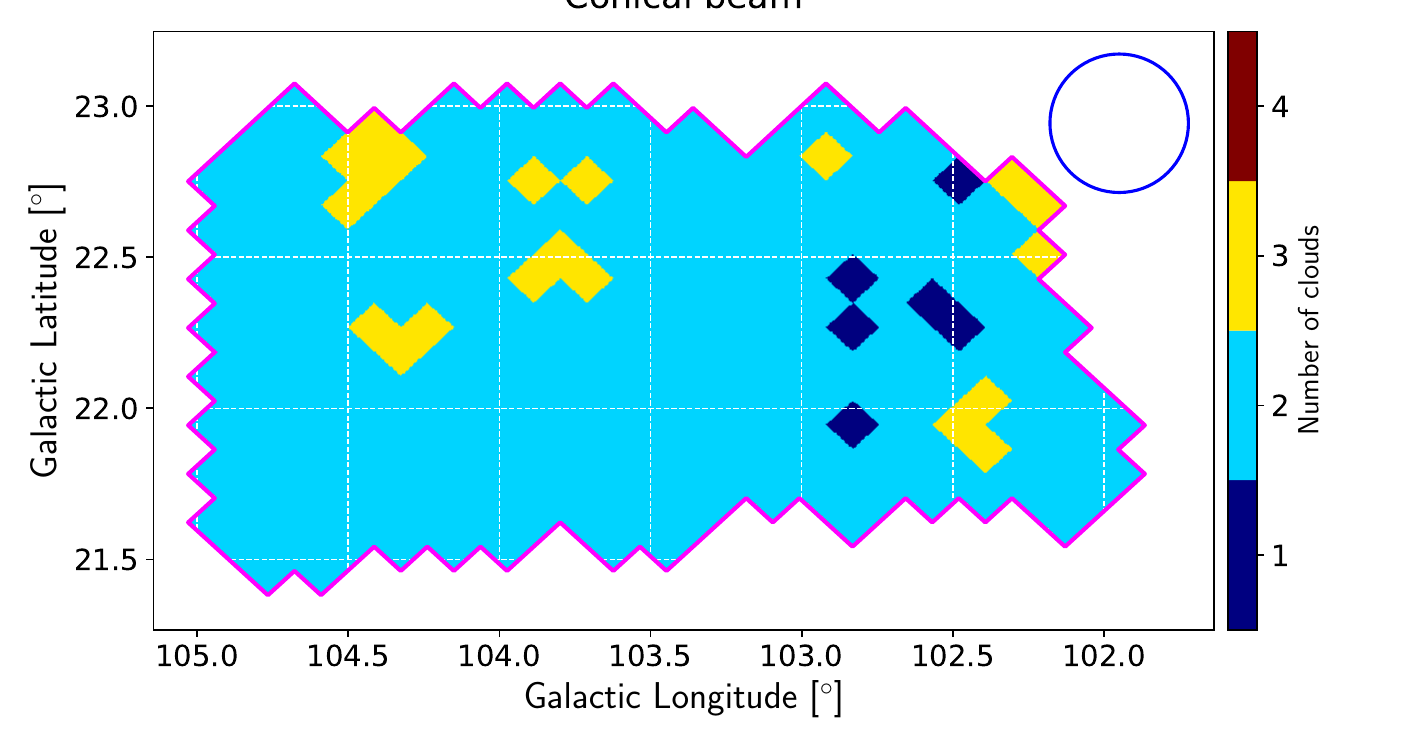}\\[-1.5ex]
    \caption{Map of the number of dust clouds along the sightlines identified at the end of Step~2 in the conical beam centered on the pixel centers. The blue circle in the top-right corner indicates the extent of the conical beam.}
    \label{fig:Ncloud_ConicalBeam}
\end{figure}

When all models have been evaluated, we proceed as in Step~1 to inspect the solutions based on the marginalized posterior on the cloud parallax and keep only the valid reconstructions, and to select the best-model which minimizes the loss of information based on the AIC criterion.
Figure~\ref{fig:Ncloud_ConicalBeam} shows the number of clouds along each of the sightlines sampling the observed regions, as in Fig.~\ref{fig:Ncloud_HybridBeam}. Again, the number of clouds ranges from one to three. A large fraction of the sightlines in this sky area intersects two clouds along the LOS. This map is further discussed in the Sect.~\ref{sec:results}.

\smallskip

The solutions of neighboring sightlines are not independent. First, the samples of stars from neighboring sightlines overlap, even for the conical beams, because we explicitly decided to oversample the observed region with a large number of sightlines and the beam size is more than twice as large as the angular separation between adjacent sightlines. Second, the solutions obtained at the end of the second step are constrained by the larger angular scales at short distances since the priors are defined from the results obtained from the hybrid beam samples. As a result, we obtain the posterior distributions for our decomposition of the POS component of the magnetic field in dusty regions for a simply connected 3D volume without gaps.

\subsection{Validation}
\label{sec:validation}
To validate the result of our tomographic reconstruction, we rely on inspection of the significance of the polarization residuals of all the stars on which the reconstruction is based. We consider that we can be confident in our reconstruction if only a few data points show significant deviations from our model prediction, and if these data points are not clustered in the 3D volume. If, on the contrary, too large a fraction of stars show significant residuals, or if they are clustered in 3D space, this would indicate flaws or limitations in our tomographic reconstruction.
In particular, the clustering of significant residuals would indicate the presence of features in the data that the model is not able to account for within uncertainties.

To estimate the significance of the residuals for all the stars taken individually, we compare the individual stellar observational data to the modeled data from the tomography results. We proceed as follows.
Since we do not have reconstruction for each LOS toward each star individually but only for each LOS toward the center of HEALPix pixels at the center of our sample beams, we first identify the pixel in which the star falls in. We then read the posterior samples of the best-model decomposition obtained earlier for the corresponding beam.

By resampling the posterior distribution, we can then evaluate the expected polarization of the star at its given distance.
According to our layer model (\citealt{Pelgrims2023}), the ISM contribution to a star polarization at distance $d_\star$ toward a given LOS is thus described by the stochastic model:
\begin{align}
    \mathbf{m}_\star &= {\rm{G}}_2(\mathbf{\bar{m}},\,C_{\rm{int}}),
\end{align}
with $\mathbf{m}_\star = (\hat{q}_\star \, \hat{u}_\star)^\dagger$ and where ${\rm{G}}_2(\mathbf{\bar{m}},\,C_{\rm{int}})$ denotes a bivariate normal distribution with mean $\mathbf{\bar{m}}$ and covariance matrix $C_{\rm{int}}$.
The values of the cloud's mean polarization and covariance matrix are obtained from the posterior samples and added, cloud wise, up to the distance of the star.
The noise covariance matrix of the polarization measurements of the star can then be added to the intrinsic-scatter covariance matrix ($\Sigma_\star = C_{\rm{obs}}^\star + C_{\rm{int}}$) and the observation ($\mathbf{s}_\star = (q_\star \,u_\star)^\dagger$) compared to the value predicted from our 3D reconstruction using the Mahalanobis distance:
\begin{align}
    d_{\rm{Maha}}^\star = \sqrt{(\mathbf{s}_\star - \mathbf{\bar{m}})^\dagger \, {\Sigma_\star}^{-1} \, (\mathbf{s}_\star - \mathbf{\bar{m}})} \; .
    \label{eq:Dmaha}
\end{align}
For each individual star, a single value of $d_{\rm{Maha}}^\star$ is obtained for each sample of the posteriors.
By resampling the posteriors of the model parameters and resampling the parallax distribution of the star, we obtain a distribution of $d_{\rm{Maha}}^\star$. This distribution informs us on the likelihood that the observed polarization is due to the dusty magnetized ISM given our 3D reconstruction. This estimate accounts for both the turbulence-induced scatter and observational noise in polarization and parallax. Large values of $d_{\rm{Maha}}^\star$ indicate significant residuals. In practice, a star is identified as an outlier (i.e., a star whose polarization is poorly predicted by our LOS model and thus showing significant residuals) if the median of its Mahalanobis-distance distribution is larger than a given threshold value.
In 2D, the square of the Mahalanobis distances is expected to follow a $\chi^2$ distribution with two degrees of freedom. Accordingly, we can compute the threshold value corresponding to a given probability ($P_{\rm{th}}$) to observe by chance a Mahalanobis distance greater than that. In 2D the threshold value is obtained as $d_{\rm{Maha}}^{\rm{\,th}} = \sqrt{-2 \, \log (P_{\rm{th}})}$, where $\log$ is the natural logarithm.

Having obtained the Mahalanobis-distance distributions for all the stars in our sample, we look at the locations of stars with significant residuals in the 3D space.
\begin{figure}
    \centering
    \includegraphics[trim={0.6cm 0.4cm 0.4cm 0.4cm},clip,width=.98\columnwidth]{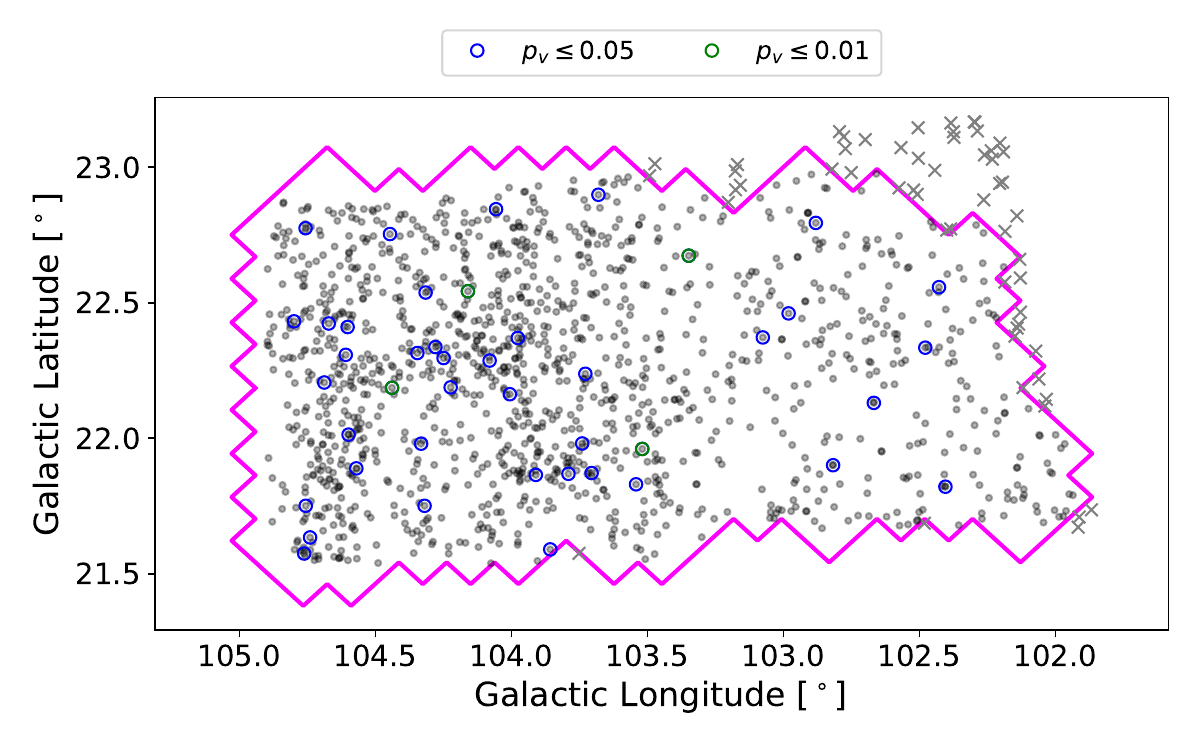}\\[-1.5ex]
    \caption{Sky map of residual significance. All the stars in the sample from which we performed our tomographic inversion are represented.
    Transparent gray dots (gray crosses) show stars that do (do not) fall in an HEALPix pixel for which we have tomography data. Blue and green circles show the stars for which the median of their distribution of Mahalanobis distances exceed a threshold values corresponding to the $p$-value indicated in the legend (see text). The lower the $p$-value, the more significant the residuals.}
    \label{fig:ResidualCheck_ISMsample}
\end{figure}
We show in Fig.~\ref{fig:ResidualCheck_ISMsample} the sky locations of the stars for which the median of their Mahalanobis-distance distributions exceed threshold values corresponding to the probabilities of 5\% and 1\% of observing by chance greater values than that.
The stars with significant residuals corresponding to lower probability of being compatible with the model are highlighted with blue and green circles, respectively.

The fractions of our star sample which show $p$-values lower than the 5\% and 1\% threshold are 3\% and 0.3\%, respectively, and no star shows a $p$-value lower than 0.02\%. These values are further discussed in Sect.~\ref{sec:IntrPolStar}.
We see from Fig.~\ref{fig:ResidualCheck_ISMsample} that the stars with significant residuals do not cluster in particular places of the sky. Visually it seems that there is an excess of significant residuals in the eastern half of the observed region but this is due to the larger stellar density in that region. However, we observed (not shown on the figure) mild preference for the stars with significant residuals to be located at large distance. This is likely an effect of the low number density of stars at large distance and of the limited angular size of our sample beam as, for example, two of the four stars with their $p$-value lower than 1\% are the farthest stars in their beam. Though, as the residuals are not very significant and that this trend is mild, we consider that this effect has no substantial effect on the present results.

This validation test makes it possible to verify the reliability of the assumptions underlying our modeling by looking at the spatial distribution of the polarization residuals. On the one hand, the violation of the thin-layer assumption would lead to systematic increases of the polarization residuals close to the reconstructed cloud distances. And, on the other hand, any significant variation of the polarization signal in the POS within the beam would lead to gradients in the residuals.
We searched for such possible trends and could not find any. This suggests that our working assumptions are appropriate to model our dataset.

\section{Results}
\label{sec:results}
The output of the 3D-inversion pipeline developed in the previous section consists of an ensemble of LOS decomposition of stellar polarization for non-independent samples. For each beam, we identified the number of components (shown in Fig.~\ref{fig:Ncloud_ConicalBeam}) and obtained the posterior distributions on their distances (cloud parallaxes, $\varpi_{\rm{C}}$) and polarization properties ($q_{\rm{C}}$, $u_{\rm{C}}$, $C_{\rm{int}}$). From these, we can now infer the properties of the POS component of the magnetic field in dusty regions.
We examine the results at the mean of the posterior distribution in Sect.~\ref{sec:basicResults} to infer the main features of the dusty magnetized ISM in the observed 3D volume. Then we build 3D maps of the posterior distributions in Sect.~\ref{sec:3DmapMaking} and visualize the main output in Sect.~\ref{sec:3DmapVisualization}. The discussion, interpretation, and validation of the results are provided in Sect.~\ref{sec:discussion} along with caveats of the method.

\subsection{Basic exploration of the output}
\label{sec:basicResults}
\begin{figure}
    \centering
    \includegraphics[trim={0.2cm 0.3cm 0.1cm 0.4cm},clip,width=1.\columnwidth]{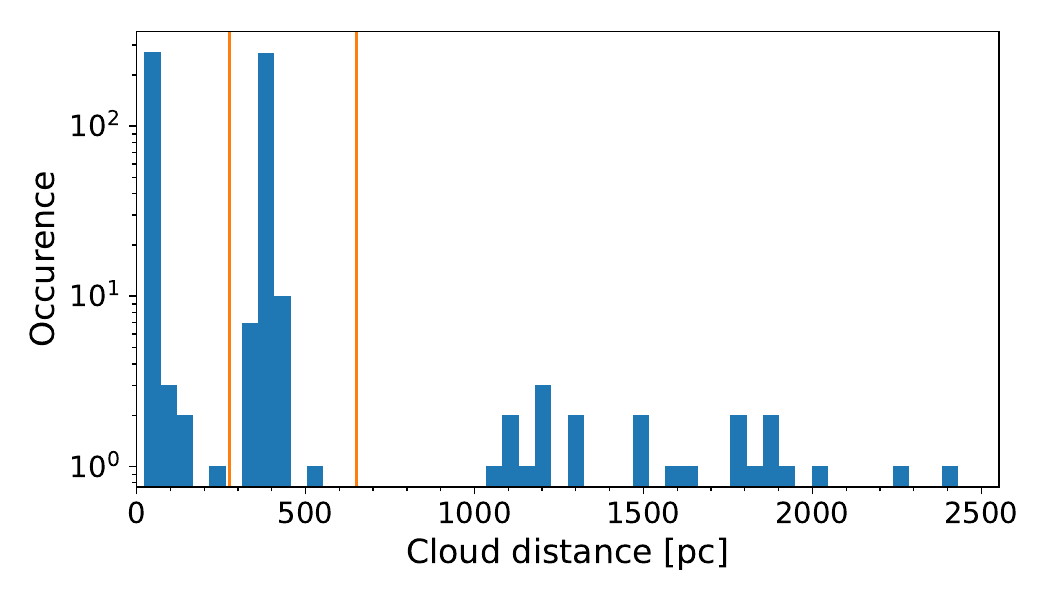}\\[-1.8ex]
    \caption{Histogram of estimated mean cloud distances ($\hat{\bar{d}}_{\rm{C}}$). 
    Distribution of mean posterior cloud distances ($\hat{\bar{d}}_{\rm{C}}$) for the best fit models determined for all sightlines.
    The vertical lines indicate the limits to define the distance bins used in Sect.~\ref{sec:basicResults}.
    }
    \label{fig:dC-hist}
\end{figure}

\begin{figure}
    \centering
    \includegraphics[trim={.0cm 1.9cm .6cm 0.cm},clip,width=1.\columnwidth]{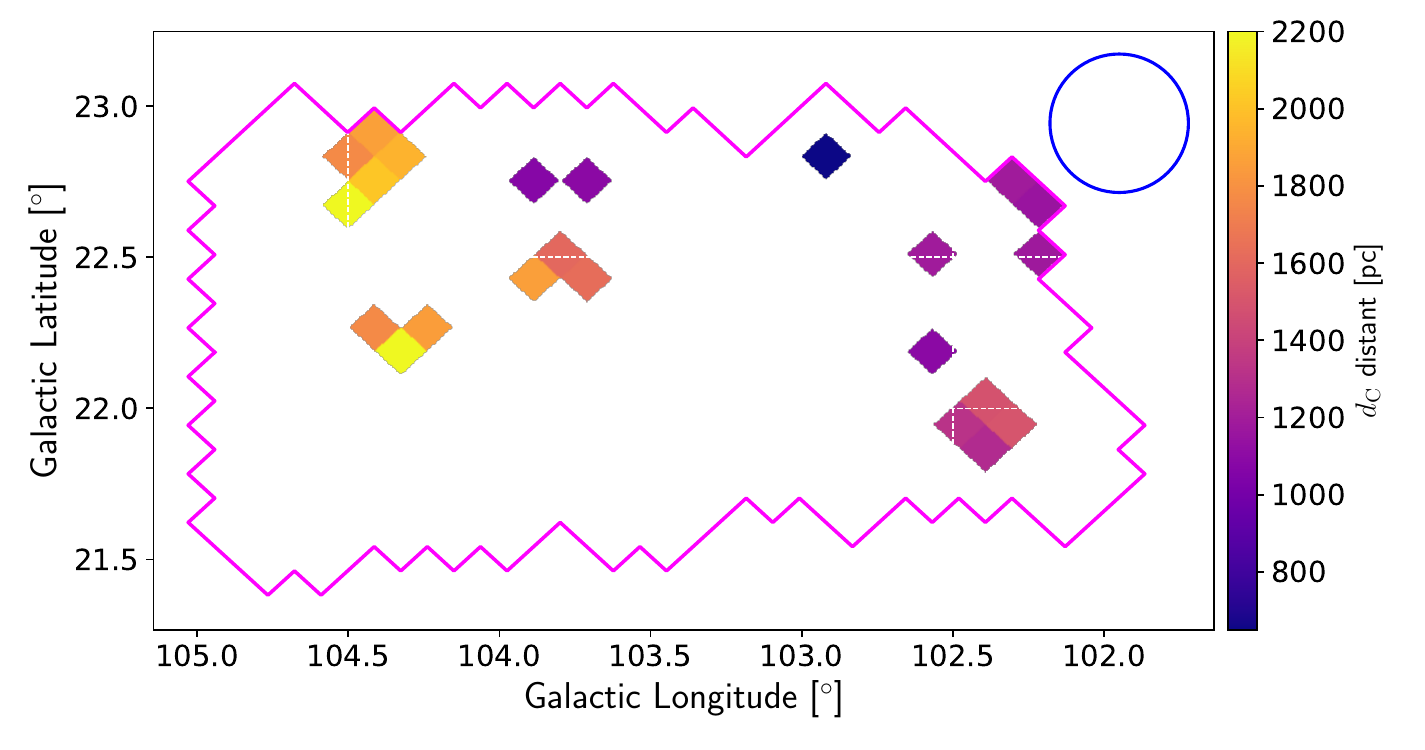}\\
    \includegraphics[trim={.0cm 1.9cm .6cm -0.2cm},clip,width=1.\columnwidth]{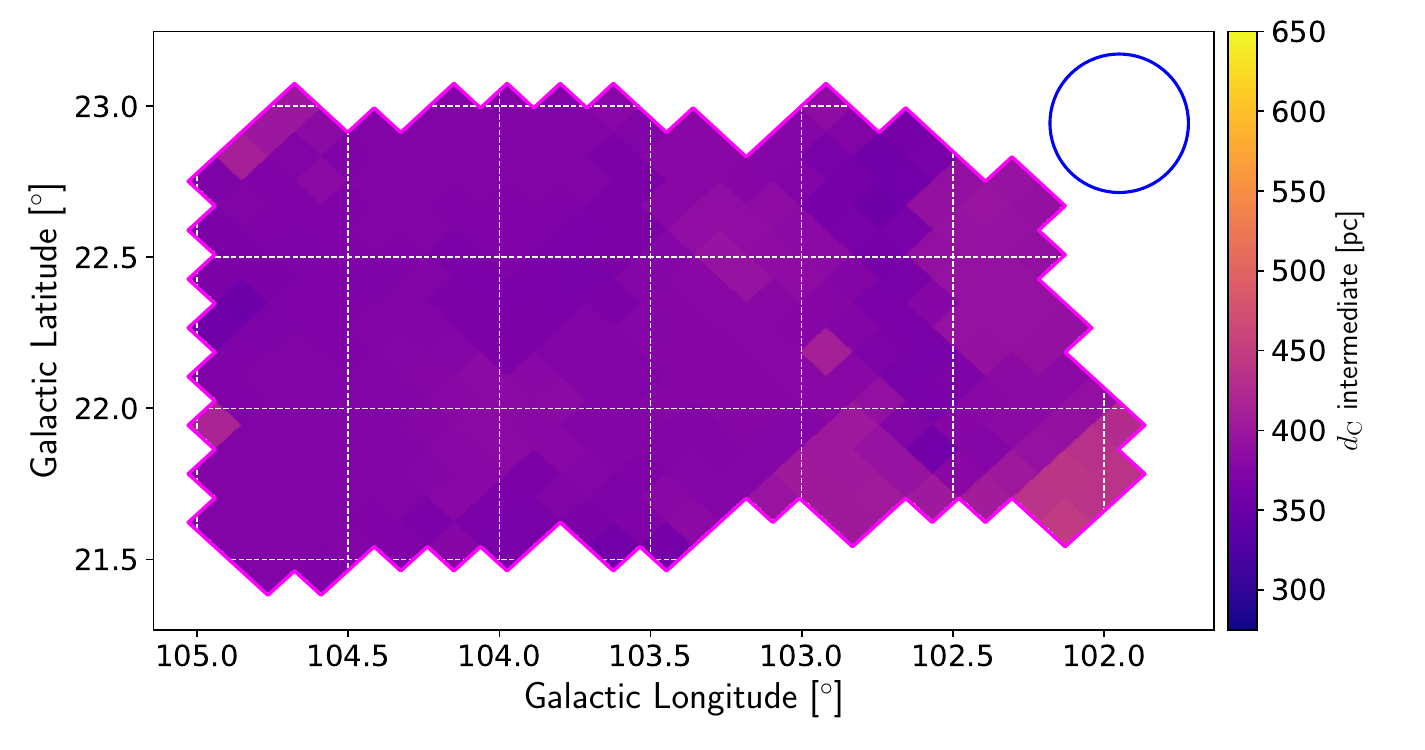}\\
    \includegraphics[trim={.0cm .5cm .6cm -0.2cm},clip,width=1.\columnwidth]{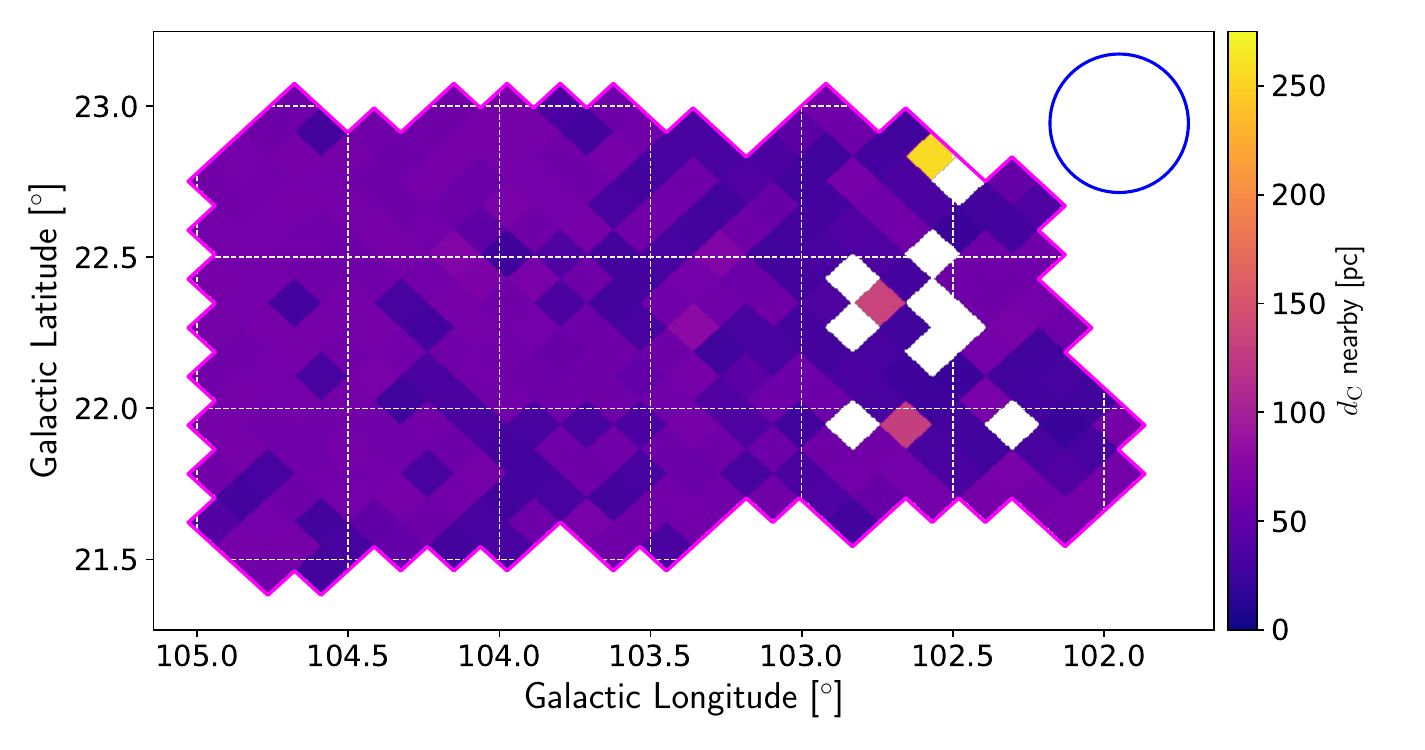}\\[-1.5ex]
    \caption{Maps of the estimated mean cloud distances in the three bins of distances identified from Fig.~\ref{fig:dC-hist}, from far away (top) to nearby (bottom). The color scales span the full range of distance in each bin. Empty pixels of the observed regions mark the absence of clouds in the specific distance range. The magenta contour and the blue circle are as in Fig.~\ref{fig:NstarMap}.
    }
    \label{fig:dC-map}
\end{figure}

\begin{figure}
    \centering
    \includegraphics[trim={.0cm 1.9cm .6cm 0.cm},clip,width=1.\columnwidth]{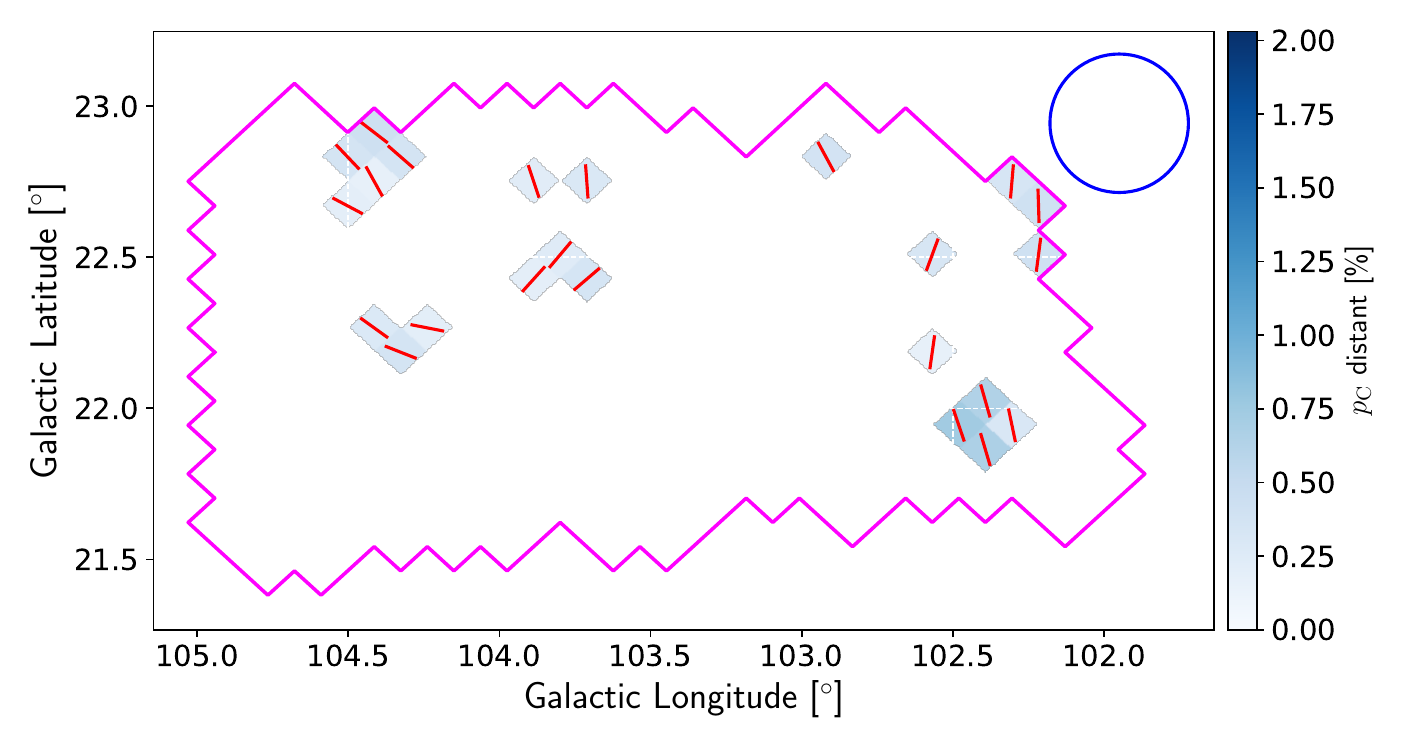}\\
    \includegraphics[trim={.0cm 1.9cm .6cm -0.2cm},clip,width=1.\columnwidth]{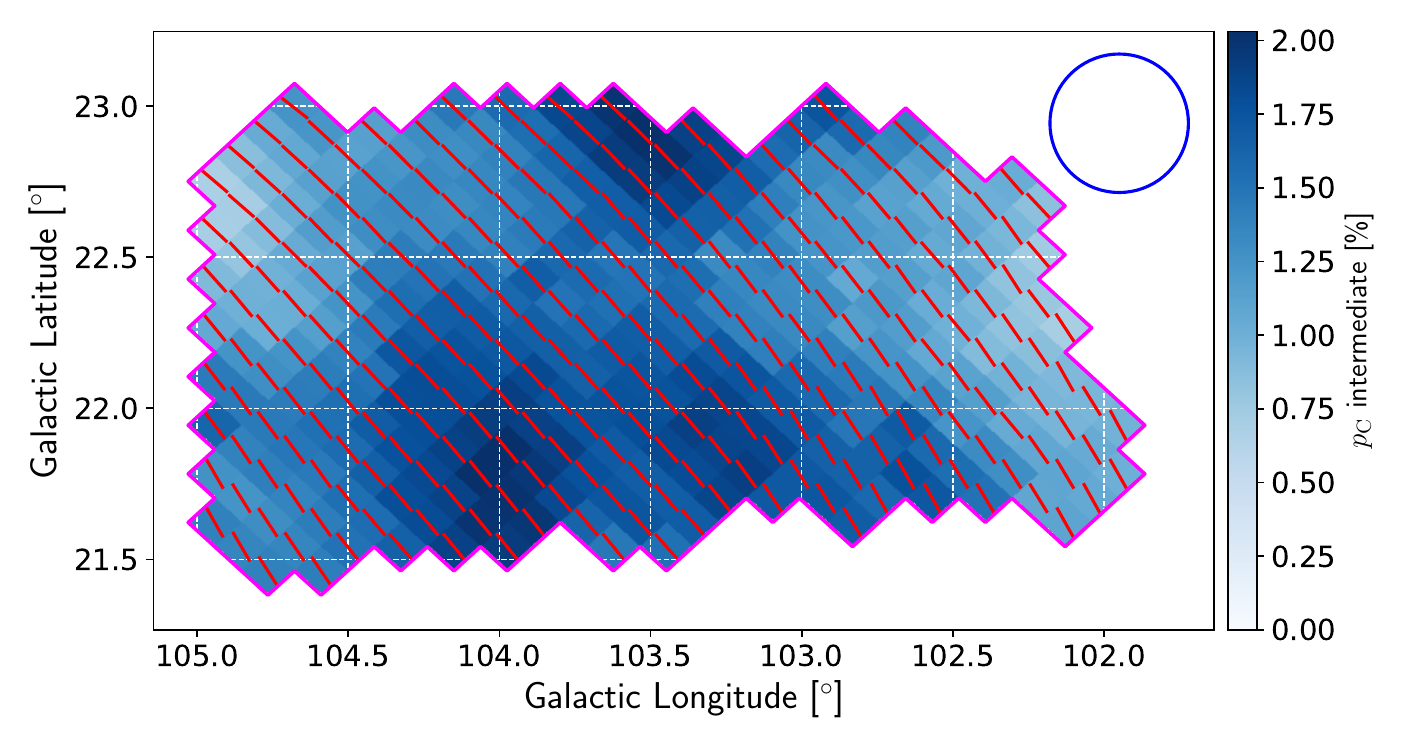}\\
    \includegraphics[trim={.0cm .5cm .6cm -0.2cm},clip,width=1.\columnwidth]{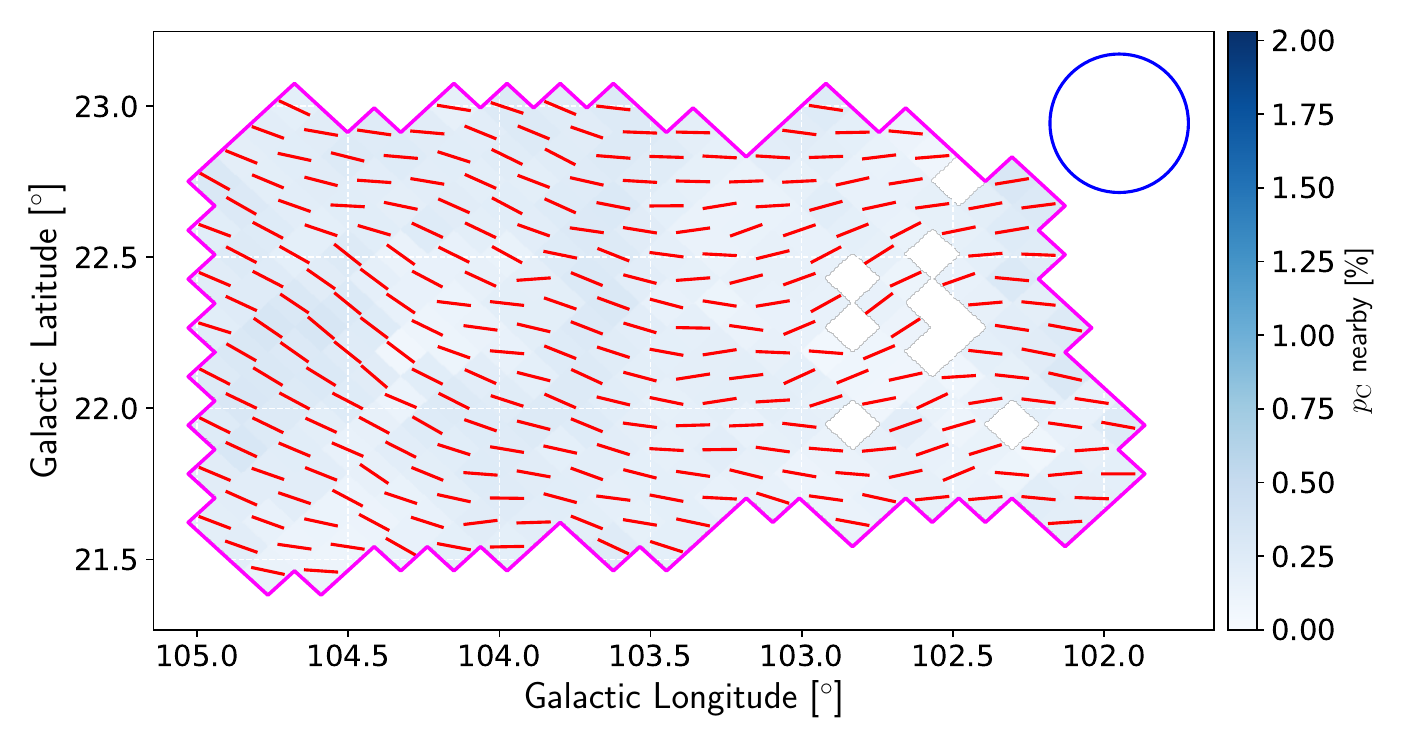}\\[-1.5ex]
    \caption{Same as for Fig.~\ref{fig:dC-map} but for the cloud mean polarization. The color scales is identical for the three maps and informs on the mean degree of polarization ($\hat{\bar{p}}_{\rm{C}}$). The red segments indicate the orientation of the mean polarization in the clouds ($\hat{\bar{\psi}}_{\rm{C}}$) which directly traces the orientation of the mean POS-component of the magnetic field.
    }
    \label{fig:pCpsiC-map}
\end{figure}

From the posterior distributions, we extract the estimated mean values for the cloud parallax ($\hat{\bar{\varpi}}_{\rm{C}}$) and mean polarization ($\hat{\bar{q}}_{\rm{C}}$ and $\hat{\bar{u}}_{\rm{C}}$) for each LOS.
In Fig.~\ref{fig:dC-hist} we show the histogram of the (estimated) mean cloud distance ($ \hat{\bar{d}}_{\rm{C}} = 1/\hat{\bar{\varpi}}_{\rm{C}}$) for all clouds and all analyzed sightlines.
This histogram shows two main separated peaks centered on 62~pc and 380~pc. A number of clouds are also found at distances larger than 1~kpc.
Relying on this observation, we divide the distance axis in three bins for nearby ($d_{\rm{C}} \leq 265$~pc), intermediate ($d_{\rm{C}} \in [265,\, 650]$~pc), and distant ($d_{\rm{C}} > 650$~pc) clouds. For each distance bin, we generate maps of the cloud mean distance and mean polarization as shown in Figs.~\ref{fig:dC-map} and~\ref{fig:pCpsiC-map}, respectively.
The polarization maps are obtained by introducing the mean degree of polarization and mean polarization angle from the mean Stokes parameters as $\hat{\bar{p}}_{\rm{C}} = (\hat{\bar{q}}_{\rm{C}} + \hat{\bar{u}}_{\rm{C}})^{1/2}$ and $\hat{\bar{\psi}}_{\rm{C}} = 0.5 \, {\rm{arctan}}2(\hat{\bar{u}}_{\rm{C}},\hat{\bar{q}}_{\rm{C}})$ where we make use of the two-argument arctangent function to handle the $\pi$-ambiguity of the arctangent.
In these maps, empty pixels of the observed regions mark the absence of detected clouds with mean distance in the specific range of distances.
We find that all sightlines intersect a cloud in the intermediate range of distances.
A large fraction of the sightlines intersect both the nearby- and intermediate-distance components. As seen in Figs.~\ref{fig:dC-map} and~\ref{fig:pCpsiC-map}, these components show a high degree of regularity in their distances and mean polarization properties, suggesting that each of these two components is related to physical entities (real interstellar clouds). The distances of detected clouds in the large distance bin are more scattered but we notice several clusterings of cloud detections in the 3D space for components with similar polarization properties.

\smallskip

Over the observed region of the sky, the mean estimated distance to the nearby cloud ranges from 30~pc to 170~pc and has a median of 58~pc. Its mean estimated degree of polarization ranges from 0.06\% to 0.33\% with a median value of 0.19\%. The estimated POS orientation of the magnetic field of this component varies smoothly across the region. It ranges from $49.4^\circ$ to $126.9^\circ$ with a mean of about $80.5^\circ$.

The mean estimated distance to the intermediate cloud ranges from 354~pc to 448~pc with a median of 374~pc. Its mean estimated degree of polarization shows spatial variation over the observed region. It is also much higher than the nearby component, spanning the range from 0.68\% to 2.03\% with a median value of 1.45\%. This is the dominant polarizing ``screen'' in the region, the effect of which was already clearly seen in Fig.~\ref{fig:outliers_qumu}. The mean estimated polarization angle of this component is nearly uniform throughout the observed region. It ranges from $28.4^\circ$ to $51.9^\circ$ and has a mean value of $39.5^\circ$. The standard deviation of all the polarization angles is $4.7^\circ$. This small dispersion indicates that the POS projection of the ordered component of the magnetic field in this cloud is nearly uniform despite the inhomogeneities seen in the mean degree of polarization.

At larger distances, we find several distinct clouds intersected by different sightlines. The distances of the distant clouds in neighboring pixels agree within uncertainties.
Most noticeably, we find a cloud toward $(l,\,b) \approx (103.8^\circ,\,22.4^\circ$) with mean distance ranging from about 1600~pc to 1850~pc and which has a mean degree of polarization of about 0.26\% and a mean polarization angle of about $136^\circ$. This is nearly perpendicular ($\approx 85^\circ$) to the mean polarization orientation of the dominant polarization screen for the same sky pixels. This small region of the sky is slightly to the northwest of the ``2-cloud LOS'' studied in (\citealt{Pan2019a}; \citealt{Clark2019}; \citealt{Pelgrims2023}) at $(l,\,b) = (104.1^\circ,\,22.3^\circ)$.
In the east-southeast of this ``2-cloud LOS'', toward $(l,\,b) \approx (104.3^\circ,\,22.2^\circ$), we detect a cloud with mean distances ranging from 1700~pc to 2300~pc which has a mean degree of polarization of about 0.28\% and a mean polarization angle of $66.2^\circ$, about $23^\circ$ away from the mean polarization angle of the dominant polarization screen in the same sky pixels.
In the southwestern part of the surveyed region, toward $(l,\,b) \approx (102.4^\circ,\,21.9^\circ$), we detect a cloud at distance between 1270~pc and 1500~pc. It has a mean degree of polarization of about 0.58\% and a mean polarization angle of $16.5^\circ$, about $20^\circ$ away from the mean polarization angle of the dominant polarization screen in the same sky pixels.

We also detect other groups of pixels with cloud detection close to the edges of the observed region. One at $(l,\,b) \approx (104.4^\circ,\,22.8^\circ$) with a somewhat large scatter in mean distance, ranging from 1760~pc to 2400~pc, and another one in the upper right corner of the observed region, with a mean distance of about 1170~pc. Other distant clouds are detected sparsely in the region.
We have checked the tomographic decompositions of all these LOS individually, and none of these cloud detections appear to be the result of a statistical flaw in our analysis.
We checked the posterior distributions of the different tested models to make sure that the automated model selection worked as expected. We additionally checked the residuals and visually compared the data and models in the $(q,\,u)-\mu$ space. In all cases, the best model was correctly identified.

We provide more discussion on our results and their reliability in Sect.~\ref{sec:discussion}. There we also relate them to known structures of the ISM that have already been studied in the literature, based on stellar extinction and \HI\ data, in particular.

\begin{figure}
    \centering
    \includegraphics[trim={.0cm .2cm .0cm 0.cm},clip,width=1.\columnwidth]{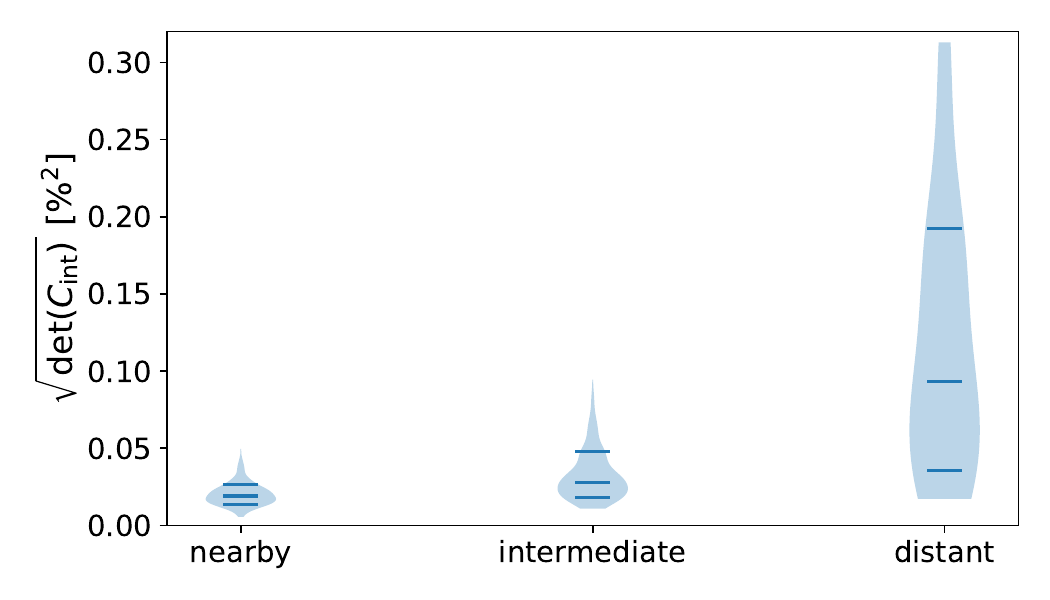}\\[-2.ex]
    \caption{Violin diagrams of the distributions of the square root of $\hat{\bar{{\rm{det}}}}(C_{\rm{int}})$ obtained from the posterior distributions and for the three distance ranges defined from Fig.~\ref{fig:dC-hist}. In each diagram, the horizontal segments indicate the 16th, 50th and 84th percentiles of the distribution.
    }
    \label{fig:detC_violin}
\end{figure}
As a last step in the basic exploration of the output, we look at the covariance matrices that encode the intrinsic scatter of polarization properties within clouds ($C_{\rm{int}}$).
We evaluate the determinant of $C_{\rm{int}}$ for each sample of the posterior distributions, for each dust layer of the best model, and for each LOS individually.
This quantity is related to the level of turbulence-induced intrinsic scatter within a cloud (see Appendix~B of \citealt{Pelgrims2023}).
We then estimate the median of the distributions of the determinants ($\hat{\bar{{\rm{det}}}}(C_{\rm{int}})$) for each layer and each LOS. We finally build the distributions of the medians splitting the sample in terms of the cloud distances, as before.
We show these distributions in Fig.~\ref{fig:detC_violin} in the form of violin diagrams.
We note that, while all the sightlines likely trace the same clouds in the nearby and intermediate distance ranges, this is unlikely to be the case for the distant distance range, as argued before. This is likely the reason for the apparent wider distribution of $\hat{\bar{{\rm{det}}}}(C_{\rm{int}})$ in this distance range than for the others.
From Fig.~\ref{fig:detC_violin}, it is seen that, at least for the clouds at nearby and intermediate distances, the intrinsic scatter is detected above observational noise.
This indicates that, in principle, subsequent analyses of the covariance matrices could lead to a detailed characterization of fluctuations in the magnetized ISM.
We will explore such an avenue in future work.

\subsection{3D-map making}
\label{sec:3DmapMaking}
In this subsection, we build 3D maps of the dusty magnetized ISM from the posterior distributions of all model parameters obtained for each LOS of the observed region.
For each LOS, we construct the probability
density function (PDF) of having a cloud at a given distance as follows. We stack the marginalized posterior distributions of the cloud parallax of each component, and obtain the distance distribution $P(d_{\rm{C}})$ by inverting every parallax value. The PDF is estimated from this distribution using a moving window of length $\Delta d$ as:
\begin{align}
    {\rm{PDF}}(d) = \frac{1}{\Delta d} \, \int_{d-\Delta d/2}^{d+\Delta d/2}
    P(d_{\rm{C}}) \, {\rm{d}} d_{\rm{C}} \; .
    \label{eq:PDFd}
\end{align}
By construction ${\rm{PDF}}(d)$ is normalized to the number of components intersected by the LOS.
Then, using the estimated ${\rm{PDF}}(d)$, we construct the distance profiles of the different polarization properties by marginalizing the posterior distributions of each polarization parameter in distance bins. We thus estimate at any distance ($d$) the differential of any of the Stokes parameters or of the elements of the intrinsic-scatter covariance matrix as:
\begin{align}
    \delta s_{\rm{C}}(d) = \left[ \int s_{\rm{C}} \, 
    P(s_{\rm{C}} \,|\, d) \, {\rm{d}}s_{\rm{C}} \right] \, {\rm{PDF}}(d) \; ,
    \label{eq:marginalization}
\end{align}
where $s_{\rm{C}}$ is any of $q_{\rm{C}}$, $u_{\rm{C}}$, $C_{{\rm{int}},qq}$, $C_{{\rm{int}},uu}$ or $C_{{\rm{int}},qu}$ and $d_{\rm{C}}$ is the distance given by the inverse of the parallax. $P(s_{\rm{C}} \,|\, d)$ is the conditional probability of having a value $s_{\rm{C}}$ given a cloud distance $d_{\rm{C}}$ at value $d$:
\begin{align}
    P(s_{\rm{C}} \,|\, d) = \frac{P(d,\,s_{\rm{C}})}{{\rm{PDF}}(d)} \; ,
\end{align}
where $P(d,\,s_{\rm{C}})$ is the 2D-marginalized posterior distribution between distance and the chosen polarization parameter $s_{\rm{C}}$, obtained by mapping the posterior $P(\varpi_{\rm{C}},\,s_{\rm{C}})$ in distance space.
The units of the differentials of the Stokes parameters, $\delta q$ and $\delta u$, are in polarization fraction per parsec, and the units of the differential of the intrinsic scatter covariance matrix, ($\delta C_{{\rm{int}},xy}$, where the subscripts $x$ and $y$ denote either $q$ or $u$) are polarization fraction per parsec squared.
In the remainder of this paper, we focus on the mean properties of the dusty magnetized ISM that we can infer from stellar polarization and refer to future work for the study and characterization of its fluctuations.

\smallskip

Due to the limited number of samples used in estimating the posterior distributions, spurious noise is observed in the distance profiles. To reduce this, we smooth the estimated distance profiles along LOS using a Gaussian kernel. The posteriors are not sampled uniformly along LOS. This is because the sampling happens in the parallax space and also because the density of stars varies significantly with distance. It is not possible to choose a constant kernel value that would both smooth the noise at large distances and not severely dilute the signal at small distances. Therefore, we choose to smooth the profiles with a Gaussian kernel that varies with distance as
\begin{align}
\sigma_k(d) = 30 + 20/(\pi/2) \, \arctan((d - 500)/50) \,,     
\end{align}
where $\sigma_k(d)$ and $d$ are given in parsec. This choice is rather arbitrary, but it effectively smooths the noise at all distances.
The choice for $\sigma_k$ is such that it is close to 10~pc at a distance of 50~pc, it increases smoothly around 500~pc, and it reaches a value of about 50~pc for all distances larger than 1~kpc. The radial profiles and subsequent visualization are not strongly dependent on this choice. However, we recommend using the posterior distributions directly rather than the profiles for any subsequent quantitative analysis.

\begin{figure}
    \centering
    \includegraphics[trim={0.5cm 0.3cm 0.2cm 0.2cm},clip,width=1.\columnwidth]{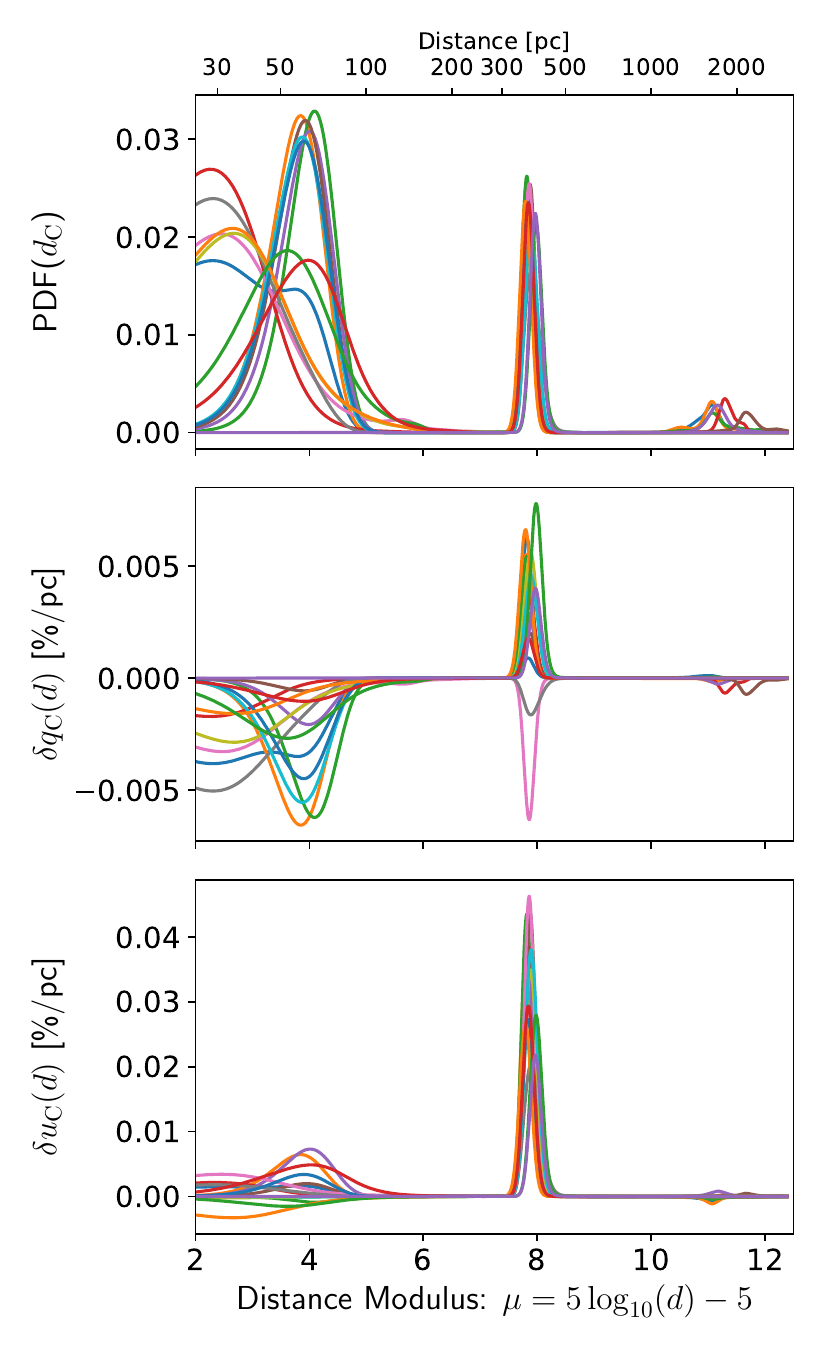}\\[-1.5ex]
    \caption{
    Variation along the distance of the normalized probability density distribution of the cloud distances (top) and of the differentials of the Stokes parameters $\delta q$ and $\delta u$ (middle and bottom) obtained from the marginalization of the full posterior distributions in distance bins for 15 sightlines as explained in the text. The moving-window length of 10~pc is adopted. The curves are smoothed with a Gaussian kernel that varies with distance as explained in the text. The same color corresponds to the same LOS across panels.
    }
    \label{fig:DistanceProfiles_dqdu}
\end{figure}

We show a set of such radial profiles of the Stokes parameters for 15 sightlines in Fig.~\ref{fig:DistanceProfiles_dqdu} (middle and bottom panels). Nine sightlines are randomly chosen in the observed regions to which we add six sightlines that intersect three components as identified in Fig.~\ref{fig:Ncloud_HybridBeam} (those with $b\leq22.5^\circ$ and $l\geq103.5^\circ$).
Similar profiles can be constructed for the three parameters $\delta C_{{\rm{int}},xy}$. The top panel of Fig.~\ref{fig:DistanceProfiles_dqdu} shows the PDF of cloud distances for the same set of sightlines. To construct these profiles, we used a moving-window width of $\Delta d = 10$~pc and sample the distance axis every parsec up to a distance of 3~kpc.

By repeating this process for all sightlines, we finally obtained the values of the differentials at any grid point sampling the 3D space corresponding to the observed sky region. These are the 3D maps that describe the dusty magnetized ISM in the observed sky region. Obtaining these maps is the main result of this paper. We present some visualizations of the maps in the next subsection.
The 3D maps come naturally in a spherical coordinate system centered on the observer.
From the 3D maps of the differential of the Stokes parameters, 3D maps of the polarization degree ($\delta p$) and of the polarization angle ($\delta \psi$) can be derived. The 3D map of the polarization degree can be interpreted as a map of the density of polarizing material through space, weighted by the geometrical factor from the local inclination of the magnetic field on the LOS and by the local polarization efficiency. The 3D map of the polarization angle informs us on the orientation of the POS component (as seen from us, the observer) of the magnetic field locally.
However, because the degree of polarization and the polarization angle are not additive quantities, as opposed to the Stokes parameters (and so of their differentials), the maps of $\delta p$ and $\delta \psi$ must not be integrated along distance.
\begin{figure}
    \centering
    \includegraphics[trim={0.5cm 0.3cm 0.2cm 0.2cm},clip,width=1.\columnwidth]{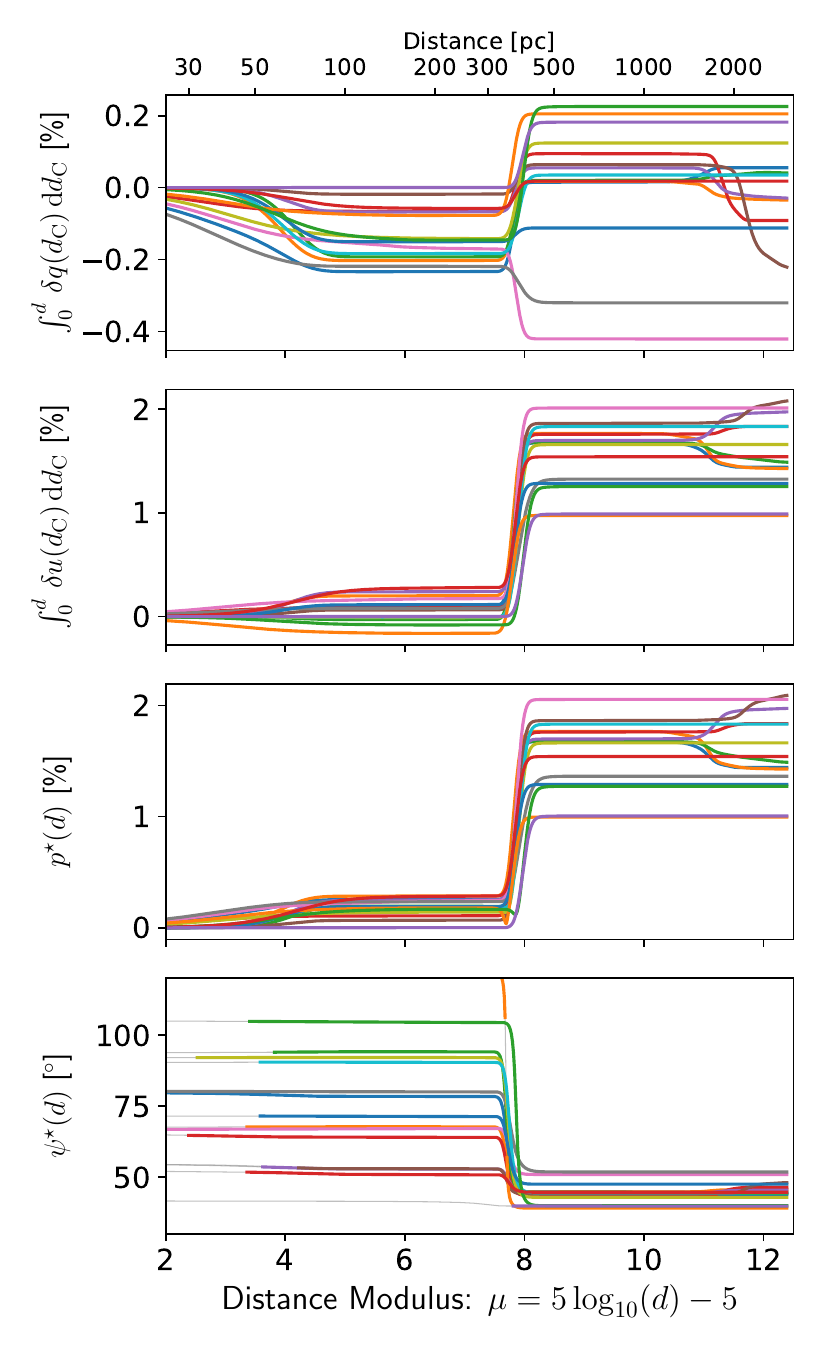}\\[-1.5ex]
    \caption{Polarization of a test star as a function of its distance for the same set of sightlines as shown in Fig.~\ref{fig:DistanceProfiles_dqdu}.
    The first and second rows show the cumulatives of the radial profiles of the differentials $\delta q$ and $\delta u$ as a function of distance.
    The third and fourth rows show the degree of polarization and polarization angle observed for the test star, as derived from the cumulatives of derived $\delta q$ and $\delta u$. For $p^\star < 0.05\%$, $\psi^\star$ is masked (shown by thin gray line).
    }
    \label{fig:CumDistanceProfiles_quppsi}
\end{figure}
In the first and second rows of Fig.~\ref{fig:CumDistanceProfiles_quppsi}, we show the cumulatives of the differentials $\delta q$ and $\delta u$ as a function of distance for the same sightlines as presented in Fig.~\ref{fig:DistanceProfiles_dqdu}. The cumulatives at a distance $d$ correspond to the Stokes parameters that would be observed for a test star at that distance if we neglect the effects from turbulence. In the third and bottom rows we show the derived (mean) degree of polarization ($p^\star(d)$) and (mean) polarization angle ($\psi^\star(d)$) that would be observed for a star at the specific distance. These quantities are obtained at any distance $d$ from the cumulatives of $\delta q$ and $\delta u$, not from the cumulative of $\delta p$ and $\delta \psi$.
The depolarization effect (decrease in $p^\star$) due to the superposition of clouds with misaligned POS component of the magnetic field is clearly observed for some of the sightlines.

\subsection{Visualization of the 3D maps}
\label{sec:3DmapVisualization}
\begin{figure*}
    \centering
    \begin{tabular}{cc}
         \includegraphics[trim={.5cm .2cm 1cm 0cm},clip,width=.98\columnwidth]{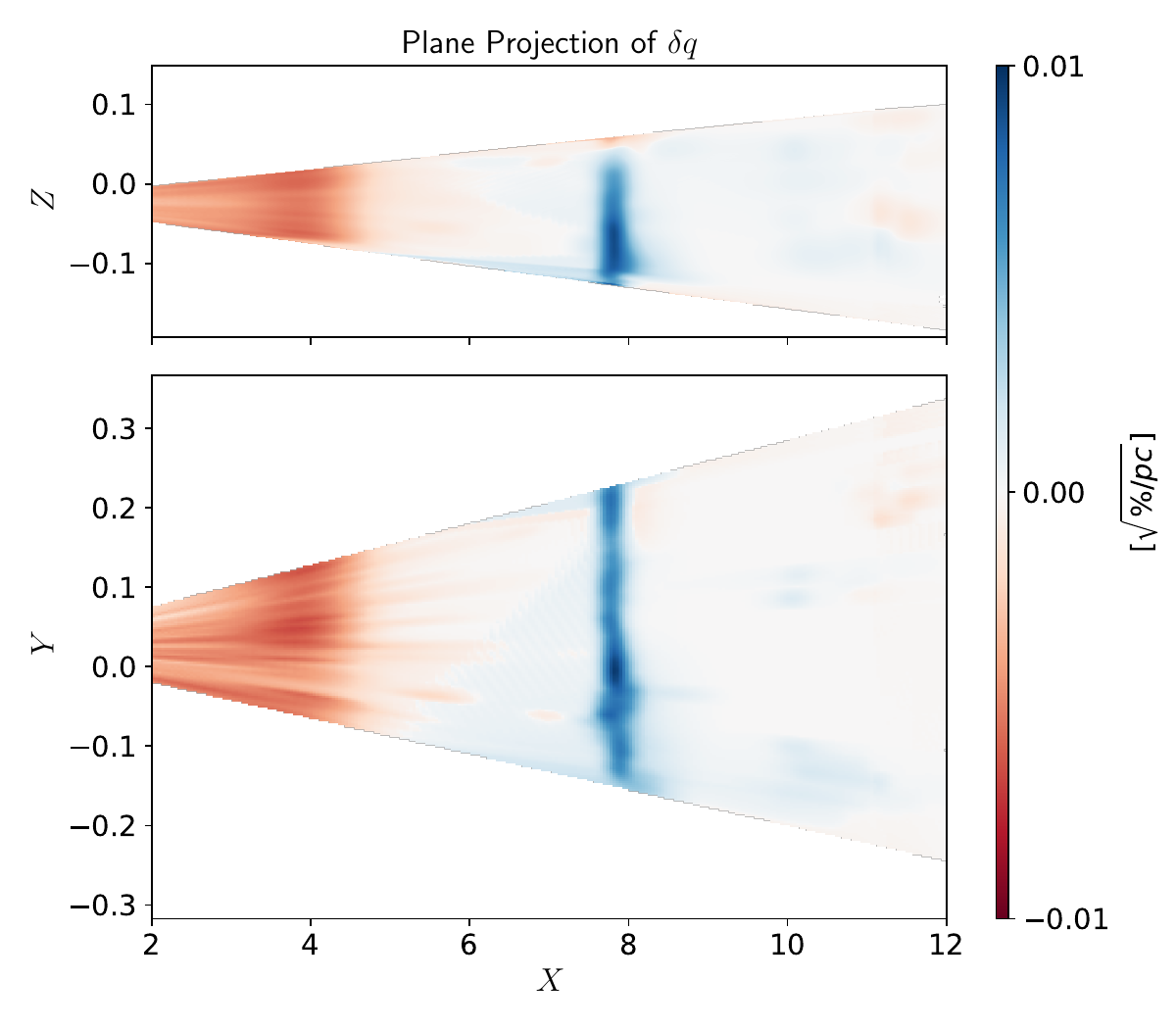}
         &  \includegraphics[trim={.5cm .2cm 1cm 0cm},clip,width=.98\columnwidth]{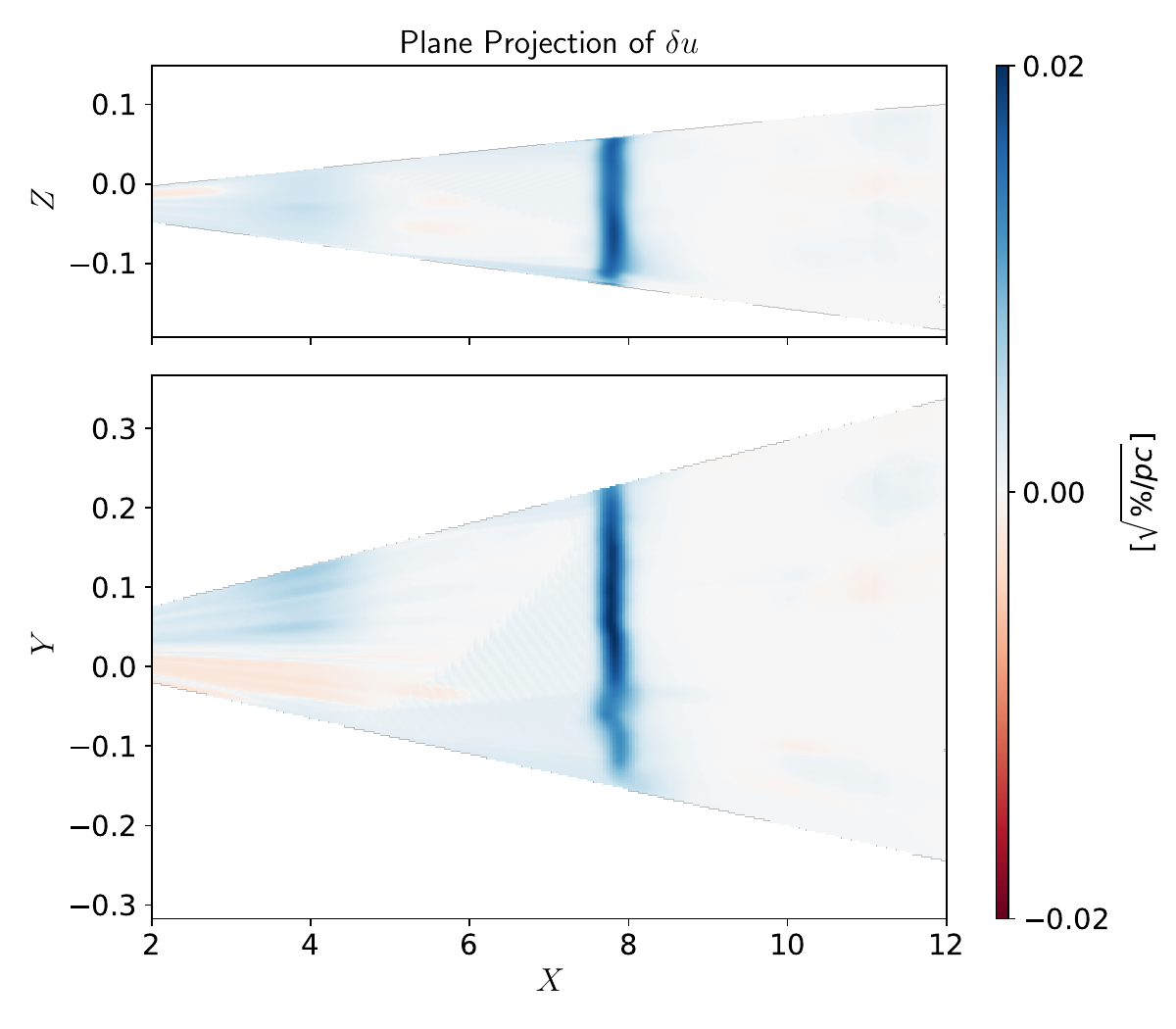}
    \end{tabular}\\[-1.7ex]
    \caption{Plane projections of $\delta q$ (left) and $\delta u$ (right) in the vertical (top) and horizontal (bottom) planes.
    The vertical and horizontal planes are defined by averaging the data cubes (rectangular pyramid) in the Cartesian grid along the $Y$-axis and the $Z$-axis, respectively. The observer (us) is on the left and distance increases to the right.
    In order to visualize faint features, the color values ($c$) are obtained from the projected values ($v$) as $c = \sign(v) \sqrt{|v|}$. The units of the color scale are therefore $\sqrt{ \% / \rm{pc}}$.
    The color scales are symmetrical about zero and the range of $\delta u$ is twice as large as that of $\delta q$, reflecting the difference in magnitude of the two quantities that is observed in Fig.~\ref{fig:DistanceProfiles_dqdu}.
    }
    \label{fig:PlaneProjection}
\end{figure*}
The portion of the Galactic space covered by our observations extends over $\approx 3.8$ square degrees across the sky and extends up to 3.5~kpc. As a result, the geometry of the 3D maps that we construct is very elongated in radial distance and resembles a ``pencil beam'' having the geometry of a rectangular pyramid. This makes it difficult to visualize the results in 3D. To circumvent this limitation, we transformed the spherical coordinate system using the distance modulus as the radial coordinate as
\begin{equation}
    \left\lbrace
    \begin{tabular}{l}
         $X = \mu(d) \, \cos{l'} \, \cos{b'}$ \\
         $Y = \mu(d) \, \sin{l'} \, \cos{b'}$ \\
         $Z = \mu(d) \, \sin{b'} \; ,$
    \end{tabular}
    \right.
    \label{eq:muSphCoord}
\end{equation}
where $\mu(d) = 5\,(\log_{10}(d) - 1)$ is the distance modulus with $d$ given in parsec and $(l',\,b')$ are the angular coordinates (longitude and latitude) defined such that $(l',\,b') = (0,\,0)$ points toward the center of the observed region. In our case, we thus have $X \approx \mu(d)$, $Y \approx \mu(d)\, l'$, and $Z \approx \mu(d)\,b'$. This coordinate system is ill-defined at small distances. However, we are not affected by this issue since all stars have a distance larger than 20~pc and that our 3D maps do not extend at distance modulus smaller than $\mu = 2$.

We thus project the 3D maps of the differentials constructed above in this coordinate system. The sampling on angular coordinates is fixed by the centers of the HEALPix pixels used to define our samples from which the LOS decompositions were performed. The sampling on radial distance is such that we have values for the differentials at every parsec for the range $\mu \in [2,\, 12.7]$. As defined, the $X$-axis runs through the center of the observed region at $(l,\,b) = (103.3^\circ,\, 22.3^\circ)$, $Y$ increases with increasing longitude and $Z$ with latitude.

Despite the use of this coordinate system, the volume is still more extended toward the $X$-axis. For visualization purposes, we thus shrink that axis by a factor of 10 as compared to the others. We then construct a regular Cartesian grid made of 256$^3$ voxels, with limits such that the volume $X/10 \times Y \times Z$ fits in. Due to the rectangular pyramid geometry of the inverted dataset, most of the volume is empty (no data). We use linear interpolation to obtain the values of the differentials at every voxel position from the 3D maps of the differentials in the modified spherical coordinate system. The data cubes can now be visualized.

\subsubsection{Plane projections}
We start by producing plane projections of the data cubes. The data cubes are integrated along the $Y$-axis and the $Z$-axis of the Cartesian grid to produce the vertical and horizontal plane projections, respectively. Figure~\ref{fig:PlaneProjection} shows the results of such projection for the differentials of the (mean) Stokes parameters. In this figure, and as explained in its caption, we tweak the color scales in order to visualize the very faint features at large $X$ values. The faint signal of the distant components seen in Fig.~\ref{fig:DistanceProfiles_dqdu} at $\mu \approx 10$ are indeed dimmed because of their small extent in space and by the integration over the full length of the data cube axes. The nearby and dominant components already seen in Fig.~\ref{fig:DistanceProfiles_dqdu} are striking on these plane projections. The dominant component appears nearly planar at constant $X$ (distance) and with very coherent polarization properties. This is reminiscent of what we already discussed in Sect.~\ref{sec:basicResults}.
The change of colors along distance indicates that the POS component of the magnetic field in the nearby, intermediate, and large distance ranges are misaligned.

Plane-projection maps reveal features that are somewhat elongated along the distance axis. This is clearly visible at small distances. Most of it is due to the use of the moving window of 10~pc and the smoothing that we use to construct the distance profiles of the differentials and due to the choice of the logarithmic scale (distance modulus) for visualization. However, it is worth emphasizing that the constructed 3D maps result from the marginalization of the posterior distributions along the distance axis. Therefore, some real extensions along the LOS exist in the maps and are related to the level of constraints we can impose on cloud distances given the stellar data. They do not inform the actual extension of a dust cloud along the LOS but rather reflect our uncertainties on cloud distances.
These features are similar to the ``finger-of-god effect'' commonly seen in 3D dust map reconstructions.
We recall that the model that we use to reconstruct the dusty magnetized ISM from stellar data in polarization and distance assumes that the clouds are thin - they have no dimension along the LOS.

\subsubsection{Visualization in 3D}
Visualizing a pseudo-vector field in 3D presents more challenges than visualizing a scalar field, such as the dust density distribution for example.
While preparing this paper, we have started addressing these challenges building up on the framework underlying \textsc{Asterion}, the tool developed in \cite{Konstantinou2022} to simulate magnetized dust clouds in 3D.
\textsc{Asterion} relies on real-time 3D visualization techniques with virtual reality capabilities. It is implemented in the real-time-engine called Unreal Engine 4 (UE4)\footnote{UE4 is a complete suite of development tools that allows for the visualization and immersive virtual worlds, multiplatform deployment, asset and plugin marketplace, among other features (\url{https://www.unrealengine.com/en-US/unreal}).} and allows for the rendering of details of the magnetized ISM and enables the user to fly through the simulated environment as is done in video games.

Extending the capabilities of this software, we are now able to visualize the main properties of our tomography map.
Yet, the current version of the software does not render the contribution from the intrinsic scatter nor does it visualize the uncertainties on our 3D reconstruction. Only the solution corresponding to the means of our posterior distributions (or at the maximum likelihood values) on the cloud distances and their mean polarization can be visualized.
Specifically, the software makes it possible to place the magnetized dust cloud in 3D and to visualize the ``differentials'' of $\delta p_{\rm{C}}$ and $\delta \psi_{\rm{C}}$ using colors, transparency and a 3D version of the line integral convolution technique (\citealt{Cabral1993}). The visualization of the 3D map obtained in this paper can be accessed online\footnote{\url{https://pasiphae.science/visualization}}.

\section{Discussion}
\label{sec:discussion}
In this section we compare the results of our 3D reconstruction of the
POS component of the magnetized ISM from starlight polarization and distance data to other datasets that inform on the complexity of the ISM. We also discuss the limitations of our results and caveats of our method. 

\subsection{Comparison with 3D dust extinction data}
We start with a comparison of our results with a 3D dust extinction map that was obtained from {\it Gaia} parallax and from both spectroscopic and photometric extinction measurements, including from {\it Gaia} and 2MASS, and which informs on the dust density distribution in 3D space in a Cartesian box of $3 \times 3 \times 0.8$~kpc$^3$ centered on the Sun (\citealt{Vergely2022}).
Our starlight-polarization-based tomography map is independent from the 3D extinction map as it relies on different observables (aside from the parallax) and as the stellar polarization sample is (much) smaller than the stellar extinction sample. Both 3D maps inform us on different properties of the ISM (extinction versus polarization). In addition, the 3D map of dust extinction and our map of stellar polarization have different spatial resolutions both along the LOS and in the POS.
The spatial resolution of the 3D dust map that we consider is 10~pc (and sampled by 3D voxels with side length of 5~pc).
Our map has a varying spatial resolution, first because the LOS inversion is not bound to a particular sampling (or gridding) of the distance, and second because the signal along each LOS has been smoothed with a distance-dependent Gaussian kernel.
Despite these differences, we wish to verify whether our approach identifies clouds at distances similar to the cloud locations in the 3D dust map. We thus construct 1D profiles of extinction as a function of distance to compare the locations of peaks in the dust distribution along the LOS with the locations of clouds in our tomography map.

For this first qualitative comparison between dust extinction and dust polarization tomography data, we sample both 3D maps with a set of sightlines that go through the surveyed sky area and set by the HEALPix tessellation with $N_{\rm{side}} = 128$.
\begin{figure}
    \centering
    \includegraphics[trim={0.4cm 0.3cm 0.3cm 0.2cm},clip,width=.98\columnwidth]{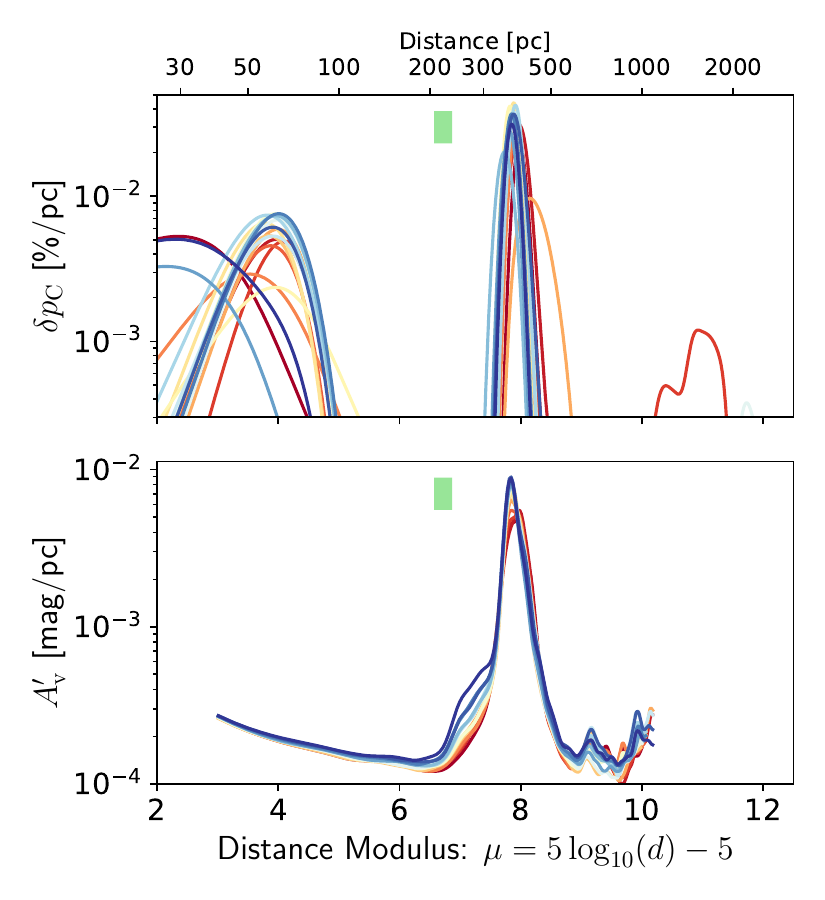}\\[-1.5ex]
    \caption{Qualitative comparison of distance profiles between dust polarization and dust extinction tomography results.
    (top) Profiles of the ``differential'' of the degree of polarization ($\delta p_{\rm{C}}$, computed locally from $\delta q_{\rm{C}}$ and $\delta u_{\rm{C}}$) as a function of distance modulus as inferred from our polarization 3D map.
    (Bottom) Differential extinction ($A'_{\rm{v}}$) as a function of distance modulus as inferred from the 3D extinction map of (\citealt{Vergely2022}).
    No extinction data is available at distance larger than about 1~kpc.
    The surveyed sky area is sampled according to an HEALPix map with $N_{\rm{side}} = 128$. Each LOS is represented with a different color, the same color is used across panels.
    The green vertical strip indicates the range of distances to the inner surface of the Local Bubble in this sky area (\citealt{Pelgrims2020}).
    }
    \label{fig:ExtAndPolProfiles}
\end{figure}
For each LOS we extract the distance profiles of differential extinction ($A'_{\rm{v}}$) and of the ``differential'' of the degree of polarization ($\delta p_{\rm{C}}$), computed locally from the differentials of the Stokes parameters ($\delta q_{\rm{C}}$ and $\delta u_{\rm{C}}$). We present the profiles in Fig.~\ref{fig:ExtAndPolProfiles} where we also indicate with a vertical strip the estimate of the distance to the inner surface of the Local Bubble as derived in (\citealt{Pelgrims2020})\footnote{\url{https://doi.org/10.7910/DVN/RHPVNC}} from the 3D extinction map of \cite{Lallement2019}.
Because of the finite size of the 3D map of (\citealt{Vergely2022}), the $A'_{\rm{v}}$ profiles do not extend over the whole distance range for which we recover information from stellar polarization.
Comparing  the profiles from dust extinction and dust polarization tomography data, we notice that both tracers indicate the presence of a very nearby component at $\mu \lesssim 5$ and a dominant component at $\mu \approx 7.9$.

It is remarkable that the stellar polarization data makes it possible to recover the very nearby component at $\mu  \lesssim 5$ although it appears to be very shallow in the extinction profiles.
The difference between extinction and polarization data in relative amplitudes between the nearby component and the dominant one could point to differences in magnetic field inclination or dust polarization properties. However, differences in 3D-map resolutions could artificially dilute more the signal of the nearby component in 3D dust density map than in polarization. Thus, this comparison will require confirmation from a dedicated analysis which goes beyond the scope of this paper.
Meanwhile, it should be noted that our polarization tomography data independently confirms the presence of a very close dust cloud identified in the 3D dust extinction map allowing us to argue that these small features in the 3D dust map are real (at least in the surveyed region).
Evidence for nearby ISM clouds within the Local Bubble are also provided by pulsar-scintillation studies (\citealt{Ocker2023} and references therein) and stellar line absorption (\citealt{Peek2011}).

The agreement between distances to the dominant peak seen in the $A_{\rm{v}}'$ and $\delta p_{\rm{C}}$ profiles at $\mu \approx 7.9$ is remarkable.
However, while this component appears as an isolated an narrow peak in the $\delta p_{\rm{C}}$ profiles, it appears broader toward lower distances, possibly featuring a second peak, in the $A_{\rm{v}}'$ profiles.
According to \cite{Pelgrims2020}, the nearby peak would correspond to the wall of the Local Bubble. The main peak (centered in $\mu \approx 7.9$) could correspond to the diffuse Cepheus Flare as several molecular clouds in this part of the sky show similar distances (see \citealt{Schlafly2014}).

However, it is unclear whether the broadening of the peak in the $A_{\rm{v}}'$ profiles actually indicates the presence of two close dust components,
or if this shape simply stems from the fact that there is too little stellar extinction data right in front of the dominant component to effectively constrain the lower distance limit of a single cloud centered at the main peak.
We recall that $A_{\rm{v}}'$ profiles, like our distance profiles in the cloud parameters, reflect the shape of the posterior distribution, which is closely related to the uneven distribution of stars along the distance axis.
Additionally, we note that we have identified the presence of a cloud with distance $\mu \approx 7$, several degrees away in the southeastern direction, outside of the surveyed region, and that the outskirt of this cloud is intersected by our region. This is observed both in 3D maps from \cite{Vergely2022} and \cite{Edenhofer2023}. It is therefore possible that the peak broadening in the $A_{\rm{v}}'$ profiles is due to the limited resolution of the 3D dust density maps, or that the broadening indicates a real dust overdensity. In the latter scenario, we understand why we are unable to find evidence for the existence of this cloud using starlight polarization by the following. If there are two dust components in the range $\mu \in [6,\,9]$, they are relatively close in distance. Thus the number of stars with polarization measurements that would be useful to identify a cloud and constrain its polarization properties in between the two peaks is very low. We have checked that, indeed, we have few polarization measurements in that distance range in our sample. Furthermore, the amplitude of the jump in degree of polarization that is induced by the dominant screen is very large. This hampers the possible identification of a counterpart of the wall of the Local Bubble in its vicinity as it would require a large number of data points.
This illustrates one of the limitations in our reconstruction which likely comes from the limited depth of our survey.
An alternative explanation would be that the nearest of these two clouds simply does not induce polarization. This could happen if this cloud is devoid of polarizing dust grains or if both its column density is very low and its permeating magnetic field lines are nearly perpendicular to the POS.

One major difference between our approach and the approach of \cite{Lallement2019} and \cite{Vergely2022} is that they use a fixed spatial kernel in the 3D space to infer the dust density in a grid while the distance to the cloud is a free parameter in our model.
Together with the thin-layer assumption in our model, this might explain the tighter constraints on cloud distances that we obtained ($378~{\rm{pc}} \pm 14~{\rm{pc}}$), compared to the fixed maximum resolution of 10~pc in the 3D dust density map of (\citealt{Vergely2022}).

Overall, the agreement between our tomography map from stellar polarization with the tomography map of dust density from stellar extinction is remarkable given the very different nature of the two datasets. We find good agreement in both the number of components and their distances to the Sun. This indicates that a systematic and dedicated analysis will help study spatial variations of the ISM properties such as the polarization efficiency of the dust grains and the inclination of the magnetic field lines with respect to the LOS. However, our qualitative comparison taught us that comparing tomography data in polarization and density obtained from different approaches (as it is the case here) and with different resolutions could lead to confusion. To benefit from the comparison of dust extinction and polarization data, it is essential to perform detailed analysis treating the different datasets in a self-consistent manner. We leave the task of devising such a framework for future work.

\subsection{Detection of distant clouds and \HI\ data}
In Fig.~\ref{fig:ExtAndPolProfiles} we observe farther away clouds in the $\delta p_{\rm{C}}$ profiles where no information exists in $A_{\rm{v}}'$ profiles, although indications for faraway clouds might be guessed in the far edge of the extinction profiles (at $\mu \approx 10$).
To corroborate our findings of faraway clouds ($\gtrsim 1$~kpc), we first turn to the inspection of \HI\ velocity spectra. As discussed in Sect.~\ref{sec:data}, we expect to observe complex \HI\ spectra with features at intermediate velocities if faraway dust clouds are present along these sightlines.

\begin{figure}
    \centering
    \includegraphics[trim={0.4cm 0.3cm 0.3cm 0.2cm},clip,width=.98\columnwidth]{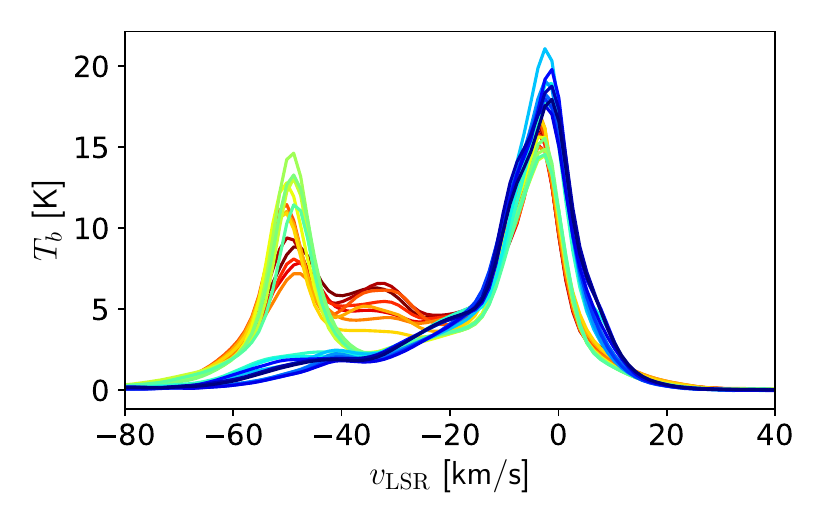}\\[-1.5ex]
    \caption{\HI\ velocity spectra for sightlines with far away clouds. The spectra are sorted according to the Galactic longitudes of their sightlines and shown with color ranging from blue to red. For visual perception, the velocity spectra are smoothed with a Gaussian kernel with width of 1.9~km/s.}
    \label{fig:HIspectra}
\end{figure}

For this purpose, we extract the \HI\ velocity spectra from the HI4PI data cube (\citealt{HI4PI2016}) in pixels corresponding to the sightlines for which we detected a cloud with mean distance in the large distance bin ($d_{\rm{C}} > 650$~pc) in Sect.~\ref{sec:basicResults}. These sightlines can be seen in the top panel of Fig.~\ref{fig:pCpsiC-map}, for example. We sorted the sightlines according to their Galactic longitude and present the spectra in Fig.~\ref{fig:HIspectra} from blue (low longitude) to red (high longitude). We exclude the range of high-velocity clouds ($v_{\rm{LSR}} \lesssim -90$~km/s) and velocities above $40$~km/s where there is no signal.

The \HI\ velocity spectra clearly show complexity with several components. All sightlines at large Galactic longitude show strong IVC components ($|v_{\rm{LSR}}| \gtrsim 35$ km/s). These IVC components do not vanish at lower longitude but are shallower. The fact that there is power in the velocity spectra for $|v_{\rm{LSR}}| > 35$~km/s for all these sightlines makes us confident in our detection of faraway clouds. However, the evidence for IVC components at low Galactic longitudes is milder than for high longitudes; the column density of these clouds is much lower in this part of the surveyed region, as already inferred from Fig.~\ref{fig:average_HI_spectrum}. The spectra corresponding to the upper right part of the region, with $l < 103^\circ$ and $b > 22.1^\circ$, are the only ones not showing local maxima at $v_{\rm{LSR}} < -20$~km/s.
At this stage we do not know if this is an indication for spurious detection of faraway clouds in our starlight-polarization based tomography.
We discuss further the reliability of our cloud detections in Sect.~\ref{sec:limitations}.

\smallskip

In their analysis of a small subset of the polarization data studied in this paper, \cite{Pan2019a} studied starlight polarization within a beam of about 9.6~arcmin radius toward $(l,\,b) = (104.1^\circ,\,22.3^\circ)$ and made the identification of a faraway dust cloud with an IVC with velocity of about $-50$~km/s.
Having enlarged the surveyed area with stellar polarization, we see that IVCs are detected in this part of the sky using starlight polarization, although not specifically in the pixel closest to the LOS studied by \cite{Pan2019a}, but in neighboring pixels (eastward and westward).
It is not clear from our polarization tomography data whether these faraway clouds form a continuous entity in 3D space or if they are independent. Assuming a common distance of 1700~pc, the detected clouds are at least $15$~pc apart.
The orientations of the POS component of the magnetic field that we find are also different in the three main clusters of pixels with faraway-cloud detection.
A different POS magnetic field orientation is however compatible with the finding of \cite{Clark2019} who used the orientation of \HI\ fibers to infer this quantity.
\begin{figure}
    \centering
    \includegraphics[trim={0.3cm 0.2cm 0.1cm 0.1cm},clip,width=.98\columnwidth]{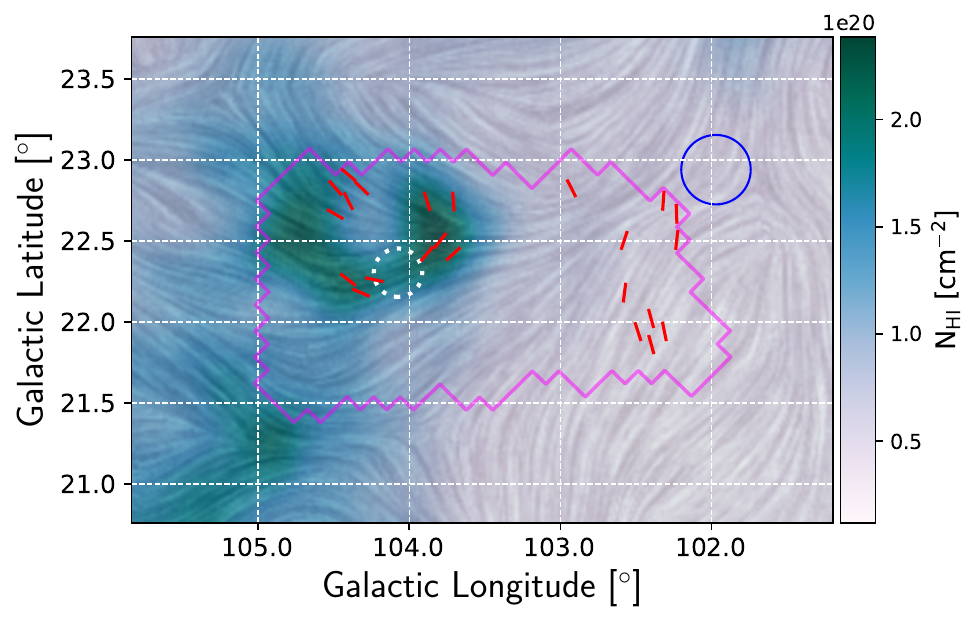}\\[-1.5ex]
    \caption{\HI-orientation data integrated in the velocity range $v_{\rm{LSR}} \in [-60.45,\,-44.99]$ km/s toward our surveyed area. The background color shows the \HI\ intensity on a linear stretch, and the texture represents the POS magnetic field orientation inferred from \HI\ fibers. The red segments indicate the mean POS magnetic field orientation obtained from starlight polarization tomography for clouds with distances larger than 1~kpc, as in Fig.~\ref{fig:pCpsiC-map}. The white-dotted circle shows the ``2-cloud region'' studied in \cite{Pan2019a}.
    The magenta contour and the blue circle are as in Fig.~\ref{fig:NstarMap}.
    }
    \label{fig:HIfibersAndStars}
\end{figure}
This is illustrated in Fig.~\ref{fig:HIfibersAndStars} where we overlaid our results from starlight-polarization-based tomography for the clouds in the large distance bin (as in the top panel of Fig.~\ref{fig:pCpsiC-map}) to the \HI-orientation data from \cite{Clark2019}. We also obtain consistent results if we instead use the \HI\ orientation maps from \cite{Halal2023}. The \HI\ data is integrated over the velocity range $v_{\rm{LSR}} \in [-60.45,\,-44.99]$~km/s. The POS component of the magnetic field { implied by the orientation of} \HI\ fibers is shown using line-integral-convolution (LIC) texture. 
{ The alignment between \HI\ structures and the magnetic field is driven by the orientation of anisotropic, cold gas structures (\citealt{Clark-etal2019}; \citealt{Peek2019}; \citealt{Murray2020}; \citealt{Kalberla2020}; \citealt{Kalberla2023}), so the LIC pattern may not trace the magnetic field well away from regions of prominent \HI\ emission.}
An IVC component, with varying POS magnetic field component, is prominent in the upper left corner of the surveyed area. 
Visual comparison of the orientation of the POS component of the magnetic field obtained from starlight-polarization tomography and \HI\ data reveals a good qualitative agreement where faraway clouds are detected, in particular toward the IVC. However, we notice that we do not detect faraway components toward all sightlines sampling the prominent IVC%
{, especially in the closest pixel toward the ``2-cloud'' region of (\citealt{Pan2019a}),}
and that our cloud detections in the western part of the region do not have counterparts in this velocity range. We further discuss these differences in the next subsection.

Overall, our comparison between starlight-polarization and \HI\ data
points toward the great potential of combining the different tracers of the magnetized ISM to study and characterize its properties in detail. Going far beyond the scope of this paper, we will explore the potential of such a synergy in future work.
Meanwhile, all of the above indicates that our inversion method for obtaining, from stellar polarization and parallax alone, a tomographic view of the POS component of the magnetic field in dusty regions leads to reliable results.

\subsection{Reliability of cloud detection and limitations}
\label{sec:limitations}
To present our results, in Sect.~\ref{sec:results}, and to compare them with other probes of the ISM, in the two subsections above, we focus on the tomographic solutions corresponding to the best model (number of clouds) obtained per LOS. However, it is interesting to look at the second-best model (defined below) and to compare the performance between the best and second-best models that we obtained. In fact this helps quantify the robustness of a solution (a given model) and, also, the reliability of a cloud detection.

In Sect.~\ref{sec:Step2}, at the end of Step~2 in the inversion process, we obtained for each LOS, the estimated maximum-likelihood values for all tested models (different number of layers). Among the tested models, the best model was identified comparing the model performances based on their AIC values (see Eqs.~\ref{eq:AIC} and~\ref{eq:ModelProb}). By ranking the AIC values, we can also identify the second-best model. The probability that this model is actually the model that minimizes the loss of information as compared to the best model (i.e., its $P_{j|\{m\}}$ value) quantifies by how much it is outperformed by the best model.

\begin{figure}
    \centering
    \includegraphics[trim={.0cm 1.9cm .6cm 0.4cm},clip,width=1.\columnwidth]{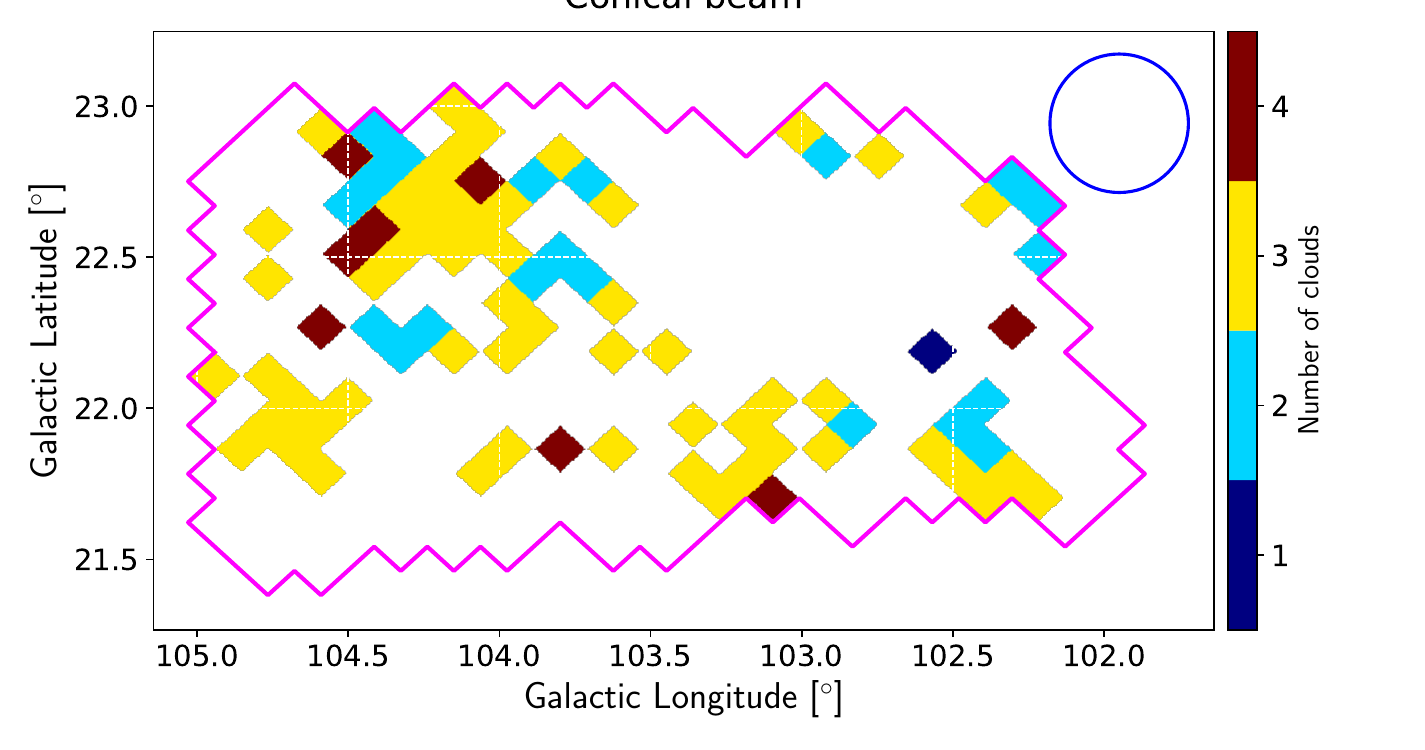}\\
    \includegraphics[trim={.0cm 1.9cm .6cm 0.3cm},clip,width=1.\columnwidth]{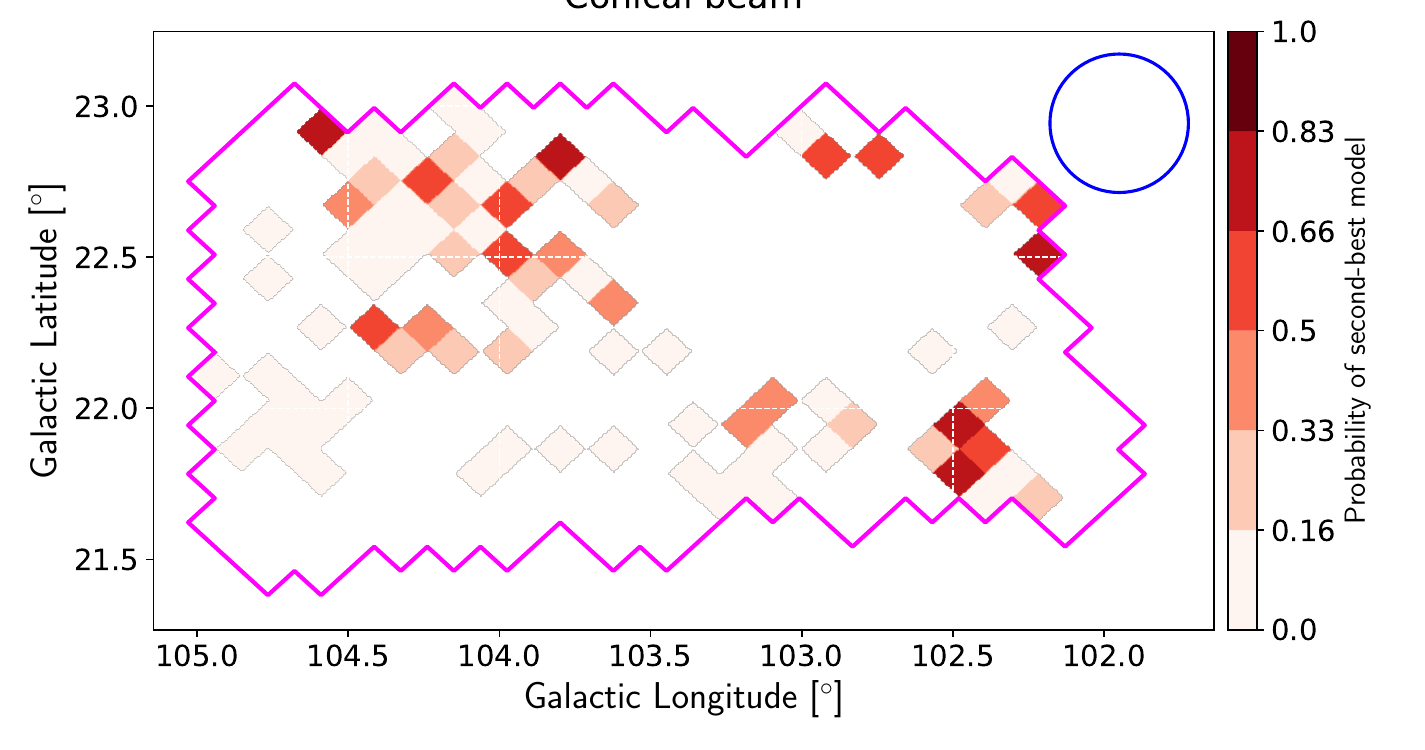}\\
    \includegraphics[trim={.0cm .4cm .6cm 0.3cm},clip,width=1.\columnwidth]{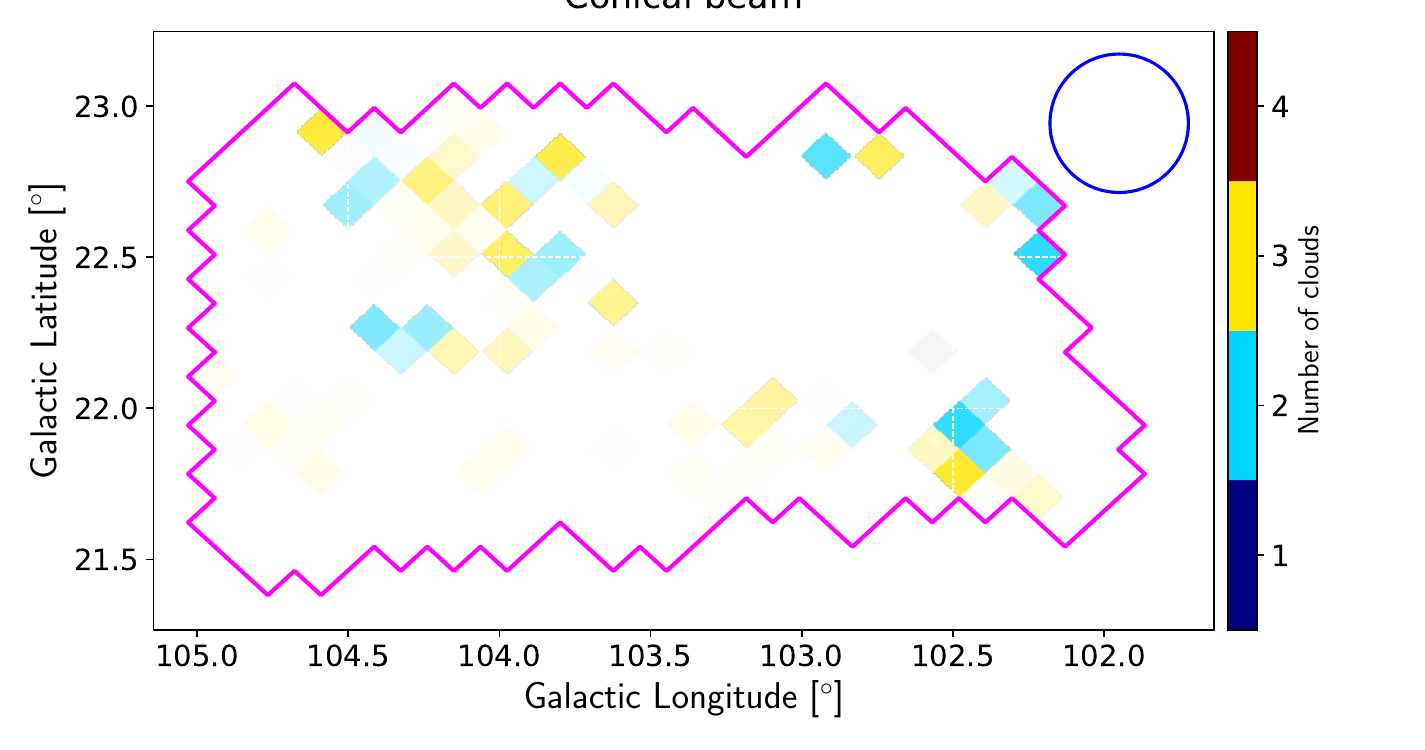}\\[-1.5ex]
    \caption{Maps of the second-best model: (top) number of clouds and (middle) probability for the second-best model to be the model that minimizes the loss of information against the best model. The bottom panel combines the information from the two upper panels. It indicates the number of clouds (color) whose opacity is given by the probability. Transparent pixels correspond to low probabilities for  second-best models.
    }
    \label{fig:2ndBest}
\end{figure}

In the top panel of Fig.~\ref{fig:2ndBest}, we show the map of the number of clouds per LOS corresponding to the second-best model. In the bottom panel of the same figure, we show the map of the second-best model probabilities. In these maps, white pixels (enclosed in the magenta outline) indicate that no second-best model can be identified. That is, there is no ``valid'' reconstruction with a different number of clouds along the LOS. In practice, this happens for several sightlines where the best model is a 2-layer model, and for which the 3-layer and 4-layer models led to bad posterior distributions (see Sect.~\ref{sec:Step1}). This means that either there is no additional cloud farther away than the second cloud, or the data is not good enough to make it possible to detect it solely based on stellar data. We recall that when a 2-layer model is selected as the best model in Step~1, the 1-layer model is not considered in Step~2.

The comparison of the number of clouds per LOS from the best model (Fig.~\ref{fig:Ncloud_ConicalBeam}) and from the second-best model (top panel of Fig.~\ref{fig:2ndBest}), taking into account its probability, suggests that for most of the sightlines where the 3-layer (2-layer) model is favored by the data, we cannot safely ignore the 2-layer (3-layer) model.
Assuming the presence of faraway clouds in the eastern (left) part of the region, as motivated by \HI\ data, this indicates that the data is not enough to allow for a strong and robust cloud detection at large distances, suggesting that we reach the limit of the cloud-detection capability given the data. In the western (right) part of the region, the probability for the 2-layer model is generally high for the 3-layer sightlines, suggesting either unreliable (spurious) detection of faraway cloud or, again, that we reach the limit of the cloud-detection capability.

The cloud-detection capability is primarily limited by the number of measurements and their uncertainties as compared to the strength of the polarization signal induced by a cloud to its background stars (see Sect.~4 in \citealt{Pelgrims2023}). This shows that the reliability of our tomographic results at large distances is limited by the current depth of our survey and that more polarization data is needed to enable a stronger model separation. 
However, given the complementary nature of stellar extinction, \HI\ and starlight polarization data, which we also highlighted above, there is the possibility to incorporate such external data in the inversion process.
In principle this would help differentiate between competing models, even in the absence of additional stellar polarization data. We will undertake this endeavor in future work.

The depth of the survey, and therefore the number density of the measurements, also sets a limit on the maximum angular resolution at which we can achieve the LOS inversion of starlight polarization data. The angular resolution is directly linked to one of the main limitations of using \texttt{BISP-1}. The method does not explicitly take into account POS variations of the polarization signal within the beam other than through the intrinsic scatter term.
We use a fixed geometry to define our beam and consider a top-hat acceptance window to include stars in our sample. As a result, while modeling the ISM along distance, stars at the edge of the beam (on the POS) contribute the same as stars at the center of the beam. This is fine as long as the polarization signal does not vary much on the POS and can be described by our model. However, artifacts (such as overestimation of the intrinsic-scatter covariance matrix and biases in the mean polarization properties) may be expected as soon as the polarization data cannot be well described by a bivariate normal distribution. This may happen in the presence of substantial POS variations at the angular scale comparable to the size of our beam or if only part of the beam intersects a cloud. Such shortcomings in polarization data modeling can naturally hamper cloud detection.

This limitation likely explains differences between the starlight-polarization and \HI\ tomographic data at large distances seen in Fig.~\ref{fig:2ndBest}, in particular toward the prominent IVC. Specifically, \HI\ data suggests substantial variation of the signal in our beam of 13.76 arcmin radius, and in particular, an abrupt change of the POS component of the magnetic field toward the ``2-cloud region'' of (\citealt{Pan2019a}).
We notice that part of these limitations could be overcome by introducing weights on the polarization data while computing the log-likelihood to account for the angular distance of each star with respect to the center of the beam, that is, of the LOS. We will test this idea in future work.

\subsection{Caveats of the inversion method}
The main caveats of our approach to obtain a 3D map of the POS component of the magnetic field in dusty regions come from the use of our Bayesian method (\texttt{BISP-1}) which works along the LOS.
This method requires the definition of beams within which the starlight polarization data is ``averaged'' on the POS and modeled as a bivariate normal distribution with a mean and covariance matrix according to the dust-layer model we rely on.

The first limitation of the designed approach is that it uses a fixed beam geometry and does not explicitly correlate solutions for different beams. We address this problem by choosing to oversample the sky with non-independent beam samples and by applying our decomposition method to each of them.
However, while the solutions for overlapping beam samples are not independent, the correlation is not quantified nor controlled through, for example, the use of density power spectrum as it is the case for 3D dust density mapping (e.g., \citealt{Green2019}; \citealt{Lallement2019}; \citealt{Leike2019}; \citealt{Vergely2022};  \citealt{Lallement2022}; \citealt{Edenhofer2023}).
Introducing such a correlation for polarization tomography is not a trivial task and would require assumptions that we wish to avoid at this stage, in particular because of the intertwined nature of matter and magnetic field, and of the inherent degeneracy between the density of polarizing dust and the inclination of magnetic field with respect to the LOS. A careful analysis of our tomography results in combination with simulation-based studies will help shed light on how to implement such correlations for the case of dust polarization. We will address this question in future work.

A second, important limitation of \texttt{BISP-1} is that it does not explicitly account for POS variations of the polarization signal. For the diffuse sightlines targeted in this work, significant POS variations within the beam do not seem to be present, as our modeling provides a good description of the data.
As mentioned in Sect.~\ref{sec:validation}, we searched for possible systematic variations of the polarization residuals within our beam and could not find any. This suggests that any POS variation of the polarization signal in our beam are successfully characterized by the intrinsic-scatter covariance matrix.
However, we expect more significant POS variations to arise toward denser regions, for example toward nearby molecular clouds, where the column density varies by an order of magnitude within tens of arcseconds (e.g., filaments measured by Herschel, \citealt{Andre2010}). Some nearby molecular clouds have been targeted with deep polarimetric surveys (e.g., \citealt{Pereyra2004}; \citealt{Santos2017}), approaching a density of 1000 measurements per square degree.
In such regions, a careful examination of the choice of beam size and the POS variations of column density would be necessary when using \texttt{BISP-1} to decompose polarization along the LOS, as envisioned by e.g. \cite{Soler2016}.
Given the existing stellar polarization data in the literature, it appears that the dataset presented here is particularly favorable for applying \texttt{BISP-1} in a moving-window scan scheme due to its combination of high number density of stars and the fact that it probes diffuse sightlines, conditions that will be encountered by the \textsc{Pasiphae} survey targeting high and intermediate Galactic latitudes.

\subsection{Astrophysical use of the output}
\label{sec:IntrPolStar}
Mapping continuously the stellar-polarization source field in 3D opens the way to tackle several science objectives that were thus far out of reach or left to the study of specific clouds or LOS.
A direct use of our results, which does not require postprocessing of the tomography map, is the production of a list of intrinsically polarized-star candidates. Other possible uses, which will however require further postprocessing and specific analysis, are mentioned in Sect.~\ref{sec:conclusion}.

\begin{figure}
    \centering
    \includegraphics[trim={0.4cm 0.3cm 0.3cm 0.2cm},clip,width=.98\columnwidth]{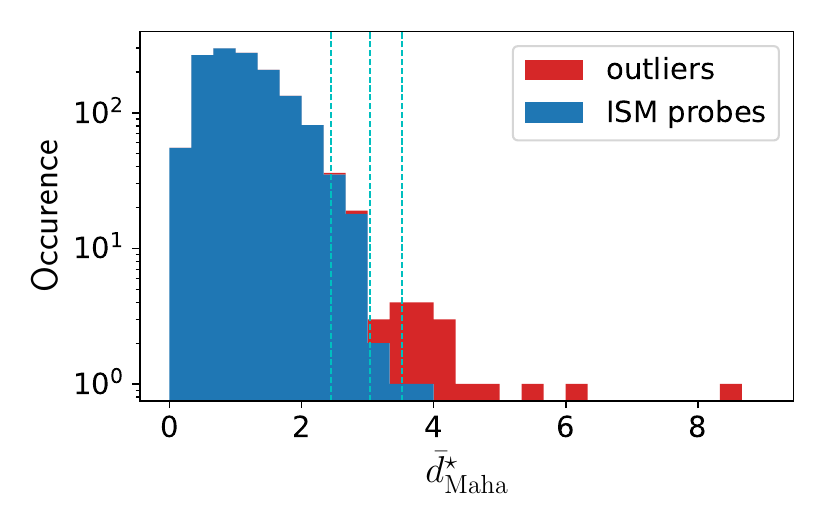}\\[-1.5ex]
    \caption{Histogram of the median of the per star $d_{\rm{Maha}}^\star$ distributions. The histograms for the ``ISM probes'' and ``outliers'' identified in Sect.~\ref{sec:outliers} are separated (in blue and red) and stacked on top of each other. Notice the logarithmic scale of the vertical axis. The dashed vertical lines correspond to the $p$-value thresholds of 5\%, 1\%, and 0.2\%. Most of the outliers show significant residuals.}
    \label{fig:ResidualsAndOutliers}
\end{figure}

Using the same procedure as in Sect.~\ref{sec:validation}, we proceed to the estimation of the significance of the residuals in polarization for all the stars making our polarization samples, with successful {\it Gaia} cross-match and which satisfy the quality criterion on parallax estimate (${\rm{RUWE}} \leq 1.4$, see Sect.~\ref{sec:Data_fullSample}). This sample is made of 1448 stars among which 18 were identified as possible outliers according to the recursive sigma-clipping approach employed in Sect.~\ref{sec:outliers}. Among the 1448 stars, 1392 fall in an HEALPix pixel for which we have a model for the magnetized ISM along distance.
For each of them we obtain a distribution of the Mahalanobis distance values ($d_{\rm{Maha}}^\star$) informing us on the likelihood that the measured polarization is compatible with our picture of the dusty magnetized ISM, taking into account all sources of uncertainties and scatter in model and observations.
We show the histogram of the median of the $d_{\rm{Maha}}^\star$ distributions for the full sample in Fig.~\ref{fig:ResidualsAndOutliers} where we separate the stars flagged as outliers in Sect.~\ref{sec:outliers} from the others. This figure confirms that most of the outliers discarded from the tomography analysis indeed show polarization properties that are not compatible with the picture of the dusty magnetized ISM that we reconstructed. In some sense, this also further validates our 3D reconstruction.

Using our tomography map to estimate the likelihood that the polarization of any star falling in the reconstructed 3D volume is solely due to the dusty magnetized ISM provides us with a robust way to identify outlier candidates and therefore intrinsically polarized-star candidates. We thus add the values of the median of the $d_{\rm{Maha}}^\star$ distribution that we obtained for each star to our published catalog (see Table~\ref{tab:polcat}).
The higher this value, the more likely the target is to be an outlier.
This may be used to plan follow-up observations to study these sources in more detail.
Only additional study will confirm whether these targets are real outlying data points or if they merely pick up fluctuations of the magnetized ISM that are unaccounted for in our reconstruction.
Our analysis also confirms that the fraction of intrinsically polarized stars in the ISM is rather low, at least at these Galactic latitudes. Only 14 stars out of 1448 have a $p$-value lower than 0.2\% for their polarization to be induced by the ISM only. That is, only about 1\% of our sample may be made of intrinsically polarized stars whereas we did not try to minimize this fraction while planning our observation, for example based on stellar type.

We must note a possible intrinsic degeneracy in our procedure.
As mentioned in Sect.~\ref{sec:outliers}, during the sigma-clipping procedure, it is possible to inadvertently exclude from the analysis data points that merely capture fluctuations in the magnetized ISM.
In this case, a too low level of intrinsic scatter would be obtained from the fit and, accordingly, any discarded points would show a large $d_{\rm{Maha}}^\star$ value in the a posteriori test described above.
The list of outlier candidates thus depends on the choice of the hyper parameters of the sigma-clipping procedure, and more specifically on the choice of the used significance threshold. This illustrates the relevance of follow-up observations and analysis of the outlier candidates.
However, if the turbulence-induced intrinsic scatter had to be systematically underestimated in our reconstruction, we would obtain a distribution of $d_{\rm{Maha}}^\star$ values that would be statistically shifted toward large values and that would not correspond to a bivariate normal distribution of the polarization residuals. This is not what we obtain and what is also shown in Fig.~\ref{fig:ResidualsAndOutliers}. On the contrary, we obtain a distribution of $d_{\rm{Maha}}^\star$ values that is slightly shifted to lower values than expected and which, therefore, suggests a small overestimation of the covariance matrices. We understand the latter as coming from unaccounted for variation of the polarization signal in the POS within our beam.

\section{Summary and concluding remarks}
\label{sec:conclusion}
In this work, we performed a survey of optical starlight polarization for a continuous region covering about four square degrees centered on $(l,\,b) = (122^\circ,\,33^\circ)$, and designed a pipeline to obtain the first 3D map of the dusty magnetized ISM based on a Bayesian analysis of the measurements in stellar polarization and distances only.
Obtaining this map, which corresponds to an extended volume in 3D space, is the main result of this paper.
Our reconstruction corresponds to a sky area of about 3.8 square degrees and extends up to 3~kpc from the Sun.
We found that the 3D volume covered by our tomography data is populated by several clouds and that a large fraction of the sightlines in the surveyed regions intersect at least two clouds, one being very close to the Sun with a distance of about 62~pc and a dominant polarizing screen at about 375~pc. Distant clouds are also detected up to a distance of about 2~kpc. We were able to corroborate our findings using a 3D dust extinction map and \HI-velocity spectra. We are thus confident that our inversion pipeline works and  that stellar data in polarization and distance alone are a powerful probe of the magnetized ISM.
Specifically, for the diffuse ISM where dust grains are expected to align their shortest axes with the ambient magnetic field lines, starlight polarization allows us to determine locally, in 3D space, the orientation of the POS component of the magnetic field (the position angle) and the amplitude of the starlight-polarization source field (the degree of polarization) which depends on the local dust density, dust grain polarization efficiency, and on the inclination of the magnetic field lines with respect to the sightlines.

We obtained our polarization tomography map by adopting a moving-window strategy to scan the surveyed region with non-independent beams and making use of the LOS-inversion method implemented in \texttt{BISP-1}. This allowed us to invert the data and reconstruct the dusty magnetized ISM in 3D while keeping the number of assumptions to its minimum. Namely, we relied on the thin-dust layer model developed in \cite{Pelgrims2023} and assumed that its assumptions are valid for the adopted beam size of 13.74 arcmin radius.
We expect that the analysis of the obtained 3D map will enable the unbiased study and characterization of the properties of the dusty magnetized ISM in 3D, such as through 3D correlation functions. This will enable us in the future to develop 3D inversion methods capable of taking into account variations and correlations in space, both along the distance and in the POS.

Finally, we expect that polarization tomography maps, as the one we obtained in this work and that provides local measurements of the POS component of the magnetic field and individual cloud polarization properties, will enable breakthroughs in the modeling of the Galactic magnetic field, in the modeling and characterization of the dusty magnetized ISM as a contaminant foreground in observations of the cosmic microwave background polarization, and in the modeling of astrophysical dust.
These research goals, along with the estimation of the strength of the magnetic field through the quantification of the variation of the POS component of the magnetic field (e.g., \citealt{Skalidis2021}; \citealt{Skalidis2021b}), will necessitate dedicated analyses and postprocessing of our 3D maps, such as to obtain proper boundaries of dust clouds. In future works we will explore these research directions based on the 3D map that we have presented here. This will set the stage for future analyses that will greatly benefit from the most awaited polarization data from the \textsc{Pasiphae} survey (\citealt{Tassis2018}).

\begin{acknowledgements}
We thank an anonymous referee for a thorough and constructive report that helped us improve the clarity of the paper.
The PASIPHAE program is supported by grants from the European Research Council (ERC) under grant agreements No. 7712821 and No. 772253; by the National Science Foundation (NSF) award AST-2109127; by the National Research Foundation of South Africa under the National Equipment Programme; by the Stavros Niarchos Foundation under grant PASIPHAE; and by the Infosys Foundation. This work was also partly supported by the ERC grant agreements No. 819478.
VP acknowledges funding from a Marie Curie Action of the European Union (grant agreement No. 101107047).
VP also thanks Philipp Frank and Sebastian Hutschenreuter for the fruitful and inspiring discussions related to this work during the program "Toward a Comprehensive Model of the Galactic Magnetic Field" at Nordita in April 2023, which was partly supported by NordForsk and Royal Astronomical Society.
VPa acknowledges support by the Hellenic Foundation for Research and Innovation (H.F.R.I.) under the “First Call for H.F.R.I. Research Projects to support Faculty members and Researchers and the procurement of high-cost research equipment grant” (Project 1552 CIRCE). VPa also acknowledges support from the Foundation of Research and Technology - Hellas Synergy Grants Program through project MagMASim, jointly implemented by the Institute of Astrophysics and the Institute of Applied and Computational Mathematics.
KT and AP acknowledge support from the Foundation of Research and Technology - Hellas Synergy Grants Program through project POLAR, jointly implemented by the Institute of Astrophysics and the Institute of Computer Science.
TG is grateful to the Inter-University Centre for Astronomy and Astrophysics (IUCAA), Pune, India for providing the Associateship programme under which a part of this work was carried out.
This work was supported by NSF grant AST-2109127 and was carried out in part at the Jet Propulsion Laboratory, California Institute of Technology, under a contract with the National Aeronautics and Space Administration.
This research has used data, tools or materials developed as part of the EXPLORE project that has received funding from the European Union’s Horizon 2020 research and innovation programme under grant agreement No. 101004214.
\end{acknowledgements}

\bibliographystyle{aa}
\bibliography{myBiblio}

\end{document}